\documentclass[showpacs,preprintnumbers,amsmath,amssymb]{revtex4-2}
\usepackage{graphicx}
\usepackage{dcolumn}
\usepackage{bm}

\begin{document}

\title{
Mirror-extended standard model
 with spontaneously broken left-right symmetry
\\
and its implementation in the course of cosmological evolution
}

\author
{ Alexander  B. Kaganovich}
\thanks{alexk@bgu.ac.il}
\address{Physics Department, Ben Gurion University of the Negev \\
and Sami Shamoon College of Engineering,\\ 
Beer
Sheva 84105, Israel}

\date{\today}

\begin{abstract}
A mirror-extended standard model (MESM) is offered, where in the left-right symmetric underlying action the sector of the standard model (SM) and its mirror copy have the same $SU(2)\times U(1)$ gauge structure and parameters; the mirror fermion counter-partners have opposite chiralities. A theory is used that allows one to obtain  MESM in  Minkowski space only if we start with an accurate account of gravity and only at the end go to the limit of  Minkowski space. Spontaneous breaking of left-right symmetry and the "wrong" sign in the mass terms of the Higgs fields potentials arise due to the initial conditions imposed on the T-model inflation. MESM allows choosing the universal Yukawa coupling constant $y$ in the underlying Lagrangian for all generations of charged leptons and up-quarks. As an example, with $y\approx10^{-3}$, it is shown that a set of additional dimensionless parameters in  width range less than 0.5 is sufficient to obtain the masses of all known charged leptons and up-quarks. In this sense, MESM bypasses the problem of the fermion mass hierarchy. With this choice of parameters, MESM predicts that  vacuum expectation value for the mirror Higgs field is $10^{14} GeV$ and masses of mirror particles are in the range $10^{8}GeV - 10^{14}GeV$. There is only gravitational interaction between particles of the SM and mirror sectors. Hence, mirror particles can constitute dark matter. The vacuum is realized as a limiting state in the cosmological evolution of  the inflaton and classical Higgs fields, in which the energy stored in them becomes minimal, providing the maximum possible contribution of these fields to  the entropy of the Universe. If further study of the model reveals the existence of an effective mechanism for preheating the Universe, this will mean that the cosmological constant problem  is absent.  Quantum corrections to MESM cannot interfere with the implementation of the new concept of vacuum proposed in this paper.

\end{abstract}


\maketitle

\section{Introduction}

It is quite expected that some problems of inflationary cosmology and mysteries of dark energy and dark matter (DM), 
on the one hand, and the unanswered questions  of the standard model of elementary particles (SM), on the other. 
can be interrelated. For adequate answers to these challenges, apparently, a theory is required  in which gravity
 would be an integral part not only at the macroscopic level, as in Einstein's general relativity (GR), 
but also at the microscopic level. This idea, as well as an idea of supersymmetry (SUSY)  have been implemented 
in supergravity and superstrings (quantum gravity). However, the question of whether supersymmetric theories
 are realistic remains unanswered\cite{Rappoccio}.

In this paper, I want to draw attention to the fact that between the SM (which can be formulated in a curved space-time),
 on the one hand, and the SUSY theories, on the other hand, there is a  lacuna that has not been noticed until now.
Namely, in the search for what is behind SM, before moving on to such a radical modification of field theory, we must 
make full use of the previously ignored property of space-time manifold, which, as shown in this paper, can be of 
fundamental importance in field theory. I mean the orientability of the space-time manifold and the 4-form used to define it. 
Since the orientation of the space-time manifold is directly related to the spatial reflection, the volume 4-form turns out to 
be an effective tool for implementing the idea of expanding  SM in such a way that it acquires  left-right symmetry.
As it will be shown in this paper, spontaneous breaking (SB) of the left-right symmetry and the $SU(2)\times U(1)$ 
gauge symmetry  have a common origin, which would be impossible without gravity. And although in local field 
interactions the gravitational field is present only as a background field, the proposed modification of the approach 
to constructing  models of field theory  really makes gravity, in a sense, an integral part of the theory.

Already in the original paper\cite{LeeYang}, where experiments were proposed to test parity conservation in weak decays,
 the idea of the existence of hypothetical mirror symmetric proton was put forward, due to which, in a broader sense,
 there is left-right symmetry.  The contradiction of the observed parity violation to our ideas about the perfection
 of the world has been a motivation for numerous attempts to retain left-right symmetry by adding mirror particles. 
The development of the idea of a mirror world began in the pioneering paper\cite{KOkP} and has a long history,
 the drama of which is  described in detail in the review paper\cite{Okun}. This paper also contains an exhaustive
 list of literature on this topic.

The lack of any evidence to support the existence of mirror particles means that they can  interact with visible matter
 only through gravity or some very weak interactions. Hence, mirror particles constitute a hidden sector and may be 
suitable candidates for DM\cite{Zeldovich}-\cite{Khlopov 2}. For a list of publications on this topic, see, for example, 
the reviews\cite{Dolgov Neutrino}-\cite{foot 2014}.

When studying possible extensions of the SM particle sector by adding a mirror sector, the question arises of what 
mirror symmetry should be and how it can be broken. The first question has been considered in Refs.\cite{Pavsic} 
and \cite{Foot Volkas}.  As for the second question, there are several models in the literature in which left-right 
symmetry is broken spontaneously. In the series of works\cite{Foot 1}-\cite{Foot 4}, cosmological and astrophysical implications were studied
in  models where  the Lagrangian includes mixing of the mirror counter-partners of the gauge fields and/or the Higgs fields, so that all symmetries of the theory are preserved. In the case of the Higgs fields mixing, depending on the parameter space,  the left-right symmetry  remains exact or  is broken spontaneously. 
Another mechanism of SB of left-right symmetry was realized in papers \cite{Mohapatra 2}-\cite{Mohapatra 4}
 where an additional pseudoscalar field is incorporated, so that all symmetries of the theory are also preserved. In these models, 
the left-right symmetry is spontaneously broken when  this pseudoscalar develops a nonzero vacuum expectation value (VEV).

In this paper, I propose a mirror-extended  SM (MESM), where the mirror sector consists of mirror partners of all SM 
particles with quantum numbers identical to the quantum numbers of the SM particles; fermions that are mirror 
counter-partners have opposite chiralities.
Using  the orientation-related attribute of a spacetime manifold  mentioned above, the  requirement of  left-right symmetry
  is imposed on the Lagrangian density of the MESM which does not contain any additional dynamical degree of freedom or  interaction term.
SB of  left-right symmetry occurs due to the scalar, which has a geometric nature and does not carry  a dynamical degree of freedom.
To simplify the presentation of the basic ideas of the theory in this paper, I restrict myself to the electroweak 
 $SU(2)\times U(1)$ gauge-invariant model. 

The organization of the paper is as follows. Section II presents the reasons for using two volume measures  in the action principle. Particular attention is paid to how this makes it possible to construct a Lagrangian density for particle fields with left-right symmetry, and also to the fact that in such a theory parity should be considered in combination with the orientation of the space-time manifold.
 Using a toy model that includes gravity and a scalar field as an example, in Sec.III we explore the general features of TMT. From what we learned in Sec.III, it becomes clear that the presence of two volume elements in the underlying action dictates the need to take into account gravity without assuming its weakness. Consequently, MESM, like the values of all parameters in the modern cosmological epoch, can only be discovered as a result of a detailed study of cosmological evolution, starting with inflation. The need to describe all aspects of the proposed model has led to a significant increase in the volume of the paper. So that at each stage of reading the proposed paper the general idea is not lost, I was forced to divide its main content into two parts. The first part (Secs.IV, V, VI) describes the MESM as a model of particle physics beyond the SM. This takes into account some of the results that will be obtained in the second part (Secs.VII, VIII),  which studies cosmological evolution and related effects, such as SB of model symmetries, the formation of a vacuum, etc. In more detail, in Secs.IV and V, the bosonic and fermionic sectors of the MESM are constructed. In Sec.VI,  MESM in Minkowski space is studied as the theory beyond the SM with predictions of possible  masses and coupling constants for mirror particles. In Sec.VII, the inflaton field is added so as to provide  T-model  inflation and we are conducting a preliminary study of a new approach to understanding what a vacuum state is. Sec.VIII is devoted to the study of the cosmological evolution of the Higgs fields, the result of which is the understanding that vacuum is a state, the transition to which is accompanied by the maximum possible increase in the entropy of the Universe by the time the vacuum is reached.
 In Sec.IX the most important results are summarized and some of open questions are discussed. The purpose of Appendix   is to help the reader better understand how the  volume 4-form allows one to construct the Lagrangian density for fermions with left-right symmetry. Therefore, I would recommend reading this appendix before reading Sec.V.

\section{Motivation for TMT and symmetries of MESM}

\subsection{Two volume elements and orientation of the space-time manifold}

It is well known\cite{manifolds}, \cite{Hawking} that the construction of the space-time manifold is carried out step by step by  equipping the corresponding topological space with a number of additional structures:
1) a differential structure that turns a topological space into a differentiable manifold;  
2) an affine connection structure (with affine connection coefficients $\Gamma^{\mu}_{\alpha\beta}$);
3) a metric structure (with a metric tensor $g_{\alpha\beta}$). Note that generically the affine and metric structures  are independent of each other.  
In my opinion, the generally accepted approach to the application of this branch of mathematics in field theory ignores some important aspects of the construction procedure described above, which, apparently, leads to the loss of very important physical consequences. This statement requires argumentation, to which I immediately turn.

Each field theory model  in curved space-time  implies the use of the action principle where a volume element (the measure of integration in the action) has to be chosen. 
The standard choice for the volume element on orientable 4D space-time manifold is $dV_g=\sqrt{-g}d^{4}x$. But it is well known (see e.g. \cite{manifolds},\cite{Hawking}) that already at the first of the three steps described above, using the differential form, one can define an alternative, metric independent volume element. 
 In the series of papers  with E. Guendelman (see e.g. \cite{GK1}-\cite{GK6}) we have studied  some consequences of using  an alternative, metric independent volume element $dV_\Upsilon=\Upsilon d^4x$, for example defined by a 4-form with the help of  4 scalar functions $\varphi_a$  ($a=1,..,4$) on the space-time manifold
\begin{eqnarray}
dV_{\Upsilon}=\Upsilon d^4x\equiv
\varepsilon^{\mu\nu\gamma\beta}\varepsilon_{abcd}\partial_{\mu}\varphi_{a}
\partial_{\nu}\varphi_{b}\partial_{\gamma}\varphi_{c}
\partial_{\beta}\varphi_{d}d^4x
\nonumber\\
=4!d\varphi_1\wedge d\varphi_2\wedge d\varphi_3\wedge d\varphi_4
\label{Phi}
\end{eqnarray}
Here $\Upsilon$ is  a scalar density, that is under general coordinate transformations with positive Jacobian it has the same transformation law as $\sqrt{-g}$.
This is usually considered a good enough reason to only use the  volume element $dV_g$ and ignore $dV_\Upsilon$. 

The use of the  volume element $dV_{\Upsilon}$ in the action integral along with  $dV_g$ means that we are expanding the arena on which the field theory is built.  If an action contains both measures of integration $dV_g$ and $dV_{\Upsilon}$, we call such a theory the Two Measures Theory (TMT).
There are {\bf two fundamental differences} between the
 two measures of integration:
\begin{itemize}
\item
I. \, The first difference  is that in the context of the action principle, contributions of $\sqrt{-g}$ and $\Upsilon$ to the variation of the action have  different dynamical effects.
 The general coordinate-invariant TMT action can be represented in the following general form
\begin{equation}
    S = \int \left(L_{1}\sqrt{-g}+L_{2}\Upsilon\right)d^{4}x
\label{S2}
\end{equation}
 where two Lagrangian densities $\mathcal{L}_1=L_{1}\sqrt{-g}$  and  $\mathcal{L}_{2}=L_2\Upsilon$ present. 
As a preliminary illustration of what this change in the form of the action leads to, let's write down the result of the variation of the action (2) with respect to
 scalar functions $\varphi_{a}$ of which $\Upsilon$ is built:
\begin{equation}
B^{\mu}_{a}\partial_{\mu}L_{2}=0 \quad \textsf{where} \quad
B^{\mu}_{a}=\varepsilon^{\mu\nu\alpha\beta}\varepsilon_{abcd}
\partial_{\nu}\varphi_{b}\partial_{\alpha}\varphi_{c}
\partial_{\beta}\varphi_{d}.
\label{varphiB}
\end{equation}
Since $Det (B^{\mu}_{a}) = \frac{4^{-4}}{4!}\Upsilon^{3}$ it follows
that if
 \begin{equation}
\quad \textsf{everywhere} \quad \Upsilon\neq 0, 
\label{Phi neq 0}
\end{equation}
the equality
\begin{equation}
 L_{2}=sM^{4} =const.
\label{varphi}
\end{equation} 
must be satisfied, where $s=\pm 1$ and $M$ is a constant of
integration with the dimension of mass.
It is assumed that, along with matter fields, $L_{1}$ and $L_2$ involve the scalar curvature in the form $\propto (16\pi G)^{-1}R$.   Following a rigorous mathematical approach to constructing a space-time manifold, of which the three steps listed above are a necessary part, the affine connection and the metric should be treated as  independent dynamical variables. Then variation of $g^{\mu\nu}$ gives 
\begin{equation}
\zeta\frac{\partial L_2}{\partial g^{\mu\nu}}+\frac{\partial
L_1}{\partial g^{\mu\nu}}-\frac{1}{2}g_{\mu\nu}L_1 =0,
\label{g-mu-nu-varying}
\end{equation}
where
\begin{equation}
 \zeta(x)\stackrel{\mathrm{def}}{=} \frac{dV_{\Upsilon}}{dV_g}\equiv\frac{\Upsilon}{\sqrt{-g}}   
 \label{zeta}
\end{equation}
 is the scalar field 
build of the scalar densities $\Upsilon$ \textsf{and} $\sqrt{-g}$. As we will see later,  the consistency of Eqs.(\ref{varphi}) and (\ref{g-mu-nu-varying}) has a form of {\em the constraint} which determines $\zeta(x)$ as a local function of matter fields.

\item

II. \, The second difference is that the volume elements  $dV_g$ and $dV_{\Upsilon}$,  being scalars with respect to general coordinate transformations with positive Jacobian, have opposite transformation law under space reflection
 (parity). 
Here it is  important to pay attention to the distinction from spatial reflection in field theory in Minkowski space. Generally speaking, it is impossible to introduce a global coordinate system on a manifold.
In order to apply the term "parity" in the context of manifold one should note that in an orientable space-time manifold there is an oriented atlas (a collection of local coordinates (charts) $x^{\mu}=(x^0,x^i)$, $i=1,2,3$).  Then by reflection $x^i\rightarrow -x^i$, $i=1,2,3$ in all charts of the oriented atlas we change both the space orientation  and the space-time orientation to the opposite (the orientation inversion). With this understanding of the space reflection operation ${\mathcal P}$ and taking into account that  $\varphi_a$ are {\underline scalars}, we conclude that $dV_g$ and $dV_{\Upsilon}$ have opposite parity since  $\hat{\mathcal P}\sqrt{-g(x)}=\sqrt{-g(x)}$ while 
\begin{equation}
 \hat{\mathcal P}\Upsilon(x)=-\Upsilon(x) \qquad \textsf{and} \qquad \hat{\mathcal P}\zeta(x)=-\zeta(x).
 \label{Pzeta}
\end{equation}
Therefore, {\bf in TMT,  the standard action of the ${\mathcal P}$-reflection on the  field theory operators has to be accompanied by
 the change of the sign of $\zeta$},  which actually means the orientation inversion.

\end{itemize}

Already at this initial stage of acquaintance with theory, it is appropriate to note the following significant differences of
TMT from scalar-tensor theories with nonminimal coupling:
\begin{itemize}
\item 
In general, the Lagrangian  $L_{2}$ in Eq.(\ref{S2}) may contain not only the scalar curvature term (or more
general gravity terms) but also all possible matter field terms.
This means that TMT {\it can modify in general both the
gravitational sector  and the matter sector in the similar fashion};

\item If $\Upsilon$ were the
fundamental (non composite) one then the variation of $\Upsilon$ would
result in the equation $L_{2}=0$ instead of (\ref{varphi}), and
therefore the dimensionful parameter $M$ would not appear.

\end{itemize}

For a better understanding of the special role that the scalar density $\Upsilon$ and the (pseudo) scalar $\zeta$ have in TMT, it is necessary to pay attention to the following three very important points:

1) Due to the structure of the scalar density $\Upsilon$, Eq.(\ref{Phi}), variation of scalar functions $\varphi_a(x)$ does not lead to a dynamical
equation for $\varphi_a(x)$. Thus, {\em functions $\varphi_a(x)$ are not dynamical variables}, but are auxiliary fields in relation to other fields of the model.
 All their dynamical effect
is displayed only in the following two ways: a) in generating the
constraint that determines $\zeta(x)$ as a local function of matter fields;  b) in possible
appearance of the scalar  $\zeta(x)$ and its gradient in
equations of motion.

2) The existence of the volume element $dV_{\Upsilon}$ satisfying the condition (\ref{Phi neq 0})  means that  {\bf  the space-time manifold is orientable}. Assuming in what follows orientability of  the space-time manifold we conclude that the matter field dynamics has to provide that the scalar function $ \zeta(x)$ do not vanish at any space-time point. Therefore $ \zeta(x)$ and $\Upsilon(x)$ are positive everywhere or both of them are negative everywhere, depending on the choice of the integration constant $sM^4$ and  the initial conditions. Accordingly, the space-time manifold get a positive or negative orientation. Originally, on the level of the action (\ref{S2}), none of the two possible orientations has preference over the other, both from a mathematical and physical point of view. But due to  the constraint which follows from equations of motion,  the sign of $\zeta$  and hence {\bf the orientation of the space-time manifold is spontaneously fixed} by choosing the integration constant in Eq.(\ref{varphi}) and initial conditions imposed on matter fields.

3) As we will see in the next section, the function $\zeta(x)=const$ and therefore the volume elements $dV_g$ and $dV_{\Upsilon}$ differ  by a "nonessential" constant factor $\zeta$ {\bf only} {\em when matter fields are in a vacuum state} or if the matter is a free massless scalar field with a fine tuned parameter in the action.  In particular, {\bf in the empty space-time the volume elements $dV_g$ and $dV_{\Upsilon}$ coincide.}

\subsection{Role of orientation inversion in left-right symmetry of  MESM}

We are going to build a model beyond the SM, in which a) there are two copies of  particles with identical quantum numbers of the $SU(2)\times U(1)$ gauge symmetry; b) the model possesses  a left-right symmetry. 
To implement the left-right symmetry in MESM, 
the spatial reflection operation described above, must be supplemented by the interchange of the mirror counter-partners
\begin{equation}
X \Longleftrightarrow \check{X}, 
 \label{mirror reflection}
\end{equation}
where we used the notation $\check{X}$ for the MESM particle, which is the mirror partner of the SM particle $X$. In the paper\cite{KOkP} it was called $A$-symmetry ($A$ from Alice, see also\cite{Okun}).
Thus, {\it we will require MESM to be invariant  under} $\textsf{{\bf (mirror}}\times\textsf{{\bf   parity)}}$ {\it transformations}  which consists of the usual  ${\mathcal P}$-reflection in field theory and the  transformations (\ref{Pzeta}) and (\ref{mirror reflection}). 

A note should be made here of how the $\textsf{(mirror}\times\textsf{parity)}$ symmetry works for chiral fermions, that is, for example, in the case of the mirror counter-partners $X=\Psi_L$ and $\check{X}=\check{\Psi}_R$ having opposite chirality. If one takes into account the action of the operator $\hat{\mathcal P}$ on the chiral states $\hat{\mathcal P}\Psi_L=\Psi_R$, $\hat{\mathcal P}\Psi_R=\Psi_L$, $\hat{\mathcal P}\check{\Psi}_R=\check{\Psi}_L$ and $\hat{\mathcal P}\check{\Psi}_L=\check{\Psi}_R$ then  the mirror reflection (\ref{mirror reflection}) applied to chiral fermions  is the result of a sequential execution of the operations $\hat{\mathcal P}$ and
$\Psi_L\Longleftrightarrow\check{\Psi}_L$ , \, $\Psi_R\Longleftrightarrow\check{\Psi}_R$.

Let us now turn to discribing the general prescription how one can implement MESM with spontaneously broken  left-right symmetry.  To avoid unnecessary complications, consider a couple of terms $\Delta L(X)$ and $\Delta L(\check{X})$  describing contributions of the mirror symmetric fields $X$ and $\check{X}$ to the underlying MESM   Lagrangians.
Here and below, it is assumed that the structures of $\Delta L(X)$ and $\Delta L(\check{X})$
 dictated by SM are identical, that is $\Delta L(X)\Longleftrightarrow\Delta L(\check{X})$ as $X\Longleftrightarrow\check{X}$.
Generically, $\Delta L(X)$ can present in the  action in the form  $\int\left(a\sqrt{-g} +b\Upsilon\right)\Delta L(X)d^{4}x$ where $a$, $b$ are real parameters. If $\Delta L(\check{X})$ enters in the action in the form $\int\left(a\sqrt{-g} -b\Upsilon\right)\Delta L(\check{X})d^{4}x$, then the part of {\it the Lagrangian density $\Delta\mathcal{L}$} defined by the equation 
\begin{equation}
\Delta S=\int d^4x\Big[\left(a\sqrt{-g} +b\Upsilon\right)\Delta L(X)+\left(a\sqrt{-g} -b\Upsilon\right)\Delta L(\check{X})\Big]=\int \Delta\mathcal{L}d^4x.
 \label{general structure l-r symm}
\end{equation}
{\it is invariant under} $\textsf{{\it (mirror}}\times\textsf{{\it   parity)}}$ {\it transformations}. 
The Lagrangian density $\Delta\mathcal{L}$  can be represented in the form
\begin{equation}
\Delta\mathcal{L}=\sqrt{-g}\Delta L(X, \check{X}; \zeta);  \quad \Delta L(X, \check{X}; \zeta)=\left(a +b\zeta\right)\Delta L(X)+\left(a -b\zeta\right)\Delta L(\check{X}),
 \label{Delta L density}
\end{equation}
where $\Delta L(X, \check{X}; \zeta)$ is  {\it the appropriate term of the Lagrangian} which  is {\it also invariant under} $\textsf{{\it (mirror}}\times\textsf{{\it   parity)}}$ {\it transformations}. Thus, regardless of properties of $\Delta L(X)$  and $\Delta L(\check{X})$ with respect to spatial reflection,  the term $\Delta L(X, \check{X}; \zeta)$ in the  Lagrangian  defined by Eq.(\ref{Delta L density}) possesses  left-right symmetry. Therefore, if the underlying MESM Lagrangian is composed of terms of a similar structure, it is  invariant under $\textsf{{\it (mirror}}\times\textsf{{\it   parity)}}$ transformations. This is true as long as $\zeta(x)$ remains arbitrary function. 
 But, as will be clear from what follows, when choosing the initial conditions and solving the field equations, due to constraint, $\zeta$ takes on certain numerical values. This manifests itself as {\it spontaneous  breaking of the} $\textsf{{\it (mirror}}\times\textsf{{\it   parity)}}$  {\it symmetry}.

We are going to compose the underlying MESM Lagrangian in which the terms corresponding to the SM particles and their mirror partners will have a structure of $\zeta$ dependence like in Eq.(\ref{Delta L density}). More exactly, if the terms in the Lagrangian corresponding to the SM particles will contain a set of signs $(\pm)$  before $\zeta$ then the terms in the Lagrangian corresponding to the mirror particles will contain an opposite set of signs, that is  $(\mp)$.
This is the key feature of the underlying MESM Lagrangian,  which ensures (mirror$\times$parity) -invariance
and  allows us to determine {\bf the new quantum number which distinguishes the SM particles and their mirror partners}. 
 Since it is related both to parity and space-time orientation let us call this new quantum number   {\bf "paritation" ($=$ parity $+$ orientation)} and  assign  $paritation = 1$ for  the set $(\pm)$, that is for  SM particles and $paritation = -1$   for  the set $(\mp)$, that is for mirror particles. It is noteworthy that the need for such a quantum number to appear in the construction of a left-right symmetric model was predicted in the original paper\cite{Pavsic}, where the so-called "parameter of internal inversion" plays a role similar to that of paritation.
It is important to emphasize that in the  underlying  MESM Lagrangian, paritation is the only quantum number that distinguishes mirror counter-partners from each other (not counting chirality). 
As a first step in the implementation of the  described idea of constructing a (mirror$\times$parity) invariant fermionic sector, in  Appendix  I consider a toy model where some technical aspects are also discussed.

\section{General view of the structure and features of TMT}

\subsection{A simple toy model}

Before exploring the model that is the subject of this paper, it is necessary to understand how the theory works using a simple model without any connection to MESM symmetries. As an example, consider a toy model including gravity and a scalar field
\begin{equation}
S=S_{gr}+S_{\phi};
\label{S-toy}
\end{equation}  
\begin{equation}
S_{gr}=\int d^4x(\sqrt{-g}+\Upsilon) \left[-\frac{1}{16\pi G}R(\Gamma,g)\right], \qquad\frac{1}{16\pi G}=\frac{M_P^2}{2},
\label{S-gr}
\end{equation}  
\begin{equation}
S_{\phi}=\int d^4x \left[(b_1\sqrt{-g}+\Upsilon)\frac{1}{2}g^{\mu\nu}\phi_{,\mu}\phi_{,\nu}
- (b_2\sqrt{-g}+\Upsilon)V(\phi)-\sqrt{-g}V_0\right],
\label{S-toy phi}
\end{equation}
where $\Gamma$ stands for affine connection; $R(\Gamma,g)=g^{\mu\nu}R_{\mu\nu}(\Gamma)$,
$R_{\mu\nu}(\Gamma)=R^{\lambda}_{\mu\nu\lambda}(\Gamma)$ and
$R^{\lambda}_{\mu\nu\sigma}(\Gamma)\equiv \Gamma^{\lambda}_{\mu\nu
,\sigma}+ \Gamma^{\lambda}_{\gamma\sigma}\Gamma^{\gamma}_{\mu\nu}-
(\nu\leftrightarrow\sigma)$. Generically  there is no reason for coefficients in the linear combinations of $\Upsilon$ and $\sqrt{-g}$ in volume elements of different terms in the action to be the same. In the gravitational term,  the common factor of  the linear combination can be absorbed by the redefinition of the Newton constant.  Thereafter, by  rescaling the fields $\varphi_a$ in Eq.({\ref{Phi}}) the volume element in the gravitational term becomes as in Eq.(\ref{S-gr}). Then in $S_{\phi}$  there is only a freedom to rescale $\phi$ and therefore the volume elements can be reduced to $\left(b_1\sqrt{-g}+\Upsilon\right)d^4x$ and $\left(b_2\sqrt{-g}+\Upsilon\right)d^4x$ where    $b_1$ and $b_2$ are dimensionless  parameters of the model.
Notice that in the Einstein's GR, i.e. without the terms with the volume element $dV_{\Upsilon}$ in the action, the constant $V_0$ would be a cosmological constant. Similar term with the volume element $dV_{\Upsilon}$  does not contribute to the equations of motion (as a total derivative).

Following the ideas outlined in Sec.II.A, the theory should take into account that generically the affine and metric structures  are independent of each other. That is why in the action principle, {\bf the metric and the affine connection, along with all mater fields,  should be considered as
independent dynamical variables. All the relations between
them follow from equations of motion}. 
The independence of the
metric and the connection in the action means that we proceed in
the first order (Palatini) formalism  and the relation between connection and
metric is not necessarily according to Riemannian geometry.

Now let's look at the equations of motion in the toy model (\ref{S-toy}).
Variation with respect to $\varphi_a$ results in Eq.(\ref{varphi}) which has now the form 
\begin{equation}
-\frac{M_P^2}{2}R(\Gamma, g)+\frac{1}{2}g^{\gamma\beta}\phi_{,\gamma}\phi_{,\beta}
-V(\phi)=sM^4
\label{Constraint-1-toy}
\end{equation}
Varying the metric we get the gravitational equation
\begin{equation}
-(\zeta+1)\frac{M_P^2}{2}R_{\mu\nu}(\Gamma)+
(b_1+\zeta)\frac{1}{2}\phi_{,\mu}\phi_{,\nu}
- \frac{1}{2}g_{\mu\nu}\left[- \frac{M_P^2}{2}R(\Gamma, g) 
 +\frac{b_1}{2}g^{\alpha\beta}\phi_{,\alpha}\phi_{,\beta} -b_2V(\phi) -V_0  \right]=0
 \label{Grav.eq.toy}
\end{equation}
 The constraint, we have mentioned in Sec.II.A, is nothing but the consistency condition of these two equations:
contraction of the last equation with $g^{\mu\nu}$ and elimination of the scalar curvature from
the system of the obtained equation  and  Eq.(\ref{Constraint-1-toy}) results in the {\em constraint} where
 $\zeta (x)$ appears as a local function of $\phi(x)$ (and of other matter fields in more general model)
\begin{equation}
\zeta+1=
2\frac{\left[sM^4+(1-b_2)V(\phi)-V_0\right]} {sM^4+V(\phi)+\frac{1-b_1}{2}g^{\alpha\beta}\phi_{,\alpha}\phi_{,\beta}}
 \label{Constraint-2-toy}
\end{equation}
 The surprising feature of the model is
 that neither Newton constant nor curvature appear in the constraint
which means that the {\it geometrical scalar field} $\zeta (x)$
{\it is determined by the matter fields configuration}  locally
and without gravitational interaction. Since $\zeta (x)$ is not a dynamical variable, when we call equality (\ref{Constraint-2-toy}) a constraint, we must bear in mind that it differs in meaning from the usual constraint in field theory models, where it describes the relationship between dynamical degrees of freedom.

Variation of the affine connection yields the equations we have solved
earlier\cite{GK2}. The result is
\begin{equation}
\Gamma^{\lambda}_{\mu\nu}=\{^{\lambda}_{\mu\nu}\}+ \frac{1}{2(\zeta+1)}(\delta^{\lambda}_{\mu}\zeta,_{\nu}
+\delta^{\lambda}_{\nu}\zeta,_{\mu}-
\zeta,_{\beta}g_{\mu\nu}g^{\lambda\beta})
 \label{GAM2}
\end{equation}
where $\{ ^{\lambda}_{\mu\nu}\}$  are the Christoffel's connection
coefficients of the metric $g_{\mu\nu}$.

If $\zeta(x)\neq const.$ the metricity condition does not hold (the covariant derivative of $g_{\mu\nu}$ with this connection is nonzero) and consequently
geometry of the space-time with the metric $g_{\mu\nu}$ is generically non-Riemannian. In this paper, I will totally ignore a possibility to incorporate the torsion tensor, which could be an additional source for the space-time  to be different from Riemannian. 

As a special case, important for understanding the theory, let's consider a  model (\ref{S-toy})-(\ref{S-toy phi}) with $V(\phi)\equiv 0$ and with  the tuned value of the parameter $b_1=1$. Then the constraint (\ref{Constraint-2-toy}) takes the form
\begin{equation}
\zeta=1-2\frac{V_0}{sM^4}= const. 
 \label{Constraint-2-toy-trivial}
\end{equation}
In this almost trivial model the affine connection coincides with the Christoffel's connection
coefficients of the metric $g_{\mu\nu}$ and the latter is the Riemannian metric. In addition, this model  clearly demonstrates a situation when the sign of $\zeta$ and its value depend on the chosen value of the integration constant $sM^4$,  and  thus the spontaneous fixation of the orientation of the space-time manifold occurs. This  situation was mentioned in Sec.II.A when the motivation was discussed.  Note also that if  $ V_0=0$ then $\zeta \equiv 1$ and  therefore $\Upsilon \equiv \sqrt{-g}>0$ which also means that the positive orientation is spontaneously fixed. Finally, it is very important to emphasize that 
 the change  $\Upsilon\rightarrow -\Upsilon$  in all terms of the action does not affect the  physical results of the theory.  Indeed, the  sign of $\Upsilon$ is associated with the sign of the space-time orientation. But if the space-time manifold is oriented, both possible orientations are indistinguishable and therefore physically equivalent.

Return to the general form of the model (\ref{S-toy})-(\ref{S-toy phi}). The scalar field $\phi$ equation reads
\begin{equation}
\frac{1}{\sqrt{-g}}\partial_{\mu}\left[(b_1+\zeta)\sqrt{-g}g^{\mu\nu}\partial_{\nu}\phi\right]
+(b_2+\zeta) V^{\prime}(\phi)=0
 \label{phi-toy-original}
\end{equation}

In the case when $\phi=const.$, the constraint (\ref{Constraint-2-toy}) shows that $\zeta =const.$ also. Then the affine connection $\Gamma^{\lambda}_{\mu\nu}$ coincides with the 
 Christoffel's connection coefficients of the metric $g_{\mu\nu}$ and the space-time is Riemannian. 
However  if $\phi\neq const.$,  the system of  equations (\ref{Grav.eq.toy}), (\ref{GAM2}),  (\ref{phi-toy-original}),  together with the constraint (\ref{Constraint-2-toy}) describe the dynamics of the model in a non-Riemannian space-time. It is easy to see that in such a case, the transformation of the metric
\begin{equation}
\tilde{g}_{\mu\nu}=(\zeta +1)g_{\mu\nu}
 \label{gmunuEin}
\end{equation}
turns  the connection $\Gamma^{\lambda}_{\mu\nu}$ into
the Christoffel connection coefficients of the metric
$\tilde{g}_{\mu\nu}$ and the space-time turns into (pseudo)
Riemannian. The set of dynamical variables using the
metric $\tilde{g}_{\mu\nu}$ we call the Einstein frame. 
It is important to note here that since this transformation should not be singular, it is necessary to require that 
\begin{equation}
\qquad \textsf{everywhere} \quad \zeta +1 > 0. 
\label{zeta+1> 0}
\end{equation}
Therefore, the model that we are going to build must ensure that this condition is met.

Gravitational equations
(\ref{Grav.eq.toy}) in the Einstein frame take canonical GR
form with the same Newton constant as in the original frame
\begin{equation}
R_{\mu\nu}(\tilde{g})-\frac{1}{2}\tilde{g}_{\mu\nu}R(\tilde{g})=8\pi GT_{\mu\nu}
\label{toy grav eq Ein}
\end{equation}
where the energy-momentum tensor reads
\begin{eqnarray}
T_{\mu\nu}&=&\frac{b_1+\zeta }{\zeta +1}\phi_{,\mu}\phi_{,\nu}
-\tilde{g}_{\mu\nu} X_{\phi}
+\tilde{g}_{\mu\nu}U(\phi,\zeta;M);  \qquad X_{\phi}\equiv\frac{1}{2}\tilde{g}^{\gamma\beta}\phi_{,\gamma}\phi_{,\beta}
 \label{Tmn-toy}
\end{eqnarray}
and the function $U(\phi,\zeta;M)$ is defined as following:
\begin{equation}
U(\phi,\zeta; M)= \frac{sM^{4}+(1-b_2)V(\phi)-V_0}{(\zeta +1)^{2}}.
 \label{U toy}
\end{equation}

The scalar $\phi$ field  
Eq.(\ref{phi-toy-original})  rewritten in the Einstein frame
reads
\begin{equation}
\tilde{\square}\phi
+\frac{1-b_1}{(\zeta+1)(b_1+\zeta)}\tilde{g}^{\alpha\beta}\partial_{\alpha}\zeta\partial_{\beta}\phi
+\frac{b_2+\zeta}{(\zeta+1)(b_1+\zeta)} V^{\prime}(\phi)=0
 \label{phi toy Ein}
\end{equation}
where
\begin{equation}
\tilde{\square}\phi\stackrel{\mathrm{def}}{=}
\frac{1}{\sqrt{-\tilde{g}}}\partial_{\mu}\left[\sqrt{-\tilde{g}}\tilde{g}^{\mu\nu}\partial_{\nu}\phi\right],
\label{square def}
\end{equation}
 prime stands for derivative with respect to $\phi$, and $\zeta$ is determined by 
 the constraint (\ref{Constraint-2-toy}) which  in the Einstein frame
  takes the form
\begin{equation}
\zeta +1= 2\frac{sM^4+(1-b_2)V(\phi)-V_0}{sM^{4}+V(\phi)+(1-b_1) X_{\phi}}.
 \label{constr toy Ein}
\end{equation}

In the Friedman-Robertson-Walker (FRW) universe, where the homogeneous scalar field $\phi$ is a function of cosmic time only, it follows from Eq.(\ref{Tmn-toy}) that the equation-of-state is 
\begin{equation}
p+\rho= \frac{b_1+\zeta}{1+\zeta}\dot{\phi}^2
 \label{Eq-of-state}
\end{equation}
where dot stands for derivative with respect to cosmic time.
In the  FRW universe, Eq.(\ref{phi toy Ein}) reads
\begin{equation}
\ddot{\phi}+\left[3H+\frac{1-b_1}{(\zeta+1)(b_1+\zeta)}\dot{\zeta}\right]\dot{\phi}
+\frac{b_2+\zeta}{(\zeta+1)(b_1+\zeta)} V^{\prime}(\phi)=0,
 \label{phi toy FRW}
\end{equation}
where $H$ is the Hubble parameter. By direct  bulky calculation, it can be shown that from Eqs.(\ref{Tmn-toy}), (\ref{U toy}), (\ref{constr toy Ein}), (\ref{phi toy FRW})  and
the Einstein equations  follows the covariant law of  energy conservation
\begin{equation}
\dot{\rho}+3H(\rho+p)=0
 \label{covar conserv}
\end{equation}

\subsection{The  TMT effective action of gravity and mater fields and the TMT procedure}

With the help of a toy model, we got an idea of how TMT works. Now we need to understand how to obtain physical results in more realistic models without violating the principles of TMT. It turns out that this is not a trivial question that requires a special discussion. 

Let's start with a closer look at the principles of TMT. 
In TMT, the following four independent structures are defined on the space-time manifold: 1) the differentiable structure and the associated volume measure $dV_\Upsilon$; 2) the affine connection; 3) the metric and the associated volume measure $dV_g$; 4) matter (scalar, vector and fermionic fields).
 The last three of them are present in models describing gravity and matter in the Palatini formalism, and they have been studied in numerous papers. 
When  we apply to TMT the requirement that the relationship between all variables in the underlying action be determined by equations arising from the variational principle, {\bf the volume measure $dV_\Upsilon$ leads to fundamental changes in the structure of the equations describing physical reality}. 

1. The main new result is the appearance of a constraint defining the scalar  $\zeta(x)$ as a function, in general, of all mater fields. Conversely, all field equations obtained from the variational principle contain $\zeta(x)$.

2. The constraint only occurs if the underlying action contains a gravity term, as in Eq.(\ref{S-gr}). But the form of the constraint in the Einstein frame  does not depend on how strong the gravitational field (produced by matter and obtained by solving Einstein's equations) is. Moreover, even in the limit of Minkowski space, the constraint does not change its form, and $\zeta$ is present in all equations of matter fields. However, if one to start with the underlying action in the limit of the Minkowski space, then the constraint does not appear at all. The constraint also does not appear if the transition to the Einstein frame   is performed in the underlying action before the application of the variational principle.

3. In the toy model, we saw that in the $\phi$-equation in the Einstein frame, Eq.(\ref{phi toy Ein}),  a term with a gradient of $\zeta(x)$ appears, and this is typical for all field equations  in TMT. As we will see later, these terms are eliminated from the equations for fermions if the transition to the Einstein's frame in addition to (\ref{gmunuEin}) also includes redefining the spinors by multiplying them by a $\zeta$-dependent factor.
 In the equations of scalar and vector fields, these terms are preserved. But in states in which $\zeta=const$, and in particular in a  vacuum, they  obviously disappear. 

4. Another interesting and typical effect is the appearance of a $\zeta$-dependent factor, as in the last term of Eq.(\ref{phi toy Ein}). This means that  instead of the original scalar field potential $V(\phi)$, the new potential appears as a result of a kind of  {\em screening effect}, which, for example, when $\zeta=const$, has the form
\begin{equation}
V_{eff}(\phi)=\frac{b_2+\zeta}{(\zeta+1)(b_1+\zeta)} V(\phi)
\label{Veff toy}
\end{equation}
As we will see, this type of effect turns out to be very important in many aspects of the model: a) in the appearance of “incorrect” sign mass terms in the potentials of the Higgs fields and, therefore,  in SSB; b)  for solving the problem of the mass hierarchy of SM fermions; c) for obtaining a giant splitting between VEV's of the SM and mirror Higgs fields and between masses of SM particles and their mirror partners. The potential (\ref{Veff toy}) is reconstructed on the basis of the equation of motion written in the Einstein frame, and it is natural to call it "the TMT effective potential". It is evident that, when $\zeta=const$,  the corresponding  TMT effective action of the scalar field  whose variation gives the $\phi$-equation (\ref{phi toy Ein}) has the standard form of the scalar field action with the potential $V_{eff}(\phi)$ 
\begin{equation}
S_{eff}(\phi)|_{\zeta=const}=\int\sqrt{-\tilde{g}}d^4x\left[\frac{1}{2}\tilde{g}^{\mu\nu}\phi_{,\mu}\phi_{,\nu}-V_{eff}(\phi)\right]
\label{Seff toy}
\end{equation}
In what follows, we will use the terms {\em TMT effective  potential}, {\em TMT effective  Lagrangian} and {\em TMT effective  action} in this sense. 
In this connection, it should be noted that the term 'TMT effective potential' must be distinguished from the effective potential obtained as a result of taking into account quantum corrections. In fact, the TMT effective potential can be regarded as the effective potential at the tree level, and the quantum corrections can be calculated. 
For $\zeta=const$ the TMT effective actions for all matter fields can be obtained similarly.
When $\zeta$,  being a local function of all matter fields, is not a constant  the problem becomes much more complicated, since all field equations acquire additional nonlinearities. Fortunately, as we will see later, in cases of physical interest, $\zeta=const$, or a very high precision constant.

It is very important that regardless of the discussed role of $\zeta$, the TMT effective gravitational action is the standard action of the Einstein's gravity with the metric tensor $\tilde{g}_{\mu\nu}$
\begin{equation}
S_{gr,eff}=-\frac{1}{16\pi G}\int \sqrt{-\tilde{g}}d^4x R(\tilde{g}_{\mu\nu}).
\label{S-gr eff}
\end{equation}
This statement follows immediately from the fact that the gravitational equations in the Einstein frame (\ref{toy grav eq Ein}) coincide with  equations of the Einstein's GR. Following the terminology described in the previous paragraph, for the  energy-momentum tensor that appears on the right hand side of these equations, we will use the term {\em  TMT effective energy-momentum tensor}. For example, in the  toy model under consideration, Eqs.(\ref{Tmn-toy}), (\ref{U toy}) describe the  TMT effective energy-momentum tensor.

Summarizing the above, it is possible to formulate a sequence of steps that must be performed in TMT in order to obtain an adequate description of the results of the studied model. The first is variation of the underlying action with respect to  $\varphi_a$, $\Gamma^{\lambda}_{\mu\nu}$, $g _{\mu\nu}$ and all matter fields. The second step consists of a) solution the equation for  
$\Gamma^{\lambda}_{\mu\nu}$ (like in Eq.(\ref{GAM2})); b) obtaining a constraint from the requirement of consistency of equations obtained by varying 
$\varphi_a$ and $g _{\mu\nu}$. At the third step, in all equations and in the constraint, a transformation is performed from initial variables to variables in the Einstein frame. At the fourth step, if necessary, at $\zeta=const$, some results can be formulated in terms of the  TMT effective potential, the  TMT effective Lagrangian, and the  TMT effective action.
In what follows {\em I will refer to the sequence of these steps as}  {\bf the TMT procedure}.
The key element is that deviations from the established procedure lead to different results.

\section{Building the underlying MESM action. Bosonic sector}

To simplify the presentation 
of the MESM let us study the mirror extended  SM with  the $SU(2)\times U(1)$ gauge invariant electroweak interactions.
The $SU(2)\times U(1)$ {\em quantum numbers of all mirror particles are identical to the quantum numbers of their partners in the SM, but paritation of the mirror counter partners is opposite}.  

\subsection{Bosonic matter content and underlying action}

Bosonic sector of MESM under consideration consists of the Higgs isodoublet $\Phi$ and its mirror partner $\check{\Phi}$
\begin{equation}
\Phi= \left(\begin{matrix} \phi^{+} \\ \phi^{0} \end{matrix}\right),  \qquad \check{\Phi}= \left(\begin{matrix} \check{\phi}^{+} \\ \check{\phi}^{0} \end{matrix}\right),
\label{Higgs  doublets}
\end{equation}
as well as the SM gauge  bosons   $\mathbf{A}_\mu$ and $B_{\mu}$ and their mirror partners $\check{\mathbf{A}}_\mu$ and $\check{B}_\mu$.
It is postulated that,  in addition to the ordinary symmetries of the  standard electroweak model, the matter sector of the underlying TMT  action is invariant under $mirror \times parity$-transformations. In the case of bosons it means that the bosonic sector of the action is invariant under transformations (\ref{Pzeta}) and (\ref{mirror reflection})  where now $X$ and $\check{X}$  denotes $\Phi$,  $\mathbf{A}_\mu$, $B_{\mu}$ and  $\check{\Phi}$, $\check{\mathbf{A}}_\mu$ 
$\check{B}_\mu$ respectively; of course, the usual  parity transformations of the  Higgs and gauge bosons are also involved into  $mirror \times parity$-transformations.

The $SU(2)\times U(1)$ gauge invariant and ($mirror \times parity$)  invariant action for the Higgs fields $\Phi$ and $\check{\Phi}$  is chosen as follows:
\begin{eqnarray}
S_{H}=\int d^4x \left[(b\sqrt{-g}+\Upsilon) 
\left(g^{\alpha\beta}\left(\mathcal{D}_{\alpha}\Phi\right)^{\dag}\mathcal{D}_{\beta}\Phi
 -\lambda |\Phi|^4\right)
 -\left(b\sqrt{-g}-\Upsilon\right)m^2 |\Phi|^2\right]
\nonumber
\\
+\int d^4x \left[(b\sqrt{-g}-\Upsilon) 
\left(g^{\alpha\beta}\left( \check{ \mathcal{D}}_{\alpha}\check{\Phi}\right)^{\dag} \check{ \mathcal{D}}_{\beta}\check{\Phi}
 -\lambda |\check{\Phi}|^4\right)
 -\left(b\sqrt{-g}+\Upsilon\right)m^2 |\check{\Phi}|^2\right],
\label{S-H 12}
\end{eqnarray}
where $|\Phi|^2=\Phi^{\dag}\Phi$, $ |\check{\Phi}|^2=\check{\Phi}^{\dag}\check{\Phi}$. The parameters $b>0$, 
$\lambda>0$ and $m^2>0$ are chosen to be positive.
It should be noted that  unlike the standard "incorrect" sign of the mass term in the Higgs potential, here the mass terms included in the action with the volume element
 $\sqrt{-g}d^4x$ (i.e. in the limiting case $\Upsilon\rightarrow 0$) are selected with the "correct" sign.
The standard definition for the operator 
\begin{equation}
 \mathcal{D}_{\mu}=\partial_{\mu}
-ig\mathbf{\hat{T} A}_{\mu}-i\frac{g'}{2}\hat{Y}B_{\mu},
\label{Nabla SM}
\end{equation}
is used while for the mirror sector the appropriate operator is defined similarly
\begin{equation}
\check{ \mathcal{D}}_{\mu}=\partial_{\mu}
-ig\mathbf{\hat{T} \check{A}}_{\mu}-i\frac{g'}{2}\hat{Y}\check{B}_{\mu}.
\label{Nabla mirror}
\end{equation}
with the same gauge coupling constants $g$ and $g'$.
 Here, as usually, $\mathbf{\hat{T}}$ stands for the three generators of the $SU(2)$ group and $\hat{Y}$ is the generator of the U(1) group.

 With standard notations $\mathbf{F}_{\mu\nu}=\partial_{\mu}\mathbf{A}_\nu -\partial_{\nu}\mathbf{A}_\mu +g\mathbf{A}_\mu\times \mathbf{A}_\nu$ and $B_{\mu\nu}=\partial_{\mu}B_\nu -\partial_{\nu}B_\mu$  for the field strengths of the isovector $\mathbf{A}_{\mu}$ and isoscalar $B_{\mu}$, and with similar notations for their mirror partners $\check{\mathbf{F}}_{\mu\nu}=\partial_{\mu}\check{\mathbf{A}}_\nu -\partial_{\nu}\check{\mathbf{A}}_\mu +g\check{\mathbf{A}}_\mu\times \check{\mathbf{A}}_\nu$  and  $\check{B}_{\mu\nu}=\partial_{\mu}\check{B}_\nu -\partial_{\nu}\check{B}_\mu$, the  gauge fields action, which is also ($mirror \times parity$) invariant, is chosen as following
\begin{eqnarray}
S_{gauge}=\int d^4x\left(b\sqrt{-g}+\Upsilon\right)\left[-\frac{1}{4} g^{\mu\alpha}g^{\nu\beta}\mathbf{F}_{\mu\nu}\mathbf{F}_{\alpha\beta}-\frac{1}{4} g^{\mu\alpha}g^{\nu\beta}B_{\mu\nu}B_{\alpha\beta}\right]
\nonumber
\\
+\int d^4x\left(b\sqrt{-g}-\Upsilon\right)\left[-\frac{1}{4} g^{\mu\alpha}g^{\nu\beta}\check{\mathbf{F}}_{\mu\nu}\check{\mathbf{F}}_{\alpha\beta}-\frac{1}{4} g^{\mu\alpha}g^{\nu\beta}\check{B}_{\mu\nu}\check{B}_{\alpha\beta}\right],
 \label{S gauge}
\end{eqnarray}
where the parameter $b$ is chosen  to be exactly equal to that in Eq.(\ref{S-H 12}).

An alternative  ($mirror \times parity$)  invariant electroweak model can be constructed if the same volume element $(b\sqrt{-g}+\Upsilon)d^4x$ is selected in Eq.(\ref{S-H 12}) for both terms in the first line and the same volume element $(b\sqrt{-g}-\Upsilon)d^4x$ for both terms in the second line. Then, to ensure SB of the $SU(2)\times U(1)$ gauge symmetry,  the mass terms entering the action with the volume element
 $\sqrt{-g}d^4x$ should be  chosen with an "incorrect" sign.
The main differences between the results of such an alternative model and the model studied in this paper will be briefly described in Sec.IX.A.

\subsection{Higgs fields equations and their vacua}

Following the TMT procedure described in detail in Sec.III.B  we have to  vary the action with respect to  bosonic fields and represent the obtained equations in the Einstein frame. With the gravitational part of the action  as in Eq.(\ref{S-gr}), the transition to the Einstein frame is carried out by the transformation (\ref{gmunuEin}).  In a vacuum of gauge fields, for the Higgs doublet  $\Phi$  and for the mirror Higgs doublet  $\check{\Phi}$ this procedure leads to the following equations
\begin{equation}
\tilde{\square}\Phi
-\frac{1-b}{(1+\zeta)(b+\zeta)}\tilde{g}^{\alpha\beta}\partial_{\alpha}\zeta\partial_{\beta}\Phi+
\frac{\partial{V^{(\Phi)}_{eff}}}{\partial{\Phi^{\dag}}}
=0 ,
 \label{Higgs eqom Ein}
\end{equation}
where  
\begin{equation}
\frac{\partial{V^{(\Phi)}_{eff}}}{\partial{\Phi^{\dag}}}\equiv \left[\frac{2\lambda}{1+\zeta}|\Phi|^2
-\frac{\zeta-b}{(1+\zeta)(\zeta+b)}m^2\right]\Phi,
           \label{Higgs 1 eff pot}
\end{equation}

\begin{equation}
\tilde{\square}\check{\Phi}
+\frac{1+b}{(1+\zeta)(b-\zeta)}\tilde{g}^{\alpha\beta}\partial_{\alpha}\zeta\partial_{\beta}\check{\Phi}+\frac{\partial{V^{(\check{\Phi})}_{eff}}}{\partial{\check{\Phi}^{\dag}}}=0 
 \label{Higgs mirror eqom Ein}
\end{equation}
where
\begin{equation}
\frac{\partial{V^{(\check{\Phi})}_{eff}}}{\partial{\check{\Phi}^{\dag}}}\equiv\left[\frac{2\lambda}{1+\zeta}|\check{\Phi}|^2
-\frac{\zeta+b}{(1+\zeta)(\zeta-b)}m^2\right]\check{\Phi}
\label{Higgs 2 eff pot}
\end{equation}
When deriving equations (\ref{Higgs eqom Ein}) and  (\ref{Higgs mirror eqom Ein}), we performed the division of the equations obtained by varying the action (\ref{S-H 12}) by $\zeta+b$ and $\zeta-b$, respectively. This is required to provide the canonical form of the Green's functions of the Higgs fields. Therefore, in what follows, we must make sure that the model guarantees that $\zeta\pm b\neq 0$.
The  $\zeta$ dependence in Eqs.(\ref{Higgs 1 eff pot}) and (\ref{Higgs 2 eff pot})   is a manifistation of the screening effect for  TMT effective (classical) potentials in the Einstein frame which was discussed in Sec.III.B.

The dependence of the Higgs TMT effective potentials  on $\zeta$, which changes during the entire cosmological evolution (from inflation to the transition to vacuum), as well as the definition of the vacuum state itself, are of particular importance in this paper and will be discussed in detail in Secs.VII and VIII. Here, at the stage of qualitative understanding the model, it is enough for us to take into account that, as will be shown in Secs.VII and VIII, the corresponding vacuum value $\zeta_0$ is positive. In addition, parameter b will be chosen so that $0<\zeta_0-b\sim 10^{-12}$.
Thus the mass terms acquire "wrong" signs, the potentials $V^{(\Phi)}_{eff}$ and $V^{(\check{\Phi})}_{eff}$ have nonzero minima and the appropriate vacuum expectation values (VEV) can be obtained as usually.
In the vicinity of vacuum, using gauge invariance, we can choose a unitary gauge in which, violating $SU(2)$ gauge invariance, the Higgs field is presented in the well known form
\begin{equation}
\Phi= \frac{1}{\sqrt{2}}\left(\begin{matrix} 0 \\ \underline{\varphi}(x) \end{matrix}\right)=\frac{1}{\sqrt{2}}\left(\begin{matrix} 0 \\\underline{v}+\underline{h}(x) \end{matrix}\right),
\label{Higgs Unitary}
\end{equation}
where
\begin{equation}
 \underline{v}^2=\frac{\zeta_0-b}{\zeta_0+b}\frac{m^2}{\lambda}.
\label{v under 2}
\end{equation}
and similarly for the mirror Higgs field
\begin{equation}
\check{\Phi}= \frac{1}{\sqrt{2}}\left(\begin{matrix} 0 \\ \check{\underline{\varphi}}(x) \end{matrix}\right)=\frac{1}{\sqrt{2}}\left(\begin{matrix} 0 \\\underline{\check{v}}+\underline{\check{h}}(x) \end{matrix}\right),
\label{Higgs Unitary}
\end{equation}
where
\begin{equation}
 \underline{{\check{v}}}^2=\frac{\zeta_0+b}{\zeta_0-b}\frac{m^2}{\lambda}.
\label{v under mir 2}
\end{equation}
 The underlines were introduced here because, as will be shown in the next subsection, the measurable values of VEV's (following from the expressions for masses of the gauge bosons) are
\begin{equation}
 v=\frac{\underline{v}}{\sqrt{1+\zeta_0}} \quad \textsf{and} \quad
\check{v}=\frac{\underline{{\check{v}}}}{\sqrt{1+\zeta_0}}.
\label{v both 2}
\end{equation}
 Due to the need to pay attention to this aspect of definitions, it will also be convenient to use redefined Higgs doublets in the future
\begin{equation}
\Phi=\sqrt{1+\zeta_0}\Phi_m, \qquad  \check{\Phi}=\sqrt{1+\zeta_0}\check{\Phi}_m,
\label{Phi and mir measured}
\end{equation}
where the subscript $m$ indicates that when using $\Phi_m$ and $\check{\Phi}_m$ near the vacuum, the {\em measurable} VEV's and quantum Higgs fields appear in standard form
\begin{equation}
\Phi_m|_{(near \, vacuum)}=\frac{1}{\sqrt{2}}\left(\begin{matrix} 0 \\\varphi\end{matrix}\right)=\frac{1}{\sqrt{2}}\left(\begin{matrix} 0 \\v+h\end{matrix}\right), \qquad  
\check{\Phi}_m|_{(near \,  vacuum)}=\frac{1}{\sqrt{2}}\left(\begin{matrix} 0 \\\check{\varphi}\end{matrix}\right)= \frac{1}{\sqrt{2}}\left(\begin{matrix} 0 \\\check{v}+\check{h}\end{matrix}\right).
\label{Higgs doublets mesured}
\end{equation}
Here the appropriate redefinitions have been used:
\begin{equation}
\underline{\varphi}=\varphi\sqrt{1+\zeta_0}, \quad  \underline{h}=h\sqrt{1+\zeta_0}; \quad and \quad \check{\underline{\varphi}}=
 \check{\varphi}\sqrt{1+\zeta_0}, \quad \underline{\check{h}}=\check{h}\sqrt{1+\zeta_0}.
\label{redefin phi h}
\end{equation}

To obtain the equations for the Higgs fields $h$ and $\check{h}$ near vacuum, it is not enough to use representations (\ref{Higgs doublets mesured}) in Eqs (\ref{Higgs eqom Ein})-(\ref{Higgs 2 eff pot}) and simply substitute $\zeta=\zeta_0$. The point is that, as will be studied in Secs.VII and VIII, the Higgs fields $\Phi$ and $\check{\Phi}$, along with the inflaton, are background fields. Therefore, they are present in the constraint, which defines $\zeta$ as a function of these fields. That is why, when representing the Higgs fields $\varphi$ and $\check{\varphi}$ near the vacuum as the sum of classical $v$ and $\check{v}$ and quantum $h$ and $\check{h}$ fields respectively, we must take into account that this entails the dependence of $\zeta$ on $h$ and $\check{h}$. It turns out that this dependence can be represented as a small addition
\begin{equation}
\zeta=\zeta_0+\delta\zeta(h, \check{h}),  \quad 0<\delta\zeta\ll \zeta_0-b\sim 10^{-12};
\label{delta zeta vac}
\end{equation} 
the  function $\delta\zeta(h, \check{h})$ will be found in Sec.VIII.D.
As a result, the second terms in Eqs.(\ref{Higgs eqom Ein}) and (\ref{Higgs mirror eqom Ein}) turn out to be quadratic forms with respect to the derivatives of $h$ and $\check{h}$. Given the smallness of $\delta\zeta$, we will neglect these terms, at least in this paper.
Then the equations for the  Higgs field $h$ and its mirror partner $\check{h}$ can be represented as following
 \begin{equation}
\tilde{\square}h+\frac{\partial V_{eff}(h;\delta\zeta(h, \check{h})}{\partial h}=0; \quad \frac{\partial V_{eff}(h;\delta\zeta(h, \check{h}))}{\partial h}=\lambda\left[2v^2 h+3vh^2+h^3-v^3\,\frac{\delta\zeta(h, \check{h})}{\zeta_0-b}\right],
\label{h eq}
\end{equation}
 \begin{equation}
\tilde{\square}\check{h}+\frac{\partial V_{eff}(\check{h};\delta\zeta(h, \check{h}))}{\partial \check{h}}=0; \quad \frac{\partial V_{eff}(\check{h};\delta\zeta(h, \check{h}))}{\partial \check{h}}=\lambda\left[2\check{v}^2 \check{h}+3\check{v}\check{h}^2+\check{h}^3+2b\check{v}^3\,\frac{\delta\zeta(h, \check{h})}{(\zeta_0-b)(\zeta_0+b)}\right].
\label{h mir eq}
\end{equation}
In Sec.VIII.D we return to the study of Higgs fields near vacuum.

\subsection{Gauge fields equations and the Higgs phenomena}
	 Similar to how this is done in the standard way in the 	
Glashow-Weinberg-Salam (GWS) model,  the gauge fields action (\ref{S gauge}) can be rewriten in terms of the physical fields $W^{\pm}_{\mu}$, $Z_{\mu}$, $A_{\mu}$ and their mirror partners $\check{W}^{\pm}_{\mu}$, $\check{Z}_{\mu}$, $\check{A}_{\mu}$.
Since all mirror  particles have the  same quantum numbers of the  $SU(2)\times U(1)$ gauge symmetry as their SM partners and the gauge couplings constants $g$ and $g^{\prime}$ are also the same, 
there is no need  to  explicitly present the expressions for the mirror bosons $\check{\mathbf{A}}_\mu$, $\check{B}_\mu$ in terms of physical 
$\check{W}^{\pm}_{\mu}$, $\check{Z}_{\mu}$, $\check{A}_{\mu}$ because they have exactly the same form as for gauge bosons of the GWS model.  Then the gauge bosons action (\ref{S gauge}) can be represented as following
\begin{equation}
S_{gauge}=S_{g}+S_{\check{g}}
\label{gauge general}
\end{equation}
\begin{equation}
S_{g}=\int d^4x\left(b\sqrt{-g}+\Upsilon\right)\left(\sum_{g}L_g^{(2)}+L^{(3)}+L^{(4)}\right),
\label{g general}
\end{equation}
\begin{equation}
S_{\check{g}}=\int d^4x\left(b\sqrt{-g}-\Upsilon\right)\left(\sum_{\check{g}}\check{L}_{\check{g}}^{(2)}+\check{L}^{(3)}+\check{L}^{(4)}\right),
\label{g mirror general}
\end{equation}
where $L^{(3)}$ and $L^{(4)}$ are the Lagrangians of the third-order and forth-order interactions of the SM gauge bosons;  $\check{L}^{(3)}$ and
 $\check{L}^{(4)}$ are the same for their mirror partners;   the free SM gauge fields Lagrangians $L_g^{(2)}$ ($g=W, Z, A$) and their mirror partners 
$\check{L}_{\check{g}}^{(2)}$ ($\check{g}=\check{W}, \check{Z}, \check{A}$) read respectivly
\begin{equation}
L_W^{(2)}= -\frac{1}{2}g^{\mu\gamma}g^{\nu\beta}\left(\partial_{\mu}W_{\gamma}^{(+)}-\partial_{\gamma}W_{\mu}^{(+)}\right)
\left(\partial_{\nu}W_{\beta}^{(-)}-\partial_{\beta}W_{\nu}^{(-)}\right)
\label{LW2}
\end{equation}
\begin{equation}
L_Z^{(2)}= -\frac{1}{4}g^{\mu\gamma}g^{\nu\beta}\left(\partial_{\nu}Z_{\beta}-\partial_{\beta}Z_{\nu}\right)\left(\partial_{\nu}Z_{\beta}-\partial_{\beta}Z_{\nu}\right)
\label{LZ2}
\end{equation}
\begin{equation}
L_A^{(2)}=-\frac{1}{4}g^{\mu\gamma}g^{\nu\beta} \left(\partial_{\mu}A_{\gamma}-\partial_{\gamma}A_{\mu}\right)\left(\partial_{\nu}A_{\beta}-\partial_{\beta}A_{\nu}\right)
\label{LA2}
\end{equation}
\begin{equation}
\check{L}_{\check{W}}^{(2)}= -\frac{1}{2}g^{\mu\gamma}g^{\nu\beta}\left(\partial_{\mu}\check{W}_{\gamma}^{(+)}-\partial_{\gamma}\check{W}_{\mu}^{(+)}\right)\left(\partial_{\nu}\check{W}_{\beta}^{(-)}-\partial_{\beta}\check{W}_{\nu}^{(-)}\right)
\label{LWcheck2}
\end{equation}
\begin{equation}
\check{L}_{\check{Z}}^{(2)}= -\frac{1}{4}g^{\mu\gamma}g^{\nu\beta}\left(\partial_{\nu}\check{Z}_{\beta}-\partial_{\beta}\check{Z}_{\nu}\right)
\left(\partial_{\nu}\check{Z}_{\beta}-\partial_{\beta}\check{Z}_{\nu}\right)
\label{LZcheck2}
\end{equation}
\begin{equation}
\check{L}_{\check{A}}^{(2)}=-\frac{1}{4}g^{\mu\gamma}g^{\nu\beta} \left(\partial_{\mu}\check{A}_{\gamma}-\partial_{\gamma}\check{A}_{\mu}\right)\left(\partial_{\nu}\check{A}_{\beta}-\partial_{\beta}\check{A}_{\nu}\right)
\label{LAcheck2}
\end{equation}
 
 Performing standard TMT procedures  we are now able to find out the  gauge fields equations in the Einstein frame. Using  Eqs.(\ref{g general})-(\ref{LAcheck2}), together with Eqs.(\ref{S-H 12})-(\ref{Nabla mirror}) and definitions (\ref{v under 2}), (\ref{v under mir 2}), (\ref{v both 2}), we obtain 
 that in the vacuum described in the previous subsection, the equations of free gauge fields in the Einstein frame are as follows:
\begin{equation}
 \frac{1}{\sqrt{-\tilde{g}}}\partial_{\gamma}\left[\sqrt{-\tilde{g}}\tilde{g}^{\mu\gamma}\tilde{g}^{\nu\beta}
\left(\partial_{\mu}W_{\nu}^{(+)}-\partial_{\nu}W_{\mu}^{(+)}\right)\right] +\frac{g^2v^2}{4}\tilde{g}^{\nu\beta}W_{\nu}^{(+)}=0,
\label{W eq}
\end{equation}
\begin{equation}
 \frac{1}{\sqrt{-\tilde{g}}}\partial_{\gamma}\left[\sqrt{-\tilde{g}}\tilde{g}^{\mu\gamma}\tilde{g}^{\nu\beta}
\left(\partial_{\mu}Z_{\nu}-\partial_{\nu}Z_{\mu}\right)\right] +\frac{(g^2+g^{\prime 2})v^2}{4}\tilde{g}^{\nu\beta}Z_{\nu}=0,
\label{Z eq}
\end{equation}
\begin{equation}
 \frac{1}{\sqrt{-\tilde{g}}}\partial_{\gamma}\left[\sqrt{-\tilde{g}}\tilde{g}^{\mu\gamma}\tilde{g}^{\nu\beta}
\left(\partial_{\mu}\check{W}_{\nu}^{(+)}-\partial_{\nu}\check{W}_{\mu}^{(+)}\right)\right] +\frac{g^2\check{v}^2}{4}\tilde{g}^{\nu\beta}\check{W}_{\nu}^{(+)}=0,
\label{W mir eq}
\end{equation}
\begin{equation}
 \frac{1}{\sqrt{-\tilde{g}}}\partial_{\gamma}\left[\sqrt{-\tilde{g}}\tilde{g}^{\mu\gamma}\tilde{g}^{\nu\beta}
\left(\partial_{\mu}\check{Z}_{\nu}-\partial_{\nu}\check{Z}_{\mu}\right)\right] +
\frac{(g^2+g^{\prime 2})\check{v}^2}{4}\tilde{g}^{\nu\beta}\check{Z}_{\nu}=0,
\label{Z mir eq}
\end{equation}
Equations of a free  electromagnetic field $A_{\mu}$ and a free mirror electromagnetic field $\check{A}_{\mu}$  have the same canonical form.

Note once again that in order to provide the canonical structure of mass terms, it became necessary to absorb the $\frac{1}{1+\zeta_0}$ factor using the redifinitions (\ref{v both 2}).  This explains why  $v$ and $\check{v}$ are measurable in particle physics.
Expressions for masses of the gauge fields follow directly from  Eqs.(\ref{W eq}) - (\ref{Z mir eq}):
\begin{equation}   
M_W=\frac{gv}{2}; \quad M_Z=\frac{M_W}{cos\theta_W}; \quad M_{\check{W}}=\frac{g\check{v}}{2}; \quad M_{\check{Z}}=\frac{\check{M}_W}{cos\theta_W},
\label{gauge masses}
\end{equation}
 where $\theta_W$ is the Weinberg angle which has the same value 
\begin{equation} 
\tan\theta_W=\frac{g^{\prime}}{g}
\label{theta}
\end{equation}
 both for  standard and mirror gauge  fields. Note that 
\begin{equation}   
 \quad M_{\check{W}}=\frac{\check{v}}{v}M_W=\frac{\zeta_0+b}{\zeta_0-b}M_W; \qquad
 M_{\check{Z}}==\frac{\check{v}}{v}M_Z=\frac{\zeta_0+b}{\zeta_0-b}M_Z,
\label{gauge masses}
\end{equation}
that is, as in the case of the mirror Higgs boson (see Eqs.(\ref{v under 2}), (\ref{v under mir 2}) and (\ref{h eq}), (\ref{h mir eq})), large masses of  the mirror gauge bosons are also realized due to the small value of $\zeta_0-b$.

It should be noted that all interactions of  $W$, $Z$ and photons with each other and with the Higgs  boson $h$ are exactly the same as in the SM.  
Interactions of mirror bosons $\check{W}$, $\check{Z}$ and mirror photons with each other and with the mirror Higgs  boson $\check{h}$ have also  the structure as in the SM, but naturally with the replacement $v\rightarrow \check{v}$. More details on interactions in MESM will be provided in Sec.VI.B.

\section{Building the underlying MESM action. Fermion sector.}

\subsection{The TMT action for 3 generations of regular leptons and their mirror partners}

  Main ideas and some technical aspects arising when fermions are included in TMT in the $\textsf{{\bf (mirror}}\times\textsf{{\bf   parity)}}$ invariant form and appropriate notations are considered and explained in the toy model  in Appendix. Here we will study a model including 3 generations of leptons
\begin{equation}
L_l=\frac{1-\gamma_5}{2} \left(\begin{matrix} \nu^{(l)} \\ l \end{matrix}\right);
\qquad l_R=\frac{1+\gamma_5}{2} l ; \qquad  \nu^{(l)}_R=\frac{1+\gamma_5}{2}\nu^{(l)}; \qquad  l=e, \mu, \tau
\label{GWS fermions 3}
\end{equation}
 and their mirror partners
\begin{equation}
\check{R}_{\check{l}}=\frac{1+\gamma_5}{2} \left(\begin{matrix} \check{\nu}^{(\check{l})}\\ \check{l} \end{matrix}\right);
\qquad \check{l}_L=\frac{1-\gamma_5}{2} \check{l}; \qquad \check{\nu}^{(\check{l})}_L=\frac{1+\gamma_5}{2}\check{\nu}^{(\check{l})}; \qquad \check{l}=\check{e}, \check{\mu}, \check{\tau}.
\label{GWS mirror fermions 3}
\end{equation}

To construct a generally coordinate invariant kinetic terms of the action for fermionic sector we have to use the following two covariant operators
\begin{equation}
\nabla_{\mu}=\mathcal{D}_{\mu}+\frac{1}{2}\omega_{\mu}^{ik}\sigma_{ik}; \qquad  \check{\nabla}_{\mu}=\check{\mathcal{D}}_{\mu}+\frac{1}{2}\omega_{\mu}^{ik}\sigma_{ik},
\label{nabla  f and mirror f}
\end{equation}
where $\mathcal{D}_{\mu}$ and $\check{\mathcal{D}}_{\mu}$ are defined by Eqs.(\ref{Nabla SM}) and (\ref{Nabla mirror}),  and they must be applied to regular and mirror fermions, respectively.  In Sec.III, we have demonstrated    how the Palatini formalism works in TMT.  Continuing here the use of the Palatini formalism, one can show that in the original frame not only the affine connection $\Gamma^{\lambda}_{\mu\nu}$, but also the spin-connection $\omega_{\mu}^{ik}$ ($i, k$ are Lorentz indeces)  differ from their standard forms in  Rimannian space-time by  terms proportional to $\partial_{\mu}\zeta$ (recall Eq.(\ref{GAM2})).

Turn now to  the Yukawa coupling terms responsible for the fermion mass generation. The standard method  is that for each generation of charged leptons, the Yukawa coupling constant $f_l$ ($ l=e, \mu, \tau$) corresponding to the experimental data is selected.
As is well known, this aspect of SM model contains a flaw indicating insufficient completeness of the theory: to obtain the electron mass, this coupling constant must be inexplicably small: $f_e\approx 3\cdot 10^{-6}$. Besides, in order to obtain the required values of the masses of charged leptons,  the corresponding Yukawa coupling constants must vary more than five orders of magnitude 
$f_e:f_{\mu}:f_{\tau}=m_e:m_{\mu}:m_{\tau}$.
Such hierarchy  looks doubtful  for a theory that claims to be a correct description of nature. The situation with the choice of the Yukawa coupling constant becomes even stranger when it is necessary to explain the value of the top quark mass. In this regard, the proposed extension of the SM offers an unexpected opportunity to circumvent these hierarchy problem. 
 It turns out that to construct a realistic MESM with three generations of leptons, it is possible to suppose that {\bf Yukawa coupling constant $y^{(ch)}$ is chosen to be universal for all charged leptons}. Then the problem of the mass hierarchy for charged leptons is solved simply by replacing one parameter $b_e$ which was in the toy model in Appendix with three parameters $b_l$, ($l=e, \mu, \tau $),  whose absolute values are close to $\zeta_0$.
 In this paper I am going to demonstrate how this idea   can be implemented for charged leptons and up quarks. 

Thus we arrive at the lepton sector action in the form
\begin{equation}
S_{3}^{(l)}=\sum_{l}S_{l}+\sum_{\check{l}}\check{S}_{\check{l}}=\int\sqrt{-g}d^4x\left(\sum_{l}\mathbb{L}_{l}+\sum_{\check{l}}\mathbb{\check{L}}_{\check{l}}\right)
\label{S3}
\end{equation}
 with the following Lagrangians for regular SM and mirror leptons respectively:
\begin{eqnarray}
\mathbb{L}_{l}=
 \frac{i}{2} (b_l+\zeta )
\left[\overline{L_l}\gamma^{\mu}\nabla_{\mu}L_l -(\nabla_{\mu}\overline{L_l})\gamma^{\mu}L_l+
\overline{l}_R\gamma^{\mu}\nabla_{\mu}l_R -(\nabla_{\mu}\overline{l}_R)\gamma^{\mu}l_R\right]
\nonumber\\
- (b_l-\zeta )y^{(ch)}\left(\overline{L}_l\Phi l_R +
\overline{l}_R\Phi^{\dag}L_l\right),
 \label{L 3}
\end{eqnarray}
\begin{eqnarray}
\mathbb{\check{L}}_{\check{l}}=
  \frac{i}{2} (b_l-\zeta )
\left[\overline{\check{R}}_{\check{l}}\gamma^{\mu}\check{\nabla}_{\mu}\check{R}_{\check{l}}
-(\check{\nabla}_{\mu}\overline{\check{R}}_{\check{l}})\gamma^{\mu}\check{R}_{\check{l}}
+\overline{\check{l}}_L\gamma^{\mu}\check{\nabla}_{\mu}\check{l}_L-(\check{\nabla}_{\mu}\overline{\check{l}}_L)\gamma^{\mu}\check{l}_L\right]
\nonumber\\
- (b_l+\zeta )y^{(ch)}\left(\overline{\check{R}}_{\check{l}}\check{\Phi} \check{l}_L +
\overline{\check{l}}_L\check{\Phi}^{\dag}\check{R}_{\check{l}}\right).
 \label{L mirror 3}
\end{eqnarray}
where $\gamma^{\mu}=V_{k}^{\mu}\gamma^{k}$, \, $\gamma^{k}$ are Dirac matrices,  $V_{k}^{\mu}$ are vierbeins, and the definition (\ref{zeta}) of $\zeta$ and the identities $(b_l\sqrt{-g}\pm\Upsilon)\equiv\sqrt{-g}(b_l\pm\zeta )$ have been used.
At this stage, we do not  write out the  terms of the Lagrangian for neutrino isosinglets $\nu^{(l)}_R$ and $\check{\nu}^{(\check{l})}_L$ describing their kinetic energies and Yukawa couplings. This will be done in Subsec.V.C. 
The action (\ref{S3})-(\ref{L mirror 3}) is invariant with respect to both  $SU(2)\times U(1)$ gauge symmetry  and ($mirror \times parity$)- transformations because each of the SM bosons and fermions and its corresponding mirror partner have the same SM quantum numbers, and because we define that the mirror transformations for fermions are as follows:
\begin{equation}
L_l\longleftrightarrow \check{R}_{\check{l}}, \qquad l_R\longleftrightarrow \check{l}_L
\label{l mirror parity trans}
\end{equation}
For invariance, it is also important that we choose the parameters $b_l$ in (\ref{L 3}) and (\ref{L mirror 3})  with the same subscripts to  be equal
(but $b_l$'s with different subscripts $l$ are different). 

\subsection{Equations of free charged leptons, TMT effective Yukawa couplings and expressions for  masses of charged leptons}

Here I am going to study the structure of mass terms for ordinary and mirror charged leptons. The interaction of fermions with gauge bosons will be discussed later. 
Although ultimately we are interested in  particle physics in Minkowski space, as we already know
  we have to follow the TMT procedure. As will become clear in Secs.VII and VIII, we are describing gauge bosons and fermions  in the background formed by inflaton and classical Higgs fields during cosmological evolution. The latter will also be explored in Sections VII and VIII.
  This approach means that gauge bosons and fermions are treated as local fluctuations on the background and their contribution to the constraint is negligible.
 
It is easy to see that omitting the interactions of fermions with gauge bosons, the  parts  of the action (\ref{S3})-(\ref{L mirror 3}) describing charged leptons $l=e, \mu, \tau$ and their mirror partners $\check{l}=\check{e}, \check{\mu}, \check{\tau}$, in unitary gauge, can be represented in the following form 
\begin{equation}
 S_{l}=\int d^4x\sqrt{-g}
\left[(b_l+\zeta )\frac{i}{2}\left(\overline{l}\gamma^kV^{\mu}_k\nabla_{\mu}^{(0)}l -(\nabla_{\mu}^{(0)}\overline{l})\gamma^kV^{\mu}_k\right)
-(b_l-\zeta )\frac{y^{(ch)}}{\sqrt{2}}\overline{l}l\underline{\varphi}(x) \right]
\label{3l action}
\end{equation}
\begin{eqnarray}
\check{S}_{\check{l}}=\int d^4x\sqrt{-g}
\left[(b_l-\zeta )\frac{i}{2}\left(\overline{\check{l}}\gamma^kV^{\mu}_k\nabla_{\mu}^{(0)}\check{l}
-(\nabla_{\mu}^{(0)}\overline{\check{l}})\gamma^kV_{k}^{\mu}\check{l}\right)
-(b_l+\zeta )\frac{y^{(ch)}}{\sqrt{2}}\overline{\check{l}}\check{l}\check{\underline{\varphi}}(x) \right]
\label{3l mirror action}
\end{eqnarray}
where
\begin{equation}
\nabla_{\mu}^{(0)}=\partial_{\mu}+\frac{1}{2}\omega_{\mu}^{ik}\sigma_{ik}
\label{nabla 0}
\end{equation}
As is usually done in SM, the lepton field $l$ is defined here as the sum of the charged lepton components of $L_l$ and $l_R$, i.e. $l=l_l+l_R$, and similarly 
 $\check{l}==\check{l}_R+\check{l}_L$.

Now our goal is to find the Dirac equations for the SM leptons $l$ and their mirror partners $\check{l}$ in the Einstein frame.  Variation of (\ref{3l action}) with respect to $\overline{l}$  leads to the equation 
\begin{equation}
iV^{\mu}_k\gamma^k\partial_{\mu}l
+\frac{i}{4}V^{\mu}_k\omega_{\mu}^{ij}\left(\gamma^k\sigma_{ij}+\sigma_{ij}\gamma^k\right)l
+\frac{i}{2\sqrt{-g}(b_l+\zeta)}\partial_{\mu}\left[\sqrt{-g}(b_l+\zeta)V^{\mu}_k\right]\gamma^kl+
\frac{\zeta -b_l }{\zeta+b_l}\frac{y^{(ch)}}{\sqrt{2}}\underline{\varphi}(x)l=0
 \label{l eq orig}
\end{equation}
The unwanted 3rd term appears in all spinor equations  and it is eliminated by the appropriate transition to the Einstein frame.  With the gravitational part of the action $S_{gr}$ as in Eq.(\ref{S-gr}), in the case of leptons $l$, the transformation of the metric tensor,  Eq.(\ref{gmunuEin})  (which means that the vierbien is transformed by $V_k^{\mu}=\sqrt{\zeta +1}\tilde{V}_k^{\mu}$),
 has to be supplemented with the spinor $l$ transformation
 \begin{equation}
\l=\frac{(\zeta +1)^{3/4}}{\sqrt{\zeta+b_l}}l^{\prime}.
 \label{l Ein}
\end{equation}
After the transition to the Einstein frame, the equation for a  regular (SM) charged lepton  $l^{\prime}$ is as follows
\begin{equation}
i\tilde{V}_k^{\mu}\gamma^k\left[\partial_{\mu}
+\frac{i}{2}\tilde{\omega}_{\mu}^{ij}\sigma_{ij}\right]l^{\prime}+
\frac{\zeta -b_l }{\zeta+b_l}\left(\frac{1+\zeta_0}{1+\zeta}\right)^{1/2}\frac{y^{(ch)}}{\sqrt{2}}\varphi(x) l^{\prime}=0,  \qquad   l^{\prime}=e, \mu, \tau 
 \label{l eq Ein}
\end{equation}
where $\tilde{\omega}_{\mu}^{ij}$ is the spin-connection in the Einstein frame and it coincides with the spin connection in the Rimannian space-time with the vierbein field $\tilde{V}_k^{\mu}$. The redefinition (\ref{redefin phi h}) of the Higgs field was also used here. In what follows we will omit prime for particles $e, \mu, \tau$.

In a similar way, varying (\ref{3l mirror action}) with respect to $\overline{\check{l}}$   we get the equation 
\begin{equation}
iV^{\mu}_k\gamma^k\partial_{\mu}\check{l}
+\frac{i}{4}V^{\mu}_k\omega_{\mu}^{ij}\left(\gamma^k\sigma_{ij}+\sigma_{ij}\gamma^k\right)\check{l}
+\frac{i}{2\sqrt{-g}(b_l-\zeta)}\partial_{\mu}\left[\sqrt{-g}(b_l-\zeta)V^{\mu}_k\right]\gamma^k\check{l}+
\frac{\zeta+b_l}{\zeta-b_l}\frac{y^{(ch)}}{\sqrt{2}}\check{\underline{\varphi}}(x)\check{l}=0
 \label{l mirror eq orig}
\end{equation}
For mirror leptons,  the spinor $\check{l}$ transformation  to  the Einstein frame,  instead of (\ref{l Ein}),  has to be  as follows
\begin{equation}
\check{l}=\frac{(\zeta+1)^{3/4}}{\sqrt{\zeta-b_l}}\check{l}^{\prime},
 \label{l mirror Ein}
\end{equation}
and the resulting Dirac equation  for the mirror charged lepton $\check{l}$ in  the Einstein frame reads  
\begin{equation}
i\tilde{V}_k^{\mu}\gamma^k\left[\partial_{\mu}
+\frac{i}{2}\tilde{\omega}_{\mu}^{ij}\sigma_{ij}\right]\check{l}^{\prime}+
\frac{\zeta+b_l }{\zeta-b_l}\left(\frac{1+\zeta_0}{1+\zeta}\right)^{1/2}\frac{y^{(ch)}}{\sqrt{2}}\check{\varphi}(x)\check{l}^{\prime}=0 \qquad \check{l}=\check{e}, \check{\mu}, \check{\tau}.
 \label{l mirror eq Ein}
\end{equation}

A fundamentally new result of the implementation of the idea of the universality of the Yukawa coupling constant in the underlying action of MESM is the emergence of the TMT effective $\zeta$-dependent Yukawa coupling parameters for SM charged leptons and their mirror partners
\begin{equation}
f_l(\zeta)=y^{(ch)}\cdot\frac{\zeta -b_l }{\zeta+b_l}\left(\frac{1+\zeta_0}{1+\zeta}\right)^{1/2}; \qquad 
f_{\check{l}}(\zeta)=y^{(ch)}\cdot\frac{\zeta+b_l }{\zeta-b_l}\left(\frac{1+\zeta_0}{1+\zeta}\right)^{1/2}
\label{fl TMT eff}
\end{equation}

In the vicinty of the vacuum of the Higgs fields (described in Sec.IV.B) where $\zeta=\zeta_0$,  \, $\varphi(x)=v+h(x)$,  \, $\check{\varphi}(x)=\check{v}+\check{h}(x)$, the Yukawa couplings produce the masses of the regular and mirror charged leptons generated by  the $SU(2)\times U(1)\rightarrow U(1)$ SSB, which are respectively
\begin{equation}
m_l=f_{l,vac}\frac{v}{\sqrt{2}}; \quad f_{l,vac}\equiv f_l(\zeta_0)=\frac{\zeta_0-b_l}{\zeta_0+b_l}y^{(ch)}, \quad l=e, \mu, \tau, \quad b_l=b_e, b_{\mu}, b_{\tau};
\label{ml and fl vac}
\end{equation}
\begin{equation}
 m_{\check{l}}=f_{\check{l},vac}\frac{\check{v}}{\sqrt{2}}; \quad     f_{\check{l},vac}\equiv f_{\check{l}}(\zeta_0)=\frac{\zeta_0+b_l}{\zeta_0-b_l}y^{(ch)} \quad \check{l}=\check{e}, \check{\mu}, \check{\tau}, \quad  \quad b_l=b_e, b_{\mu}, b_{\tau}.
\label{mirror ml and fl vac}
\end{equation}

For the ratio of the masses of the mirror counter-partners we obtain
\begin{equation}
\frac{m_{\check{l}}}{m_l}=\frac{\check{v}}{v}\cdot\left(\frac{\zeta_0+b_l}{\zeta_0-b_l}\right)^2,
\label{ratio masses}
\end{equation} 
which   can be made as large as necessary by simply choosing parameters $b_l$ so that $\zeta_0-b_l\ll \zeta_0+b_l$.

It is interesting that the following {\em 'multiplication rule'} holds
\begin{equation}
m_em_{\check{e}}=m_{\mu}m_{\check{\mu}}=m_{\tau}m_{\check{\tau}}=\frac{\left(y^{(ch)}\right)^2}{2}v\check{v}.
\label{rule}
\end{equation} 
Therefore, in the family of charged mirror leptons, the mirror electron $\check{e}$  is the heaviest, and the mirror
 tau-lepton $\check{\tau}$ is the lightest.

\subsection{Neutrino masses. New approach to the problem}

By developing an approach for neutrinos similar to that for charged leptons, the idea of a “seesaw” mechanism can naturally be realized without the addition of a specially selected hypothetical sterile heavy neutrino. To simplify this first attempt to implement the intended idea, in this paper I will completely ignore neutrino mixing and limit myself to considering only the electron neutrino $\nu_{e}$ and its mirror partner ${\check \nu}_{e}$. For brevity, in what follows I will omit the index e in $\nu_{e}$. Keepping all symmetries of the Lagrangians (\ref{L 3})-(\ref{L mirror 3}), we have to add the isosinglet kinetic neutrino terms and the terms of the  neutrino-Higgs Yukawa couplings 
\begin{equation}
\mathbb{L}_{\nu,(isosinglet)}=
 \frac{i}{2} (b_{\nu}+\zeta )
\left[\overline{\nu_R}\gamma^{\mu}\nabla_{\mu}\nu_R -(\nabla_{\mu}\overline{\nu_R}\gamma^{\mu}\nu_R\right]
-(b_{\nu}-\zeta)y^{(\nu)}\left(\overline{L}\Phi_c \nu_R +
\overline{\nu_R}\Phi_c^{\dag}L\right)
 \label{nu 3}
\end{equation}
\begin{equation}
\mathbb{\check{L}}_{\check{\nu},(isosinglet)}=
 \frac{i}{2} (b_{\nu}-\zeta )
\left[\overline{\check{\nu}_L}\gamma^{\mu}\nabla_{\mu}\check{\nu}_L -(\nabla_{\mu}\overline{\check{\nu}_L}\gamma^{\mu}\check{\nu}_L\right]
-(b_{\nu}+\zeta)y^{(\nu)}\left(\overline{\check{R}}\Phi_c \check{\nu}_L +
\overline{\check{\nu}_L}\check{\Phi}_c^{\dag}\check{R}\right)
 \label{nu mirror 3}
\end{equation}
 Here $\Phi_c$ and $\check{\Phi}_c$ are the charged conjugated to the Higgs doublets (\ref{Higgs  doublets}). Notice that there is no reason to expect that Yukawa coupling constant in the underlying action is the same  for charged leptons and for neutrinos. This is why in Eqs.(\ref{nu 3}) and (\ref{nu mirror 3}) the neutrino Yukawa coupling constant $y^{(\nu)}$ is introduced.

In the previous subsection, we showed how to avoid the need to introduce an inexplicably large variation in the values of the Yukawa coupling constants for different generations of leptons. 
To generate huge mass differences in members of the same fermion generation, such as between the masses of electron and electron neutrino,   one can choose both the appropriate value of the parameter $b_{\nu}$ and the neutrino Yukawa coupling $y^{(\nu)}$.   The differences, i.e. $b_{\nu}\neq b_l$ and $y^{(\nu)}\neq y^{(ch)}$,  does not affect the symmetries of the theory.

Further, with neutrino fields, we must perform all the procedures that were made in previous subsection for charged leptons, and go to the Einstein frame. For neutrino fields, instead of transformations (\ref{l Ein}) and (\ref{l mirror Ein}) we have to define
\begin{equation}
\nu=\frac{(\zeta+1)^{3/4}}{\sqrt{\zeta+b_{\nu}}}\nu^{\prime};  \qquad
\check{\nu}=\frac{(\zeta+1)^{3/4}}{\sqrt{\zeta-b_{\nu}}}\check{\nu}^{\prime},
 \label{nu and nu mirror Ein}
\end{equation}
where, as usually, $\nu$ and $\check{\nu}$ are sums of the appropriate isodublet $\nu_L$,  $\check{\nu}_R$ and isosinglet $\nu_R$, $\check{\nu}_L$ components: $\nu=\nu_L+\nu_R$ and $\check{\nu}=\check{\nu}_R+\check{\nu}_L$.
After the transition to the Einstein frame, the equations for  neutrino $\nu^{\prime}$ and mirror neutrino $\check{\nu}^{\prime}$ are similar to those for the charged leptons (\ref{l eq Ein}) and (\ref{l mirror eq Ein}) with the TMT effective $\zeta$-dependent Yukawa couplings
\begin{equation}
f_{\nu}(\zeta)=y^{(\nu)}\cdot\frac{\zeta -b_{\nu}}{\zeta+b_{\nu}}\left(\frac{1+\zeta_0}{1+\zeta}\right)^{1/2}; \qquad 
f_{\check{\nu}}(\zeta)=y^{(\nu)}\cdot\frac{\zeta+b_{\nu}}{\zeta-b_{\nu}}\left(\frac{1+\zeta_0}{1+\zeta}\right)^{1/2}
\label{fnu TMT eff}
\end{equation} 
In the vacuum ($\zeta=\zeta_0$,   $\varphi=v$,  $\check{\varphi}=\check{v}$), we obtain the following expressions for masses of the SM electron neutrino and its mirror partner
\begin{equation}
m_{\nu}=f_{\nu,vac}\frac{v}{\sqrt{2}}; \qquad  f_{\nu,vac}\equiv f_{\nu}(\zeta_0)=\frac{\zeta_0 -b_{\nu}}{\zeta_0+b_{\nu}}y^{(\nu)};
\label{mnu fnu vac}
\end{equation}
\begin{equation}
 m_{\check{\nu}}=f_{\check{\nu},vac}\frac{\check{v}}{\sqrt{2}}; \qquad  f_{\check{\nu},vac}\equiv f_{\check{\nu}}(\zeta_0) = \frac{\zeta_0+b_{\nu}}{\zeta_0-b_{\nu}}y^{(\nu)}.
\label{mirror mnu fnu vac}
\end{equation}
For the ratio of their masses we obtain
\begin{equation}
\frac{m_{\check{\nu}}}{m_{\nu}}=\frac{\check{v}}{v}\left(\frac{\zeta_0+b_{\nu}}{\zeta_0-b_{\nu}}\right)^2.
\label{nu ratio masses}
\end{equation} 

\subsection{Quarks in MESM: some pecularities}

As will be shown in the next section, with an appropriate choice of small deviations of the absolute values of the parameters $b$, $b_i$ ($i=e, \mu, \tau, \nu$) from the vacuum value $\zeta_0$ of the function $\zeta$, the desired hierarchy between the corresponding quantities in the SM sector and in the mirror sector can be realized: 1) between the VEV's  and the masses of  the Higgs bosons; 2) between the masses of the gauge bosons; 3) between the masses of leptons. 
These results are possible due to the appropriate (opposite) choice of  ($\pm$) signs in linear combinations ($b_i\sqrt{-g}\pm\Upsilon$) in the underlying action for SM fields and their mirror partners. 

However, we have not yet touched on how to describe quarks in the mirror-extended electroweak SM.
In order to preserve the general principles of building the model, we must assume that, like leptons, the Yukawa coupling constants in the underlying action must also be universal: one for all up-quarks and, probably, another for all down-quarks (with the corresponding modification caused by the Kobayashi-Maskawa mixing  matrix). Therefore, in order to obtain the mass of $u$-quark, the appropriate TMT effective Yukawa coupling constant must be sufficiently small. But then we will faced with the problem of obtaining the top quark mass, which is $m_t\approx\frac{v}{\sqrt{2}}$. A possible solution to this problem may be to take into account the arbitrariness in the choice of signs in ($b_q\sqrt{-g}\pm\Upsilon$), ($q=u, d, c, s, t, b$), in the underlying action. Indeed, if we choose the signs in ($b_q\sqrt{-g}\pm\Upsilon$) for quarks opposite to those for leptons, then, as will be shown in the next section,  with the universal Yukawa coupling constant one can simultaneously obtain both the masses of light quarks and  the mass $m_t\approx 173 GeV$ of the top quark. 

The implementation of such an approach to describing quarks with allowance for quark mixing would require too many detailed calculations, which could become an obstacle to demonstrating the main issues of the model under study. Therefore, I will focus here on open questions of principle. The description of the SM up-quarks mass generation and the estimated prediction of the mirror up-quarks mass is just one of them.
Thus, we start with an action  $S_q$ for three generations of quarks,  which has a structure similar to the action (\ref{S3})-(\ref{L mirror 3}) for leptons, but with opposite signs in ($b_q\sqrt{-g}\pm\Upsilon$)  compared to the lepton case. In addition, instead of the universal Yukawa constant for charged leptons $y^{(ch)}$, the universal Yukawa constant $y^{(up)}$ for up-quarks appears now.
Repeating calculations similar to those performed in Sec.V.B for three generations of charged leptons and performing
  transition to the Einstein frame  by transforming the metric tensor, Eq.(\ref{gmunuEin}),
   complemented by the  transformation of the  spinors $q$
 \begin{equation}
q=\frac{(\zeta +1)^{3/4}}{\sqrt{\zeta-b_q}}q^{\prime} \qquad   q= u, c, t,
 \label{q Ein}
\end{equation}
we obtain the equations for SM up-quarks   $q^{\prime}$  (where again interactions with gauge bosons are omitted)
\begin{equation}
i\tilde{V}_k^{\mu}\gamma^k\left[\partial_{\mu}
+\frac{i}{2}\tilde{\omega}_{\mu}^{ij}\sigma_{ij}\right]q^{\prime}+
\frac{\zeta+b_q}{\zeta-b_q}\left(\frac{1+\zeta_0}{1+\zeta}\right)^{1/2}\frac{y^{(up)}}{\sqrt{2}}\varphi q^{\prime}=0.  
 \label{q eq}
\end{equation}
Similarly, for mirror quarks $\check{q}$,
after the transition to the Einstein frame including the transformation of the  spinors $\check{q}$
\begin{equation}
\check{q}=\frac{(\zeta +1)^{3/4}}{\sqrt{\zeta+b_q}}\check{q}^{\prime} \qquad   q= u, c, t,
 \label{q mirror Ein}
\end{equation}
we obtain the equations for  mirror up-quarks $\check{q}^{\prime}$
\begin{equation}
i\tilde{V}_k^{\mu}\gamma^k\left[\partial_{\mu}
+\frac{i}{2}\tilde{\omega}_{\mu}^{ij}\sigma_{ij}\right]\check{q}^{\prime}+
\frac{\zeta-b_q}{\zeta+b_q}\left(\frac{1+\zeta_0}{1+\zeta}\right)^{1/2}\frac{y^{(up)}}{\sqrt{2}}\check{\varphi}\check{q}^{\prime}=0.  
 \label{q mirror eq}
\end{equation}
It follows from Eqs.(\ref{q eq}) and (\ref{q mirror eq}) that the TMT effective Yukawa coupling constants (in the tree aproximation) for quarks and their mirror partners depend on $\zeta$ as follows
\begin{equation}
f_q(\zeta)=y^{(up)}\frac{\zeta+b_q}{\zeta-b_q}\left(\frac{1+\zeta_0}{1+\zeta}\right)^{1/2}, \qquad f_{\check{q}}(\zeta)=y^{(up)}\frac{\zeta-b_q}{\zeta+b_q}\left(\frac{1+\zeta_0}{1+\zeta}\right)^{1/2}.
 \label{fuk eff q zeta}
\end{equation}

 In the vacuum, i.e. when $\zeta=\zeta_0$, $\varphi=v$, $\check{\varphi}=\check{v}$,  we get the following expressions for the masses of three generations of the SM up-quarks and their mirror partners
\begin{equation}
m_q=f_{q,vac}\frac{v}{\sqrt{2}};  \quad f_{q,vac}\equiv f_q(\zeta_0)=\frac{\zeta_0 +b_q }{\zeta_0-b_q}y^{(up)},  \quad q=u, c, t; \quad  
 b_q=b_u, b_c, b_t 
\label{fq mq vac}
\end{equation}
\begin{equation}
 m_{\check{q}}=f_{\check{q},vac}\frac{\check{v}}{\sqrt{2}};  \quad f_{\check{q},vac}\equiv f_{\check{q}}(\zeta_0)=\frac{\zeta_0-b_q}{\zeta_0 +b_q}y^{(up)}, \quad \check{q}=\check{u}, \check{c}, \check{t}; \quad b_q=b_u, b_c, b_t.
\label{mirror fq mq vac}
\end{equation}
For the ratio of their masses we obtain
\begin{equation}
\frac{m_{\check{q}}}{m_q}=\frac{\check{v}}{v}\left(\frac{\zeta_0-b_q}{\zeta_0+b_q}\right)^2.
\label{q ratio masses}
\end{equation}

\section{MESM as a theory beyond the Standard Model}


\subsection{A typical example of the implementation of hierarchy in VEV's,  TMT effective Yukawa coupling constants and particle masses.}

In the next section, we will find that in  the  vacuum the  scalar function  $\zeta$ has a value $\zeta_0\approx 0.248$. Using this result, we can, along with the masses of the SM particles, obtain the masses of the mirror partners of the SM particles in a fairly wide range due the appropriate selection of the parameters $b$ and  $b_i$, ($i=e, \mu, \tau, u, c, t$).  In this section, I am going to demonstrate how this can be done when the absolute values of all these parameters are close to $\zeta_0$. Here and further in this paper, only this numerical example and its consequences are studied. Of course, with a more systematic study of the model and comparison with observational data, it may be necessary to select other values of the parameters.

\subsubsection{VEV's and masses of the Higgs and gauge bosons}

Let us choose the value of the parameter $b$ by 
\begin{equation}
b=\zeta_0 -1.3\cdot 10^{-12}.
 \label{bh choice}
\end{equation}
First of all, it should be noted that with this choice of the parameter $b$ it follows from Eqs.(\ref{v under 2}), (\ref{redefin phi h}), (\ref{h eq}) that
\begin{equation}
v^2=\frac{\zeta_0-b}{(1+\zeta_0)(\zeta_0+b)}\frac{m^2}{\lambda}\approx 0.8\cdot 10^{-12}\frac{m^2}{\lambda}
 \label{v2 with choice}
\end{equation}
and for the square of the mass of the SM Higgs boson in MESM we get
\begin{equation}
m_h^2=2\lambda v^2 \approx 2\cdot 0.8\cdot 10^{-12}m^2.                         
 \label{m h2}
\end{equation}
Comparing this result of MESM with the corresponding result in SM $(m_h^2=2m^2$) we see that the relation between the values of the parameter $m^2$ in these models is as follows
\begin{equation}
\frac{\zeta_0-b}{(1+\zeta_0)(\zeta_0+b)}{m^2}|_{MESM}\approx 0.8\cdot 10^{-12}m^2|_{MESM}= m^2|_{SM}
\label{m  with choice}
\end{equation}
Within MESM, as in SM, there is no theoretical prediction for the value of the Higgs mass, since the parameter $m$ remains undefined. The presence of an extremely small multiplier $\zeta -b$ in expression (\ref{v2 with choice}) for $v^2$ is compensated by the corresponding redefinition of the  parameter $m$.
Thus, the difference between  (\ref{v2 with choice}) and the standard expression $v^2=\frac{m^2}{\lambda}$ does not affect the results in any way, at least at the tree level.
Since the gauge coupling constants $g, g'$ appear  in the SM sector of MESM exactly as in the SM, we can follow the standard approach  for obtaining VEV of the SM Higgs field in MESM. Indeed, using the standard expression for  the W-boson mass found in Eq.(\ref{gauge masses}), and the relation with the Fermi constant $G_F$
\begin{equation}
\frac{g^2}{8M_W^2}=\frac{G_F}{\sqrt{2}}
\label{MW F}
\end{equation}
obtained in the low-energy limit (see also below in Sec.VI.C), we obtain that, as expected, VEV of  the SM Higgs field is $v\approx 246 GeV$.

Now, using Eqs.(\ref{v under mir 2}), (\ref{v both 2}), (\ref{redefin phi h}), (\ref{h mir eq}) and the SM Higgs mass $m_h\approx 125 GeV$, we obtain the following VEV of the mirror Higgs field and   the value  (not final) of  the mirror Higgs boson:
\begin{equation}
\check v\approx  10^{14} GeV  \qquad  \textsf{and} \qquad  m_{{\check h},0}\approx 5\cdot 10^{13} GeV,
\label{check v v}
\end{equation}
where the subscript 0 in $m_{{\check h},0}$ indicates that the possible contribution to the mass of the mirror Higgs boson arising from $\delta\zeta$ in 
Eq.(\ref{h mir eq}) is omitted. As we will see in Sec.VIII.D, the corresponding contribution to the  mass of the SM Higgs boson in Eq.(\ref{h eq}) is negligible.
 
Using  Eq.(\ref{gauge masses}) and the values of the masses of the SM bosons $W$ and $Z$,  we obtain that
the masses of the mirror bozons ${\check W}$ and ${\check Z}$   equal to
\begin{equation}
M_{\check W}\approx  3.1\cdot 10^{13} GeV, \qquad M_{\check Z}\approx  3.5\cdot 10^{13} GeV.
\label{M W G f}
\end{equation}

\subsubsection{Masses of charged leptons}

Now let us turn to the demonstration of how, by choosing the values of the parameters $b_l$ ($l=e, \mu, \tau$), one can simultaneously achieve the necessary hierarchy of masses of  charged SM leptons and obtain predictions for the masses of their mirror partners.
Using results of Sec.V.B, based on the idea that  the  Yukawa coupling constant  $y^{(ch)}$  has the same universal  value for all generations of charged leptons, let us parameterize it as follows 
\begin{equation}
y^{(ch)}=p_l\cdot 10^{-n_l}, 
\label{fl parametrization}
\end{equation}
where the parameter $p_l\sim \mathcal{O}(1)$ and $n_l$ is positive integer. Using the  mass value  of the SM electron we obtain from 
Eq.(\ref{ml and fl vac}) that
\begin{equation}
b_e= \zeta_0\frac{p_l-2.9\cdot 10^{-(6-n_l)}}{p_l+2.9\cdot 10^{-(6-n_l)}}.
\label{be} 
\end{equation}

From Eq.(\ref{mirror ml and fl vac}), using (\ref{be}), we get that  the TMT effective Yukawa coupling of the mirror electron  reads
\begin{equation}
f_{\check{e},vac}\equiv f_{\check{e}}(\zeta_0)\approx 0.35p_l^2\cdot 10^{6-2n_l}.
\label{fe mirr vac} 
\end{equation}
To provide the possibility of using the loop expansion for calculating quantum corrections to the tree approximation, it is necessary to choose the parameter $n_l$ so that $f_{\check{e},vac}<1$, that is $n_l\geq 3$. Let us choose $n_l=3$, i.e.
\begin{equation}
y^{(ch)}=p_l\cdot 10^{-3}. 
\label{fl parametrization final}
\end{equation}
Then 
\begin{equation}
b_e\approx \zeta_0\left(1-\frac{5.8}{p_l}\cdot 10^{-3}\right); \quad f_{e,vac}\approx 2.9\cdot 10^{-6}; \quad f_{\check{e},vac}\approx 0.35p_l^2
\label{be fcheck e}
\end{equation}
and
\begin{equation}
m_{\check{e}}\approx 2.4 p_l^2\cdot 10^{13}GeV.
\label{me mirror param} 
\end{equation} 

With $y^{(ch)}$, Eq.(\ref{fl parametrization final}), using the mass value of the SM muon and proceeding in a similar manner, we obtain
\begin{equation}
b_{\mu}= \zeta_0\frac{p_l-0.6}{p_l+0.6}.
\label{bmu} 
\end{equation}
and hence
\begin{equation}
f_{\mu,vac}\approx 0.6\cdot 10^{-3}; \quad f_{\check{\mu},vac}\approx 1.7p_l^2\cdot 10^{-3}; \quad m_{\check{\mu}}\approx 1.2 p_l^2\cdot 10^{11}GeV.
\label{f and m mu mirror vac} 
\end{equation} 

Using the mass value of the SM $\tau$-lepton and repeating similar evaluations we obtain
\begin{equation}
b_{\tau}\approx -\zeta_0\left(1-\frac{1}{5}p_l\right).
\label{btau} 
\end{equation}
Then the corresponding TMT effective Yukawa couplings   and the mass of the mirror $\tau$-lepton are 
\begin{equation}
f_{\tau,vac}\approx 10^{-2}; \quad f_{\check{\tau},vac}\approx p_l^2\cdot 10^{-2}; \qquad m_{\check{\tau}}\approx \frac{p_l^2}{\sqrt{2}}\cdot 10^{10}GeV.
\label{f and m tau mirror vac} 
\end{equation} 

The multiplication rule for masses of charged SM leptons and their mirror partners is as follows:
\begin{equation}
m_e m_{\check{e}}=m_{\mu}m_{\check{\mu}}=m_{\tau}m_{\check{\tau}}=\left(\frac{y^{(ch)}}{\sqrt{2}}\right)^2\cdot v\check{v}\approx (1.1p_l\cdot 10^5 GeV)^2.
\label{rule num l}
\end{equation}

\subsubsection{Masses of quarks}

Similar to what was done in the case of charged leptons, let us parameterize the universal Yukawa coupling constant for up-quarks as follows 
\begin{equation}
y^{(up)}=p_q\cdot 10^{-n_q}, 
\label{fq parametrization}
\end{equation}
where the parameter $p_q\sim \mathcal{O}(1)$ and $n_q$ is positive integer. Using the expression for the mass of the $u$-quark in 
Eq.(\ref{fq mq vac}) and its known value $m_u\approx 2.2MeV$, \cite{ParticleDataGroup}, we get
\begin{equation}
\frac{\zeta_0+b_u}{\zeta_0-b_u}\approx \frac{1.3}{p_q}\cdot 10^{n_q-5}
\label{bu expr} 
\end{equation}
Then  it follows from  Eq.(\ref{mirror fq mq vac}) that  the TMT effective Yukawa coupling constant for the mirror $\check{u}$-quark reads
\begin{equation}
f_{\check{u},vac}\equiv f_{\check{u}}(\zeta_0)\approx 0.8p_q^2\cdot 10^{5-2n_q}.
\label{fu  mirror vac} 
\end{equation}
To ensure $f_{\check{u},vac}<1$, the condition $n_q>2.5$ must be satisfied. Let us choose $n_q=3$. Then
\begin{equation}
b_u\approx -\zeta_0\left(1-\frac{2.6}{p_q}\cdot 10^{-2}\right);  \quad f_{u,vac}\approx 1.3\cdot 10^{-5}; \quad f_{\check{u},vac}\approx  0.08p_q^2;  \qquad  .
\label{bu}
\end{equation}
and
 \begin{equation}
m_{\check{u}}=f_{\check{u},vac}\cdot\frac{\check{v}}{\sqrt{2}}\approx 5.7p_q^2\cdot 10^{12}GeV.
\label{m u mir}
\end{equation}

Using Eq.(\ref{fq mq vac})  for the $c$-quark, the value of its  mass  $m_c\approx 1.28 GeV$, and Eq.(\ref{mirror fq mq vac}) we obtain 
 \begin{equation}
b_c\approx \zeta_0\frac{7.3-p_q}{7.3+p_q}; \quad f_{c,vac}\approx 7.3\cdot 10^{-3}; \quad  f_{\check{c},vac}\approx 1.4p_q^2\cdot 10^{-4}
\label{bc} 
\end{equation}
and
\begin{equation}
m_{\check{c}}=f_{\check{c},vac}\cdot\frac{\check{v}}{\sqrt{2}}\approx p_q^2\cdot 10^{10}GeV.
\label{m c mir}
\end{equation}

By using Eqs.(\ref{fq mq vac}) and (\ref{mirror fq mq vac}) and carrying out the same calculations for  the top-quark  
with the known value of its mass $m_t\approx 173GeV$, \cite{ParticleDataGroup}, we get
\begin{equation}
b_t\approx \zeta_0\left(1-2p_q\cdot 10^{-3}\right); \quad  f_{t,vac}=\sqrt{2}\frac{m_t}{v}\approx 1; \quad f_{\check{t},vac}
\approx 0.7p_q^2\cdot 10^{-6}
\label{bt} 
\end{equation}
and
\begin{equation}
m_{\check{t}}=f_{\check{t}}^{(vac)}\cdot\frac{\check{v}}{\sqrt{2}}\approx 0.7p_q^2\cdot 10^{8}GeV.
\label{m u mir}
\end{equation}

The multiplication rule for masses of the SM up-quarks and their mirror partners is as follows:
\begin{equation}
m_um_{\check{u}}=m_{c}m_{\check{c}}=m_{t}m_{\check{t}}=\left(\frac{y^{(up)}}{\sqrt{2}}\right)^2\cdot v\check{v}\approx (1.1p_q\cdot 10^5 GeV)^2.
\label{rule num q}
\end{equation}
This approach to the quark sector leads to the prediction that the masses of the mirror quarks are less than the masses of the mirror leptons.
Special attention should be paid to a very interesting result: there are no theoretical obstacles to the choice of $p_l=p_q=p$, which means that the same Yukawa coupling constant in the underlying action can be universal for both three generations of charged leptons and three generations of up-quarks, i.e. $y^{(ch)}=y^{(up)}=p\cdot 10^{-3}$. Then  the multiplication rules  for masses of the charged leptons (\ref{rule num l}) and for up-quarks  (\ref{rule num q}) coincide.

\subsubsection{Masses of neutrinos}

With the results of Sec.V.C, the approach used  for charged leptons can also be applied to numerically estimate the masses of  mirror neutrinos.
In the absence of reliable data on the visible neutrino masses, the application of this approach  can give a very rough estimate. For this reason, it makes no sense to take into account both  the existence of three types of neutrinos and neutrino mixing  in the framework of such estimates. Therefore, in this paper it makes sense to limit ourselves to demonstrating the possibility of obtaining an effect usually reached by a see-saw  mechanism. 

Apparently, there is no reason to assume that the Yukawa coupling constants for charged leptons and neutrinos are equal. 
With the known upper constraint\cite{ParticleDataGroup} on the electron neutrino mass, let's take $m_{\nu}\approx 1\, eV$. 
Then it follows from Eq.(\ref{mnu fnu vac}) that $y^{(\nu)}\cdot (\zeta_0-b_{\nu})\approx 3\cdot 10^{-12}$.
 Substituting this into Eq.(\ref{mirror mnu fnu vac}), we get
\begin{equation}
m_{\check{\nu}}=f_{\check{\nu},vac}\frac{\check{v}}{\sqrt{2}}\approx 10^{25}\left(y^{(\nu)}\right)^2GeV.
\label{m mir via f2}
\end{equation}
Parametrization 
$y^{(\nu)}=p_{\nu}\cdot 10^{-n_{\nu}}$ \, ($p_{\nu}\sim \mathcal{O}(1)$) leads to
\begin{equation}
m_{\check{\nu}}\approx p_{\nu}^2\cdot 10^{25-2n_{\nu}}GeV.
\label{m mir via nnu}
\end{equation}
To ensure $m_{\check{\nu}}\lesssim 10^{17}GeV$, it is necessary that $n_{\nu}\geq 4$. 
If a mirror neutrino is the lightest of the mirror leptons (and, therefore, stable), then, given the masses of charged leptons  predicted in the previous subsections, we get $n_{\nu}\geq 8$ and  
\begin{equation}
f_{\check{\nu},vac}\lesssim 10^{-4}, \quad \zeta_0-b_{\nu}\geq 3\cdot 10^{-4}, \quad m_{\check{\nu}}\lesssim 10^{9}GeV. 
\label{f and m mirr vac }
\end{equation}

\subsection{Structure of  interactions  in MESM}

In Secs.IV and V, we constructed the underlying action of MESM
\begin{equation}
S_{MESM}=S_{H}+S_{gauge}+S_3^{(l)}+S_{\nu}+S_3^{(q)},
\label{S MESM}
\end{equation} 
where the actions for Higgs bosons $S_H$, gauge bosons  $S_{gauge}$ and charged leptons $S_3^{(l)}$ are defined by 
Eqs.(\ref{S-H 12}), (\ref{S gauge}) and (\ref{S3}) respectively. The structure of the underlying actions for the neutrino $S_{\nu}$
 and for up-quarks $S_3^{(q)}$ was explained in Secs.V.C and V.D. Focusing there on studying the equations of free fields, 
we obtained expressions for the VEV's of the Higgs fields and particle masses. To this end, in the explicitly written equations 
in the Einstein frame for fermions and gauge fields in the vacuum of Higgs fields, we ignored the terms describing the field 
interactions; only selfinteractions of Higgs fields were obtained and their structure is the same as in the SM, up to corrections
 that could arise from the term with $\delta\zeta$
 (see Eqs.(\ref{h eq}) and (\ref{h mir eq})). To see the structure of interactions for both SM and mirror particles, a more
 complete analysis is needed. This must be done according to the TMT procedure described in Sec.III.B. In general, when 
$\zeta\neq const.$, the implementation of this program is very complex.
However, in reality we are interested in the correspondence between MESM and SM. In this case, it is sufficient to obtain the 
TMT effective  action of the MESM near the vacuum states (\ref{Higgs doublets mesured}) of the Higgs boson and its mirror 
partner, where $\zeta=\zeta_0= const.$.

Detailed calculations show that, as in Secs.IV and V, all $\zeta$-dependent factors are absorbed by the previously obtained
 expressions for  masses  and due to the redifinitions (\ref{v both 2}), (\ref{redefin phi h}). The resulting TMT effective 
MESM action is the sum of two actions, one for the SM particles and the other for their mirror partners. The action for SM 
particles is exactly the same as the action of the GWS electroweak model. 

In the TMT effective action for mirror particles, the gauge coupling constants $g$ and $g^{\prime}$, the Weinberg angle $\theta_W$, 
and the electric charge of the mirror electron are the same as in the SM. On the whole, the structure of this action is similar 
to that in the SM, but with {\em two important differences}: 
1) weak currents have the $(V+A)$ structure instead of the 
standard $(V-A)$; 
2) the values of particle masses are different.
In addition, there are differences in the values of the coupling 
constants of the Higgs mirror field $\check{h}$ with fermions and vector bosons. 
The first is related to the fact that the Yukawa 
coupling constants are the TMT effective Yukawa parameters  for mirror fermions at
$\zeta=\zeta_0$ studied in Sec.V. The last difference is directly related to the fact that the VEV of the Higgs mirror field
 is $\check{v}$, not $v$. But the corresponding term in the Lagrangian, which also gives rise to the  masses of the mirror 
vector bosons, has the same form as in the SM:
\begin{equation}
\frac{g^2}{8}\tilde{g}^{\mu\nu}\left(2\check{W}_{\mu}^{(+)}\check{W}_{\nu}^{(-)}+
\frac{1}{\cos^2\theta_W}\check{Z}_{\mu}\check{Z}_{\nu}\right)(\check{v}+\check{h})^2
\label{gauge to higgs eff mirr coupling}
\end{equation}
As a result, the coupling constants of interactions $\check{W}\check{W}\check{h}$ and 
$\check{Z}\check{Z}\check{h}$ are $\check{v}/v$ times greater than the corresponding coupling constants in the SM.

Since I have not given here long and routine calculations, it seems appropriate to give 
a qualitative explanation of why the structure  of the TMT effective action of MESM near the vacuum turns out to
 be so good. As far as fermions are concerned, the procedure for obtaining equations 
in the Einstein frame for the SM fermions and their mirror partners differs from that carried out 
 in Sec.V, only in that the terms with gauge fields must be preserved in the covariant derivatives.
 After this, the construction of the corresponding TMT  effective Lagrangian, the variation of which leads 
to the obtained equations, is obvious.

The greatest doubts can arise in the bosonic sector. To demonstrate what happens when obtaining
 equations, it is worthwhile to write the corresponding part of the underlying action again 
(see first lines in Eqs.(\ref{S-H 12}) and (\ref{S gauge})).
 The corresponding underlying Lagrangian density in the sector of the SM bosons is the following
\begin{eqnarray}
\sqrt{-g}\mathbb{L}_{bos}^{(SM)}=\left(b\sqrt{-g}+\Upsilon\right)\Big(-\frac{1}{4} 
g^{\mu\gamma}g^{\nu\beta}\mathbf{F}_{\mu\nu}\mathbf{F}_{\gamma\beta}-\frac{1}{4} 
g^{\mu\gamma}g^{\nu\beta}B_{\mu\nu}B_{\gamma\beta}\Big)
\nonumber
\\
+\left(b\sqrt{-g}+\Upsilon\right)g^{\mu\nu}
\Big[\left(\partial_{\mu}-ig\mathbf{\hat{T} A}_{\mu}
-i\frac{g'}{2}\hat{Y}B_{\mu}\right)\Phi\Big]^{\dag}\left(\partial_{\nu}-ig\mathbf{\hat{T}A}_{\nu}-i\frac{g'}{2}\hat{Y}B_{\nu}\right)\Phi
\label{L bos SM orig}
\end{eqnarray}
As $\zeta=\frac{\Upsilon}{\sqrt{-g}}=\zeta_0$ the terms 
with gradient of $\zeta$ disappear from equations obtained 
by variation of the gauge bosons. And what is really crucual is that 
 $b+\zeta_0$ is the common  factor in all terms of these equations and it cancels 
out without affecting the equations of the gauge fields. This becomes possible due
 to the choice of the same volume measure $\left(b\sqrt{-g}+\Upsilon\right)d^4x$
 in the gauge invariant terms of the underlying action describing the gauge bosons
 and the kinetic term of the Higgs boson. The same happens with the mirror gauge 
bosons where $\left(b\sqrt{-g}-\Upsilon\right)d^4x$ is chosen as the same volume 
measure in the two corresponding terms of the underlying action 
(see second lines in Eqs.(\ref{S-H 12}) and (\ref{S gauge})).

At the same time, it is worth recalling that it is precisely thanks to these factors 
and the corresponding factors in the potentials of the Higgs fields that huge splittings
 in the values of VEV's and in the masses of the SM particles and their mirror partners become possible.

\subsection{Fermi constant for  $V+A$ four-fermionic weak interaction of mirror leptons. Typical decay widths}

Using the results formulated in the previous subsection, it is easy to conclude that
 in the limit of Minkowski space, the scattering amplitudes of $\nu_e e\rightarrow \nu_e e$ and 
 $\check{\nu}_{\check{e}} \check{e}\rightarrow \check{\nu}_{\check{e}} \check{e}$
 in the low-energy limit are respectively 
\begin{equation}
\mathcal{M}= \frac{g^2}{2M_{W}^2}\left(\bar{e}_L\gamma^k\nu^{(e)}_L\right)\left(\bar{\nu}^{(e)}_L\gamma_ke_L\right); \qquad
\check{\mathcal{M}}= 
\frac{g^2}{2M_{\check{W}}^2}\left(\bar{\check{e}}_R\gamma^k\check{\nu}^{(\check{e})}_R\right)
\left(\bar{\check{\nu}}^{(\check{e})}_R\gamma_k\check{e}_R\right).
\label{e nu ampl}
\end{equation} 
These amplitudes have to be compared with those in the effective  four fermionic descriptions
\begin{equation}
\mathcal{M}= \frac{G_F}{\sqrt{2}}\left(\bar{e}\gamma^k(1-\gamma_5)\nu^{(e)}\right)
\left(\bar{\nu}^{(e)}\gamma_k(1-\gamma_5)e\right); \qquad
\check{\mathcal{M}}= \frac{\check{G}_F}{\sqrt{2}}\left(\bar{\check{e}}\gamma^k(1+\gamma_5)\check{\nu}^{(\check{e})}\right)
\left(\bar{\check{\nu}}^{(\check{e})}\gamma_k(1+\gamma_5)\check{e}\right).
 \label{e nu 4-ferm ampl}
\end{equation}
where $G_F$ is the Fermi constant for regular $V-A$ four-fermionic weak interaction and $\check{G}_F$ 
is the appropriate Fermi constant for  $V+A$ four-fermionic weak interaction of mirror leptons. 
By using the expressions for the $W$ and $\check{W}$ masses in Eq.(\ref{gauge masses}) we obtain\cite{Lipmanov} 
\begin{equation}
\frac{\check{G}_F}{G_F}=\frac{M_W^2}{M_{\check{W}}^2}=\frac{v^2}{\check{v}^2}\approx 6\cdot 10^{-24}.
\label{Gf mirror to Gf}
\end{equation}

To get an idea of the degree of instability of mirror fermions, it is sufficient to estimate the order
 of the ratio of the width $\check{\Gamma}$ of weak decays of mirror leptons to the corresponding
 decay width $\Gamma$ of SM leptons.
A rough estimate of this ratio can be presented in the following form
\begin{equation}
\frac{\check{\Gamma}}{\Gamma}\sim\frac{\check{G}_F^2\check{\Delta}^5}{G_F^2\Delta^5},
 \label{Gamma to Gamma}
\end{equation}
where $\Delta$ and $\check{\Delta}$ are typical  mass differences of the decaying and produced leptons, respectively
 for the SM  and mirror particles. For the SM leptons  $\Delta\sim 0.1\div 1 GeV$. Using the values of  the masses of  
charged mirror leptons obtained in Sec.VI.A.2 one can estimate $\check{\Delta}\sim 10^{10}\div 10^{13} GeV$, where  
it is also assumed that mass of the stable mirror neutrino $m_{\check{\nu}}\lesssim 10^9 GeV$, (see the end of Sec.VI.A.4). 
Then we get that the estimated interval is
\begin{equation}
\frac{\check{\Gamma}}{\Gamma}\sim10^3\div 10^{23}.
 \label{Gamma to Gamma estim}
\end{equation}

\section{Gravity, inflaton and the  Higgs fields at the  classical level}

\subsection{Formulation of the problem. The inflaton model}

The main goal of this paper is to build MESM. But as we indicated in Secs.IV-VI, the Higgs fields vacuum is realized when $\zeta(x)$  reaches a special value 
$\zeta_0$ corresponding to the vacuum. 
The TMT is structured in such a way that even in the case when we are interested in particle physics in the approximation 
of complete disregard of gravity, we must first take into account gravity {\bf exactly} and only at the end consider 
the limit transition to the Minkowski space (if it is possible). This is the result of the special role of the scalar $\zeta(x)$  
of the geometric nature: it is present in all equations of motion, and, due to the constraint,  its value at each space-time
 point is determined by the (classical) mater fields at this point. For this reason, for example, the need to give answer 
to the cosmological constant  problem is built into the very essence of the theory\cite{GK1}-\cite{GK4}, \cite{GK6}.
 This circumstance radically distinguishes TMT from the usual approach to the standard particle model, where
 the contribution to vacuum energy resulting from SSB, along with the bare cosmological constant,
 is commonly considered as physical quantities that are irrelevant in particle physics. 

In Secs.IV-VI, we began to study the basic features of particle physics in  MESM on the background 
consisting of the gravitational field and  classical fractions of the Higgs fields, focusing on what 
happens near the vacuum. This was done without sufficient clarity as to how accurately the state 
of the vacuum was determined. In accordance with the stated in the previous paragraph, we must
 understand how the vacuum state is  reached in the course of  cosmological evolution.
Therefore, in order to study the cosmological evolution of the background fields, we will work 
in this section with an action involving gravity,
 inflaton and Higgs fields.
 To describe inflation, we will use a type of T-model potential motivated by theories based on 
SUSY and conformal ivariance\cite{KL T-model}. 
 The underlying TMT action designed to provide the desired cosmological scenario is chosen as follows
\begin{equation}
S=S_{gr}+S_{\phi}+S_{H}+S_{(nonmin)},
\label{S-background}
\end{equation}
where $S_{gr}$ and $S_{H}$ are defined by Eqs.(\ref{S-gr}) and (\ref{S-H 12}); the action $S_{\phi}$
 for the inflaton $\phi$ has the form
\begin{equation}
S_{\phi}=\int d^4x  \left[\left(\sqrt{-g}+\Upsilon\right)\left(\frac{1}{2}g^{\alpha\beta}\phi_{,\alpha}\phi_{,\beta}
-qM_P^4 \tanh^2\left[\frac{\phi}{\sqrt{6}M_p}\right]\right)-\sqrt{-g}V_1-\frac{\Upsilon^2}{\sqrt{-g}}V_2\right],
\label{S-inflaton}
\end{equation}
where $M_p=m_{pl}/\sqrt{8\pi}$, $m_{pl}= 1.121\cdot 10^{19} \, GeV$, and  $q>0$ is a parameter whose value will be evaluated later. 
The last  two terms added  to the inflaton action have a sense of vacuum contributions. If the first of them  was present in Einstein's GR,  
$V_1$ would be a cosmological constant. 
The term with $V_2$ was first introduced in Ref.\cite{GK2}. The first reason for adding the term with $V_2$ is that the term
 $\propto \Upsilon d^4x= \zeta\sqrt{-g}d^4x$, which can be expected as the contribution of quantum-gravitational effects
 to a vacuum-like action, does not contribute to the equations of motion. Therefore, the next  in powers of $\zeta$ 
vacuum-like term can be of the form $\propto \zeta^2\sqrt{-g}d^4x= \Upsilon^2/\sqrt{-g} d^4x$. 
The second reason 
is more pragmatic. It turns out that thanks to this term it is possible 
to realize both T-model inflation, and, after its termination,
 a transition to a vacuum with small or equal to zero energy density.  
Assuming that both of these vacuum-like terms are rudiments
of the Planck era, it will be natural to suppose that the values of the parameters
 $V_{1}^{1/4}$ and  $V_{2}^{1/4}$ 
are not many 
orders of magnitude less than  the Planck mass. As the given model parameters, let's choose
\begin{equation}
V_1=0.300\cdot  10^{-10}M_P^4, \qquad V_2=4.880\cdot  10^{-10}M_P^4.
 \label{V1 V2 values}
\end{equation}
In order for the SM and its mirror copy to be renormalizable in the gravitational background, it is
 necessary to add non-minimal couplings of the Higgs fields with the curvature scalar. They are
 described by the last term in  Eq.(\ref{S-background}), and to ensure the
 $\textsf{{\it (mirror}}\times\textsf{{\it   parity)}}$ symmetry, we choose it in the form
\begin{equation}
S_{(nonmin)}=\int d^4x \, \xi R(\Gamma, g)\Bigl[\left(c\sqrt{-g}+
\Upsilon\right)|\Phi|^2 +\left(c\sqrt{-g}-\Upsilon\right) |\check{\Phi}|^2\Bigr], 
\label{S nonmin}
\end{equation}
where $c$ is another parameter added to the $b'$s parameter set previously introduced in MESM. 
The common sign is chosen so that in the absence of $\Upsilon$ the value of the non-minimal 
coupling constant $\xi=\frac{1}{6}$ would correspond to the conformal coupling
 (with our definitions of the curvature tensor, see notations after Eq.(\ref{S-toy phi})).

Acting according to the TMT procedure, we start with a variation
 of the action with respect to  scalars $\varphi_a$ ($a=1,..,4$) from which $\Upsilon$ is build, and we obtain 
\begin{eqnarray}
&&-\frac{1}{2}M_P^2R(\Gamma,g)+\frac{1}{2}g^{\alpha\beta}\phi_{,\alpha}\phi_{,\beta}
-qM_P^4 \tanh^2\left(\frac{\phi}{\sqrt{6}M_{p}}\right)-2\zeta V_2
\nonumber
\\
&&+g^{\alpha\beta}\Phi^{\dag}_{,\alpha}\Phi_{,\beta}
-g^{\alpha\beta}\check{\Phi}^{\dag}_{,\alpha}\check{\Phi}_{,\beta}-\lambda\left(|\Phi|^4-|\check{\Phi}|^4\right)+\left[m^2+
\xi R(\Gamma, g)\right] \left(|\Phi|^2-|\check{\Phi}|^2\right)=\mathcal{M},
\label{varphi constr real model}
\end{eqnarray}
where $\mathcal{M}$ - constant of integration of dimension $(mass)^4$.
 
Variation with respect to $g^{\mu\nu}$ yields  the equations
\begin{eqnarray}
&&(1+\zeta)\left[-\frac{1}{2}M_P^2R_{\mu\nu}(\Gamma)+
\frac{1}{2}\phi_{,\mu}\phi_{,\nu}\right]+(b+\zeta)\Phi^{\dag}_{,\mu}\Phi_{,\nu}
+(b-\zeta)\check{\Phi}^{\dag}_{,\mu}\check{\Phi}_{,\nu}
\nonumber
\\
&&+\xi R_{\mu\nu}(\Gamma)\Bigl[\left(c+\zeta\right)|\Phi|^2 +
\left(c-\zeta\right) |\check{\Phi}|^2\Bigr]
\nonumber
\\
&-&\frac{1}{2}g_{\mu\nu}\Biggl[-\frac{1}{2}M_P^2R(\Gamma, g)+
\frac{1}{2}g^{\alpha\beta}\phi_{,\alpha}\phi_{,\beta}-
qM_P^4 \tanh^2\left(\frac{\phi}{\sqrt{6}M_{p}}\right)-V_1+\zeta^2 V_2 
\nonumber
\\
&&+bg^{\alpha\beta}\left(\Phi^{\dag}_{,\alpha}\Phi_{,\beta}
+\check{\Phi}^{\dag}_{,\alpha}\check{\Phi}_{,\beta}\right)-
b\lambda\left(|\Phi|^4+|\check{\Phi}|^4\right)
-\left[bm^2 -c\xi R(\Gamma, g) \right]\left(|\Phi|^2+|\check{\Phi}|^2\right)\Biggr]=0,
\label{gmn Varying}
\end{eqnarray}
trace of which is the following
\begin{eqnarray}
&-&2\zeta^2 V_2+(\zeta-1)\left[-\frac{1}{2}M_P^2R(\Gamma, g)+\frac{1}{2}g^{\alpha\beta}\phi_{,\alpha}\phi_{,\beta}\right]
+(\zeta -b)g^{\alpha\beta}\Phi^{\dag}_{,\alpha}\Phi_{,\beta}
-(\zeta+b)g^{\alpha\beta}\check{\Phi}^{\dag}_{,\alpha}\check{\Phi}_{,\beta}
\nonumber
\\
&+&2qM_P^4 \tanh^2\left(\frac{\phi}{\sqrt{6}M_{p}}\right)+2V_1
+2b\lambda \left(|\Phi|^4+|\check{\Phi}|^4\right)
+2b m^2 \left(|\Phi|^2+|\check{\Phi}|^2\right)
\nonumber
\\
&+&\xi R(\Gamma, g)\bigl[\left(\zeta-c\right)|\Phi|^2-\left(\zeta +c\right)|\check{\Phi}|^2\bigr] =0,
\label{trace gmn Varying}
\end{eqnarray}
Substituting the expression for $-\frac{1}{2}M_P^2 R(\Gamma, g)$ from (\ref{varphi constr real model}) into (\ref{trace gmn Varying})
 one can see that the term  $2\zeta^2V_2$ is canceled.  The resulting  equation is the constraint 
describing $\zeta(x)$ as the local function of the inflaton and Higgs fields. 

Before proceeding, I must make a note that will simplify the calculations. 
Unlike the Higgs inflation models\cite{Higgs infl 1}, \cite{Higgs infl 2}, 
the inflation scenario with the T-model potential  that we are going to implement
 in this paper does not require a large non-minimal coupling of the Higgs fields to gravity. 
This is why I can choose the non-minimal coupling constant $|\xi|\lesssim 1$, and in Sec.VIII.A I assume that $0<\xi<1$.
In addition, it should be noted that at the tree approximation, the maximum possible values of the pair
 $(|\Phi|^2,  \, |\check{\Phi}|^2)$ 
 are of order $\check{v}^2=(10^{14}GeV)^2$. 
Comparing the latter with $M_P^2\approx (2.4\cdot 10^{18}GeV)^2$, 
we see that relative corrections to Eqs.(\ref{varphi constr real model}) - (\ref{trace gmn Varying}), 
which can be caused by non-minimal couplings, are less than $ 10^{-8} $.
 As will be shown below, the corrections directly from the Higgs fields  (that is, at $\xi =0$) to the results of interest  also have the same order.
Moreover, as it turns out, in this paper there is no need to take these corrections into account at all.
Therefore, in order to avoid unnecessarily cumbersome equations, further in this section I will omit the terms with non-minimal couplings.
With this in mind, as a result of the substitution described in the previous paragraph we get the constraint which
after transition to the Einstein frame, Eq.(\ref{gmunuEin}), can be represented in the following convenient form
\begin{equation}
\zeta+1=\frac{2B}
{D-qM_P^4 \tanh^2\left(\frac{\phi}{\sqrt{6}M_{p}}\right)
-(1-b)X_{\Phi}+(1+b)X_{\check{\Phi}}},
\label{constraint z+1 Ein}
\end{equation}
where we used notations
\begin{equation}
B=V_1+V_2-\mathcal{M}-P_2(|\Phi|^2)+\check{P}_2(|\check{\Phi}|^2)
\label{B}
\end{equation}
\begin{equation}
D=2V_2-\mathcal{M}-P_1(|\Phi|^2)+\check{P}_1(|\check{\Phi}|^2)
\label{D}
\end{equation}
\begin{equation}
P_1(|\Phi|^2)=\lambda |\Phi|^4-m^2|\Phi|^2, \qquad P_2(|\Phi|^2)=(1-b)\lambda |\Phi|^4-(1+b)m^2|\Phi|^2
\label{P1 P2}
\end{equation}
\begin{equation}
\check{P}_1(|\check{\Phi}|^2)=\lambda |\check{\Phi}|^4-m^2|\check{\Phi}|^2, \qquad 
\check{P}_2(|\check{\Phi}|^2)=(1+b)\lambda |\check{\Phi}|^4-(1-b)m^2|\check{\Phi}|^2
\label{P1 P2 mirror}
\end{equation}
\begin{equation}
X_{\Phi}=\tilde{g}^{\alpha\beta}\Phi^{\dag}_{,\alpha}\Phi_{,\beta}; \qquad 
X_{\check{\Phi}}=\tilde{g}^{\alpha\beta}\check{\Phi}^{\dag}_{,\alpha}\check{\Phi}_{,\beta}
\label{X Phi X mirror}
\end{equation}

The gravitational equations in the Einstein frame have the canonical form (\ref{toy grav eq Ein}) of the Einstein's GR. 
The features of TMT are manifested in the structure of the TMT effective energy-momentum tensor, which I present here, neglecting the correction given by the non-minimal Higgs fields couplings, which, as well as  in the case of the constraint, has a relative order of $10^{-8}$.
 After using Eqs.(\ref{toy grav eq Ein}), (\ref{gmn Varying})-(\ref{X Phi X mirror}), the TMT effective  energy-momentum tensor  takes the form
\begin{eqnarray}
T_{\mu\nu}&=&\phi_{,\mu}\phi_{,\nu}-\tilde{g}_{\mu\nu}X_{\phi}+\frac{\zeta+b}{1+\zeta}\left(\Phi^{\dag}_{,\mu}\Phi_{,\nu}+h.c.\right)
-\tilde{g}_{\mu\nu}X_{\Phi}
-\frac{\zeta-b}{1+\zeta}\left(\check{\Phi}^{\dag}_{,\mu}\check{\Phi}_{,\nu}+h.c\right)+\tilde{g}_{\mu\nu}X_{\check{\Phi}}
\nonumber
\\
&+&\tilde{g}_{\mu\nu}\mathcal{W}(\phi,|\Phi|^2,|\check{\Phi}|^2;\zeta), 
 \label{Tmn-scalar}
\end{eqnarray}
where
\begin{equation}
X_{\phi}=\frac{1}{2}\tilde{g}^{\alpha\beta}\phi_{,\alpha}\phi_{,\beta},
 \label{X Higgs}
\end{equation}

\begin{equation}
\mathcal{W}(\phi,|\Phi|^2,|\check{\Phi}|^2;\zeta)=V_2-\frac{B}{(1+\zeta)^2}=\mathcal{U}(\phi,|\Phi|^2,|\check{\Phi}|^2,X_{\Phi},X_{\check{\Phi}})
 \label{Wmn}
\end{equation}
and  the second equality was obtained making use the constraint (\ref{constraint z+1 Ein}) with the following result:
\begin{eqnarray}
\mathcal{U}(\phi,|\Phi|^2,|\check{\Phi}|^2,X_{\Phi},X_{\check{\Phi}})= U_{eff}(\phi,|\Phi|^2,|\check{\Phi}|^2)&-&K_1(\phi,|\Phi|^2,|\check{\Phi}|^2)\left[(1+b)X_{\check{\Phi}}-(1-b) X_{\Phi}\right]
\nonumber
\\
&-&K_2(\phi,|\Phi|^2,|\check{\Phi}|^2)\left[(1+b)X_{\check{\Phi}}-(1-b) X_{\Phi}\right]^2,
\label{cal U}
\end{eqnarray}
\begin{equation}
U_{eff}(\phi,|\Phi|^2,|\check{\Phi}|^2)
=V_2-\frac{1}{4B}\left[D-qM_P^4 \tanh^2\left(\frac{\phi}{\sqrt{6}M_{p}}\right)\right]^2
\label{Ueff}
\end{equation}
\begin{equation}
K_1(\phi,|\Phi|^2,|\check{\Phi}|^2)= \frac{1}{2B}\left[D-qM_P^4 \tanh^2\left(\frac{\phi}{\sqrt{6}M_{p}}\right)\right]
\label{K1}
\end{equation}
\begin{equation}
 K_2(|\Phi|^2,|\check{\Phi}|^2)= \frac{1}{4B}
 \label{K2}
\end{equation}
The appearance of a quadratic dependence on the kinetic terms  is typical for TMT and it leads to the k-essence type model considered in detail in 
Ref.\cite{GK3}.
 In the model that we are studying here, an additional complication of the structure arises due to the presence  of terms with opposite signs in front of $\zeta$ in $S_H$.

 Varying the action (\ref{S-inflaton}) with respect to $\phi$ and making the transition to the Einstein frame we obtain the inflaton $\phi$ equation as follows
\begin{equation}
\tilde{\square}\phi
 +\frac{qM_P^4}{1+\zeta}\frac{d}{d\phi}\left[\tanh^2\left(\frac{\phi}{\sqrt{6}M_{p}}\right)\right]
=0
 \label{phi  Ein}
\end{equation}
Recall that $1+\zeta$ must be positive (see Eq.(\ref{zeta+1> 0})) and we will see later that this condition is satisfied. In  next subsections we will need to know the sign of the derivative of $\zeta$ with respect to $\phi$
\begin{equation}
\frac{\partial\zeta}{\partial\phi}=\frac{4qM_p^3B}
{\sqrt{6}\left[D-qM_P^4 \tanh^2\left(\frac{\phi}{\sqrt{6}M_{p}}\right)
+(1+b)X_{\check{\Phi}}-(1-b)X_{\Phi}\right]^2}\frac{\tanh\left(\frac{\phi}{\sqrt{6}M_{p}}\right)}{\cosh^2\left(\frac{\phi}{\sqrt{6}M_{p}}\right)},
\label{derivative z in general}
\end{equation}

\subsection{A general look at the cosmological scenario: vacuum and  inflation}

In the constructed model, let us study some of the characteristic features of the cosmological evolution of the FRW universe filled with the classical homogeneous inflaton and Higgs fields. In this section, we will focus first on studying the features of the final state of evolution (vacuum) and then on the initial stage of evolution (inflation).

\subsubsection{The vacuum energy postulate as a necessary element in the construction of the  model}

In Sec.IV.B, following the generally accepted definition of VEV we have defined the VEV's of the SM Higgs field
 and its mirror partner as the values of these fields in the minima of their TMT effective potentials. 
However, TMT has a feature that makes {\em the definition of vacuum fundamentally different from the standard one}.
In the studied TMT model, there is a continuum of the Higgs fields values in which their potentials have minima,
at least local ones. As can be seen from Eqs.(\ref{Higgs 1 eff pot}) and (\ref{Higgs 2 eff pot}), 
{\em this continuous set of minima is parameterized by the scalar $\zeta$}, which, in turn, depends on the inflaton 
and  Higgs fields contributions  to the constraint (\ref{constraint z+1 Ein}). Therefore, the standard definition
 of vacuum must be supplemented by a certain requirement to the value of $\zeta$ in the vacuum.
It is clear from Eq.(\ref{phi  Ein}) that the vacuum value of the inflaton is $\phi=0$. Then, using the notations 
of Sec.IV.B, one can represent the constraint (\ref{constraint z+1 Ein}) in the vacuum as follows
\begin{equation}
1+\zeta_0=2\frac{V_1+V_2-\mathcal{M}-P_2(v^2)+\check{P}_2(\check{v}^2)}
{2V_2-\mathcal{M}-P_1(v^2)+\check{P}_1(\check{v}^2)}.
\label{constraint z+1 vac}
\end{equation}
Recall that the VEV's $v$ and $\check{v}$ of the Higgs fields    
included in the constraint depend on the still unknown value of $\zeta_0$.
 But the contribution of the Higgs fields to the constraint turns out to be subdominant 
in comparison with the parameters $V_1$ and $V_2$ (see Eq.(\ref{V1 V2 values})), 
which plays an important role in what follows. In addition, the constraint contains the integration constant $\mathcal{M}$.

At first glance, the presence of the integration constant $\mathcal{M}$ in the TMT 
effective potential  resembles a situation typical for unimodular gravity 
models\cite{Ng}-\cite{Shaposhnikov}, where the integration constant has 
the status of a cosmological constant. Indeed, by appropriately choosing 
the value of $\mathcal{M}$ in the TMT effective potential, one can obtain 
any desired vacuum energy density $\rho_{vac}$. But since $\mathcal{M}$
 is also contained in the constraint, the choice of its value will  affect  value 
of $\zeta$ in the vacuum and, consequently, the VEV's of the Higgs fields 
and the masses of all particles of the MESM. Therefore, the model we are
 studying is fundamentally different from unimodular models of gravity. 
Instead of acting by selecting a constant of integration, it is more constructive
 to formulate an alternative approach to the problem:
on the obtained system of the field equations  and the constraint, we impose
 {\em an additional requirement on the value of the vacuum energy density}.
In fact accepting this idea we introduce a new concept in understanding 
the cosmological constant issue. Let us call this concept {\em "the vacuum energy postulate"}. 

 The generally accepted approach to the cosmological constant problem is as follows: 
vacuum is defined as a state with the values of  fields (such as inflaton and Higgs) at the minima of their potentials.
To achieve the required tiny or zero vacuum energy density in theories of this type,
 a fine tuning of the model parameters is required\cite{Weinberg1}.
 Otherwise there will be an insurmountable gap between Einstein's GR 
and the particle field  theory.
 In other words, the problem is to try to answer the question 
{\em why the vacuum energy density  is tiny}, or, which is the same, 
why we live almost in Minkowski space.
Instead, when there is a continuous set of states of Higgs fields 
at the minima of their potentials, it is possible, assuming that we
 now live in almost Minkowski space, {\em define vacuum as a state 
from this continuous set in which the energy density is  tiny}. 
In fact, this is only the first step in the proposed change in the approach
 to the cosmological constant  problem. In section VIII.C, 
after examining the features of the transition of Higgs fields to the vacuum state, 
I will present a rationale from which it becomes clear that 
{\em  the vacuum energy postulate acts simply as a working tool},
 which allows us, as we progress in knowledge of particle physics 
and the evolution of the universe, gradually increase the accuracy and 
come closer to a complete description and understanding of the vacuum.

Within the framework of the constructed model, let's proceed to the practical 
implementation of the outlined idea. We could consider scenarios with different
 vacuum energy density $\rho_{vac}$. For example,  the value of $\rho_{vac}$ 
could be chosen: of the SM energy scale or less; tiny or equal to zero.
 For the sake of simplicity, let us start to explore a scenario with $\rho_{vac}=0$. 
As will be clear from what follows, any other choice satisfying the condition
 $\rho_{vac}\ll V_2$ leads to small corrections compared to what we get if  $\rho_{vac}=0$.
So, considering the first equality in Eq.(\ref{Wmn}), we impose the following condition
\begin{equation}
\rho_{vac}=0  \Longrightarrow (1+\zeta_0)^2=\frac{1}{V_2}\Bigl[V_1+V_2-\mathcal{M}-P_2(v^2)+\check{P}_2(\check{v}^2)\Bigr],
\label{rho vac 0}
\end{equation}
where $\zeta_0$ is a  value of $\zeta$ in the vacuum state ($\phi =0, \, v, \, \check{v}$).
The consistency of the condition (\ref{rho vac 0}) with  the constraint 
(\ref{constraint z+1 vac}) leads to the following system of two equations 
with two unknowns $\zeta_0$ and $\mathcal{M}$
\begin{equation}
\mathcal{M}=-2V_2\zeta_0 +\check{P}_1(\check{v}^2)-P_1(v^2),
\label{M4 vac}
\end{equation}
\begin{equation}
\zeta_0^2=\frac{1}{V_2}\left[V_1 +\left(\check{P}_2(\check{v}^2)-\check{P}_1(\check{v}^2)\right)-\left(P_2(v^2)-P_1(v^2)\right)\right],
\label{z02 eq}
\end{equation}
where $v^2$ and $\check{v}^2$ also depend on $\zeta_0$  according to Eqs.(\ref{v under 2}), (\ref{v under mir 2}),  (\ref{v both 2}).
Some algebraic calculations can show that the solution  with $\zeta_0<0$ does not allow obtaining 
a desirable form of the  T-model inflationary potential. As will be shown in the next subsection, a positive solution for $\zeta_0$ allows this idea to be realized.
  With the choice (\ref{V1 V2 values}) of the parameters $V_1$, $V_2$ and with already used $\zeta_0 -b=1.3\cdot 10^{-12}$, (see Eq.(\ref{bh choice})).
 one can see that the following is hold: 
$P_1(v^2)\sim P_2(v^2)\sim \lambda v^4$, 
$\check{P}_1(\check{v}^2)\sim \check{P}_2(\check{v}^2)\sim \lambda\check{v}^4$
 and $10^{46}\lambda v^4\sim  \lambda\check{v}^4\sim 10^{-8}\sqrt{V_1V_2}$. 
Therefore, after some algebra, the positive solution for $\zeta_0$ can be   represented with high accuracy in the form 
\begin{equation}
\zeta_0=\sqrt{\frac{V_1}{V_2}}+b\frac{\lambda \check{v}^4}{8V_2}\left(1+\frac{1}{2}\sqrt{\frac{V_2}{V_1}}
+\frac{1}{2}\sqrt{\frac{V_1}{V_2}}\right)+\mathcal{O}\left(\frac{\lambda v^4}{V_2}\right)\approx 0.248
+\mathcal{O}\left(\frac{\lambda\check{v}^4}{\sqrt{V_1V_2}}\right),
\label{zeta 0  final}
\end{equation}
where $\zeta_0$ dependence of $\check{v}^2$ and $v^2$ was also taken taken into account explicitly.
Then  with the same accuracy, we obtain the value of  the integration constant 
\begin{equation}
\mathcal{M}=-2\sqrt{V_1V_2}-\frac{b}{4}\lambda\check{v}^4\left[1-\sqrt{\frac{V_1}{V_2}}+\frac{1}{2}\sqrt{\frac{V_2}{V_1}}\right]
+\mathcal{O}(\lambda v^4)=-2.42\cdot 10^{-10}M_P^4\cdot \biggl [1+ \mathcal{O}\left(\frac{\lambda\check{v}^4}{\sqrt{V_1V_2}}\right)\biggr]
\label{M4 vac final}
\end{equation}
Using this expression for $\mathcal{M}$ and neglecting the relative corrections of the order of 
$10^{-8}$ we get 
\begin{equation}
V_1+V_2-\mathcal{M}\approx(\sqrt{V_1}+\sqrt{V_2})^2.
 \label{V1 V2 M}
\end{equation}

If the vacuum energy density  $\rho_{vac}$ is chosen to be non-zero, 
but $0<\rho_{vac}\ll V_2$,  then Eq.(\ref{rho vac 0}) is modified. 
In this case we again use Eq.(\ref{Wmn}) but instead of equating $V_2-B/(1+\zeta)^2$ to zero,
 we must equate it to $\rho_{vac}\neq 0$ obtaining
\begin{equation}
\rho_{vac}\neq 0  \Longrightarrow (1+\zeta_0)^2=
\frac{1}{V_2}\Bigl[V_1+V_2-\mathcal{M}-P_2(v^2)+
\check{P}_2(\check{v}^2)\Bigr]\cdot \left(1+ \frac{\rho_{vac}}{V_2}\right).
\label{rho vac 0 nonzero}
\end{equation}  
This leads to very small,
of order $\rho_{vac}/V_2\ll 1$, changes in the values of $\zeta_0$ and $\mathcal{M}$. 

Thus, imposing an additional requirement on the vacuum energy density $\rho_{vac}\ll V_2$, 
we have obtained expressions for $\zeta_0$ and $\mathcal{M}$
 in terms of the model parameters $V_1$, $V_2$ and the VEV's   $v$ 
 and  $\check{v}$ of the SM Higgs field and its mirror partner respectively.
And vice versa, for the found value of $\zeta_0$, for the selected in
   Eq.(\ref{bh choice}) the value of the parameter $ b $, and using the observed value $v$ (see Sec.VI.A.1),
   the well-defined VEV of the mirror Higgs field $\check{v}\approx 10^{14}GeV$ was also found in Sec.VI.A.1.

The  influence of the mirror Higgs field on the found values of $\zeta_0$ and $\mathcal{M}$ turned out to be suppressed by a factor of the order of $10^{-8}$ in comparison with contributions of $V_1$  and $V_2$. As we already mentioned, the corresponding corrections  from non-minimal couplings of the Higgs fields with the curvature scalar are suppressed by a factor of the same order.
The SM Higgs boson corrections are tens of orders less. 

It is very important to emphasize that there are no strict restrictions on the choice of the model parameters $V_1$, $V_2$  made in 
 Eq.(\ref{V1 V2 values}). Indeed, due to the vacuum energy postulate, another choice of the values of $V_1$, $V_2$ simply leads to a corresponding change in  the values of $\mathcal{M}$ and $\zeta_0$. This means that {\it the  vacuum energy postulate is implemented without the need  to fine-tune the model parameters}.

The expressions found for $\zeta_0$ and $\mathcal{M}$ allow one to obtain information about the inflaton. 
In fact, it follows from Eq.(\ref{phi  Ein}) that in the vicinity of the vacuum state ($\phi =0, \, v, \, \check{v}$), the homogeneous inflaton field obeys the equation
\begin{equation}
\ddot{\phi}+3H\dot{\phi} +\frac{qM_P^2}{3(1+\zeta_0)}\phi=0.
 \label{phi  cosm vac eq}
\end{equation}
From this we obtain an expression for the inflaton mass in terms of $q$ and $\zeta_0$:
\begin{equation}
m_{\phi}^2=\frac{qM_P^2}{3(1+\zeta_0)},
 \label{phi  mass}
\end{equation}
Note that the value of the parameter $q$ is not yet known.

\subsubsection{Inflationary era}

The purpose of this section is to demonstrate that when applying the model to the study of the inflationary era, we are dealing with a 
T-model\cite{KL T-model} that can provide agreement with the Planck observational data\cite{Planck}.

In the next section, we will see that contributions of the Higgs fields to both the constraint and the energy-momentum tensor are negligible during inflation; here we will assume that these conditions are satisfied.
First of all, let us show that under these conditions the considered model has a typical structure of the T-model.
Indeed, if we use Eqs.(\ref{M4 vac final}), (\ref{V1 V2 M}) then, omitting the contributions of the Higgs fields, the potential 
$U_{eff}$ that was defined by Eq.(\ref{Ueff}),  is reduced to  the T-model type potential
\begin{equation}
U_{eff}(\phi)
=qM_P^4\frac{\sqrt{V_2}}{\sqrt{V_1}+\sqrt{V_2}}\tanh^2\left(\frac{\phi}{\sqrt{6}M_P}\right)
\left[1-\frac{qM_P^4}{4\sqrt{V_2}(\sqrt{V_1}+\sqrt{V_2})}\tanh^2\left(\frac{\phi}{\sqrt{6}M_{p}}\right)\right],
\label{Ueff phi T-model}
\end{equation}
where the vacuum term (at $\phi =0$) disappears, since we  have already used that $\rho_{vac}=0$ .

For $\phi\gg M_P$,  if we neglect higher degrees of $e^{-\sqrt{\frac{2}{3}}\phi/M_{p}}$ in the expansion of $\tanh^2\left(\phi/\sqrt{6}M_{p}\right)$, this potential takes the usual form of the simplest T-model potential in the inflationary epoch\cite{KL T-model}
\begin{equation}
U_{eff}^{(infl)}(\phi)=V_{*}\left(1-4e^{-\sqrt{\frac{2}{3}}\phi/M_{p}}\right),
\label{Ueff infl}
\end{equation}
where
\begin{equation}
V_{*}=\gamma M_P^4; \quad \gamma=q\frac{\sqrt{V_2}}{\sqrt{V_1}+\sqrt{V_2}}\left(1-\frac{qM_P^4}{4\sqrt{V_2}(\sqrt{V_1}+\sqrt{V_2})} \right)
=\frac{4}{5}q\left(1-\frac{q}{24}10^{10}\right).
\label{V star}
\end{equation}
 To satisfy the Planck data we must demand\cite{KL T-model} $\gamma\approx 1.1\cdot 10^{-10}$.
Solving the quadratic  equation for $q$ we get
\begin{equation}
q\approx 1.5\cdot 10^{-10}.
\label{q}
\end{equation}
With the value $\zeta_0\approx 0.25$ found in (\ref{zeta 0 final}),   from the expression (\ref{phi  mass}) we obtain that the inflaton mass is
 \begin{equation}
m_{\phi}\approx 1.5\cdot 10^{13}GeV.
\label{mphi}
\end{equation}

In the inflationary epoch, the  equation for the homogeneous inflaton field $\phi(t)$ is obtained using Eq.(\ref{phi  Ein}) 
\begin{equation}
\ddot{\phi}+3H\dot{\phi} +4\sqrt{2/3}\frac{qM_P^3}{1+\zeta}e^{-\sqrt{\frac{2}{3}}\phi/M_{p}}
=0.
 \label{phi  cosm infl eq}
\end{equation}
Here $\zeta(t)$ is determined by the constraint (\ref{constraint z+1 Ein}). First, rewrite it by omitting the Higgs field contributions and using the results of the previous subsection
\begin{equation}
1+\zeta=\frac{2(\sqrt{V_1}+\sqrt{V_2})^2}
{2\sqrt{V_2}(\sqrt{V_1}+\sqrt{V_2})-qM_P^4\tanh^2\left(\frac{\phi}{\sqrt{6}M_{p}}\right)}.
 \label{zeta T-model}
\end{equation}
For $\phi\gg M_P$, the constraint can be used in the form
\begin{equation}
\zeta|_{infl}\approx 0.42-0.77\cdot e^{-\sqrt{\frac{2}{3}}\phi/M_{p}},
 \label{zeta infl}
\end{equation}
and we again neglected higher degrees of $e^{-\sqrt{\frac{2}{3}}\phi/M_{p}}$. For the same reason, $\zeta(t)$ in Eq.(\ref{phi  cosm infl eq})
can be taken equal to the initial value $\zeta_{in}\approx 0.42$, that is, at the very beginning of inflation.

Using Eqs.(\ref{derivative z in general}), (\ref{B})  and  (\ref{V1 V2 M}) we see that  $\frac{\partial\zeta}{\partial\phi}>0$ not only during inflation  but also after the inflation terminates.  Consequently,  $\zeta$  decreases monotonically along with the inflaton field both during inflation and some time after its end;  this is not the case in the transition to the vacuum state, when the inflaton field oscillates around $\phi =0$.

As we have seen, the presence of $\tanh^2\left(\frac{\phi}{\sqrt{6}M_P}\right)$ in the potential used in the underlying action (\ref{S-inflaton}) is not enough to get  the resulting TMT effective potential in the form typical for the T-models.  To realize this kind of  effective TMT potential, we were forced to involve into the theory: a) two vacuum-like terms with parameters $V_1$, $V_2$ in the underlying action (\ref{S-inflaton}); b) the vacuum energy postulate with the choice $\rho_{vac}=0$. The latter is used here instead of the usual assumption that the cosmological constant is somehow tuned \cite{KL T-model}.

 Summarizing the results of this and the previous subsections, it is important to pay attention to the following two fundamental aspects.

\begin{itemize}

\item

By setting the values of the parameters $V_1$ and $V_2$ and adding to them the value of the parameter $V_*$ following from the Planck observations, we have found the values of  $\zeta_0$,  the integration constant $\mathcal{M}$  and  the inflaton $m_{\phi}$ (or, equivalently, $q$).

\item

Having imposed the initial conditions for the chosen inflationary scenario, we found the value of $\zeta$ at the very beginning of inflation: $\zeta_{in}\approx 0.42$. This appearance of a definite positive value of $\zeta$ means that the orientation of the spacetime manifold is spontaneously fixed (see the related discussion in Secs.II.A and III.A). In the course of cosmological evolution,  $\zeta$ decreases from $\zeta_{in}$ to the vacuum value $\zeta_0\approx 0.248$, keeping positive value. Consequently, the sign of orientation of the space-time manifold, spontaneously fixed at the very beginning of cosmological evolution, remains unchanged until the end of evolution.

\end{itemize}

\section{Cosmological evolution of the Higgs fields. Tree approximation}

\subsection{Inlationary and postinflationary stages}

 Let us analyze the cosmological evolution of the Higgs fields using  the parameter values given by Eqs.(\ref{bh choice}) and (\ref{V1 V2 values}). In an inflationary era, the non-minimal couplings of the Higgs fields with the curvature scalar have a significant impact. Therefore the appropriate action 
(\ref{S nonmin}) has to be added to the action (\ref{S-H 12}). Then, instead of Eqs.(\ref{Higgs  eqom Ein}) and (\ref{Higgs mirror eqom Ein}), we get the following equations for the homogeneous Higgs fields in the FRW universe
\begin{equation}
\ddot{\Phi}+3H\dot{\Phi}
+\frac{1-b}{(1+\zeta)(\zeta+b)}\dot{\zeta}\dot{\Phi}+
 \biggl[\frac{2\lambda}{1+\zeta}|\Phi|^2
-\frac{1}{\zeta+b}\biggl(\frac{\zeta-b}{1+\zeta}m^2+(\zeta+c)\xi \tilde{R}\biggr)\biggr]\Phi
=0 ,
 \label{Higgs  eqom Ein cosm}
\end{equation}
\begin{equation}
\ddot{\check{\Phi}}+3H\dot{\check{\Phi}}
+\frac{1+b}{(1+\zeta)(\zeta-b)}\dot{\zeta}\dot{\check{\Phi}}+
\biggl[\frac{2\lambda}{1+\zeta}|\check{\Phi}|^2
-\frac{1}{\zeta-b}\biggl(\frac{\zeta+b}{1+\zeta}m^2+(\zeta-c)\xi \tilde{R}\biggr)\biggr]\check{\Phi}=0, 
 \label{Higgs mirror eqom Ein cosm}
\end{equation}
where $\tilde{R}\equiv R(\tilde{g}_{\mu\nu})$ is the scalar curvature in the Einstein frame and $H$ is the Hubble parameter; the dot means derivative with respect to cosmic time.

The appearance of a spontaneously fixed value $\zeta=\zeta_{in}$ at the very beginning of inflation not only means that the orientation of the space-time manifold is spontaneously fixed, but also manifests itself in a spontaneous violation of the (mirror$\times$parity) symmetry of the underlying MESM action (\ref{S MESM}). 

Moreover, together with the chosen value of $b$, Eq.(\ref{bh choice}), we have 
$b<\zeta_0<\zeta_{in}$. Then the ratio
\begin{equation}
\frac{\zeta-b}{\zeta+b}>0
\label{ratio z-b z+b}
\end{equation}
 remains positive throughout the entire process of the evolution under study. 
Consequently, if the non-minimal coupling constant $\xi$ were equal to zero, then the mass terms in the TMT (classical) effective Higgs potentials would have “wrong” signs in the entire process of cosmological evolution.
As a result, nonzero minimum positions of TMT effective Higgs potentials would arise, that is, cosmological evolution would begin with the phase of the broken $SU(2)\times U(1)$ gauge symmetry.

However, in order to provide the conditions for renormalizability of the model, we must take into account non-minimal couplings.
To analyze and compare the behavior of the inflaton and Higgs fields at the very beginning of inflation, one could do this using the TMT effective potential 
$U_{eff}(\phi,|\Phi|^2,|\check{\Phi}|^2)$, which is present in the TMT effective energy-momentum tensor, Eq.(\ref{Tmn-scalar}). But in order to avoid unnecessary complications, in Sec.VII.A, when calculating the TMT effective energy-momentum tensor, we neglected the contribution of the non-minimal coupling. This drawback can be easily overcome if we take into account that the information of interest to us about the curvature of the TMT effective potential 
$U_{eff}(\phi,|\Phi|^2,|\check{\Phi}|^2)$ in the directions of the inflaton and Higgs fields can be obtained from the equations of these fields (\ref{phi  cosm infl eq}),  (\ref{Higgs  eqom Ein cosm}), (\ref{Higgs mirror eqom Ein cosm}). These equations contain the derivatives of the TMT effective potential in the corresponding directions.  The presence of $\zeta(t)$ in the derivatives of these effective potentials does not complicate the calculation of their curvature, since at the initial stage of inflation $\zeta(t)$ remains constant with high accuracy.

As it was mentioned in the discussion paragraph after Eq.(\ref{trace gmn Varying}), I assume that $0<\xi<1$. In the initial period of the T-model inflation, space-time can be described as de-Sitter space. Then $\tilde{R}=-12H^2$ and $H\approx 10^{13}GeV$. Below, after discussing the various possibilities for the value of the parameter $c$, in item Stage 2, the final choice $\zeta_0<c<\zeta_0 + 0.006$ will be made.
 Estimating the value of $m$ at the tree level using Eq.(\ref{m h2}) gives $m\sim 10^{8}GeV$. Then, if $\xi$ is not too  small,  the non-minimal couplings are able to provide positive effective squared mass terms in the TMT effective Higgs potentials not only  from the very beginning of inflation, but even at later stages of evolution. 
Hence, during this entire period of evolution, the only  minima of the TMT effective Higgs potentials will be $\Phi=0$ and $\check{\Phi}=0$ (at least at the tree level). 

It follows from Eqs.(\ref{Higgs  eqom Ein cosm}) and (\ref{Higgs mirror eqom Ein cosm}) that 
 the curvatures of the TMT effective  potentials  both for the SM Higgs field and for the mirror Higgs field at the very beginning of inflation approximatly equal to $12\xi (10^{13}GeV)^2$. Curvature in the direction of inflaton
is the curvature of the inflaton potential (\ref{Ueff infl}), which is almost flat for $\phi\gg M_P$.
Thus, we expect that at the very beginning of inflation, the Higgs fields roll down to their zero minima. Therefore, at the initial stage of the evolution of the Universe, {\em the $SU(2)\times U(1)$ gauge symmetry of the MESM remains unbroken}. In this respect, the model resembles what takes place in the hybrid inflation model\cite{hybrid}, but in the model under study this is due to non-minimal couplings of the Higgs fields.

Phase transitions begin  after the effective squared mass terms in Eqs.(\ref{Higgs  eqom Ein cosm}) and (\ref{Higgs mirror eqom Ein cosm})  cross zero, changing sign from positive to negative. 
The latter is due to the action of two factors: decreasing $|R|$ and decreasing $\zeta$. The distinctions in the structures of the effective squared mass for $\Phi$ and $\check{\Phi}$ leads to an interesting phenomenon: {\em phase transitions of the SM and the mirror Higgs fields start at different stages of the cosmological evolution}. The factor 
$(\zeta -c)$ in front of $\xi R$ in Eq.(\ref{Higgs mirror eqom Ein cosm}) can cause a phase transition when $|R|$ is still relatively large. For example, depending on the selected value of the parameter $c$, this can happen at the last stage of inflation or after its end. Conversely, the presence of the factor $(\zeta-b)$ in front of $m^2$ in Eq.(\ref{Higgs  eqom Ein cosm}) can affect so that the phase transition for the SM Higgs field will be possible only when $|R|$ is very small. These second order phase transitions and their influence on cosmology deserve detailed study. In this paper I will limit myself to a preliminary qualitative analysis.

The novelty of the cosmological model proposed in this paper is mainly in the role played by the scalar $\zeta(t)$. In a sense, this is reminiscent of the role of temperature in a second-order phase transition in cosmological scenarios based on more standard theories. But, as we will see, there is a significant difference that leads to fundamentally new results.

 The division of cosmological evolution into stages depending on the values of $\zeta(t)$ turns out to be very informative. Here I offer a first attempt at implementing this idea by dividing the interval 
$\zeta_0\approx 0.248\leq\zeta(t)\lesssim 0.42\approx \zeta_{in}$ into five parts.

\begin{itemize}

\item
Stage 1 is the primordial  inflationary era, when the inflaton potential has the form (\ref{Ueff infl}), and $\zeta(\phi)$ is described by Eq.(\ref{zeta infl}). During an initial  period when $\phi\gtrsim 8M_P$,  the expantion of the Universe is governed mainly by the almost constant energy density $V_*$, and  
$\zeta(t)\approx 0.42$ remains almost constant.  In the interval  $4.5M_P\lesssim\phi\lesssim 8M_P$, where  the effect of  the $\phi$ dependent term becomes significant, $\zeta(t)$  decreases from $\zeta\approx 0.42$ to   $\zeta\approx 0.4$. 

\item

Stage 2 is the late inflationary era, when in the interval   $0.5 M_P\lesssim\phi\lesssim 4.5 M_P$ one should use the constraint in the form as in Eq.(\ref{zeta T-model}). Accordingly, $\zeta$  decreases from $\zeta\approx 0.4$ to   $\zeta\approx \zeta_0+ 0.006$. If one to select the value of the parameter $c$ so that $\zeta_0 + 0.006<c<0.4$, then the phase transition  for the mirror Higgs field $\check{\Phi}$ will begin at stage 2. But I will continue the analysis, assuming the choice of $c$ so that
$\zeta_0<c<\zeta_0 + 0.006$. In this case, the minima of the TMT effective Higgs potentials remain at $\Phi=0$ and $\check{\Phi}=0$, and therefore the gauge symmetry of MESM is not broken during the entire inflation process. 

\item

Stage 3 is the epoch after the end of inflation in the interval   $0<\phi\lesssim 0.5 M_P$.
In  the case of zero vacuum energy density. the  constraint (\ref{constraint z+1 Ein}) reduces to the following
\begin{equation}
1+\zeta=\frac{2(\sqrt{V_1}+\sqrt{V_2})^2+N_1(\varphi,\check{\varphi})}
{2\sqrt{V_2}(\sqrt{V_1}+\sqrt{V_2})-\frac{q}{6}M_P^2\phi^2+N_2(\varphi,\check{\varphi})}.
 \label{zeta stage 3}
\end{equation}
\begin{equation}
N_1(\varphi,\check{\varphi})=(1+\zeta_0)\left[(1+\zeta_0)(1+b)\frac{\lambda}{4}\check{\varphi}^4-(1-b)\frac{m^2}{2}{\check{\varphi}}^2
-(1+\zeta_0)(1-b)\frac{\lambda}{4}\varphi^4+(1+b)\frac{m^2}{2}\varphi^2\right]
\label{N}
\end{equation}
\begin{equation}
N_2(\varphi,\check{\varphi})=(1+\zeta_0)\left[(1+\zeta_0)\frac{\lambda}{4}\check{\varphi}^4-\frac{m^2}{2}\check{\varphi}^2-(1+\zeta_0)\frac{\lambda}{4}\varphi^4+\frac{m^2}{2}\varphi^2+(1+b)\frac{1}{2}\dot{\check{\varphi}}^2-(1-b)\frac{1}{2}\dot{\varphi}^2\right],
\label{D}
\end{equation}
where, in order to apply the constraint also at stage 4, the unitary gauge is used. 
It is easy to see that with the beginning of  stage 3, that is, when $\zeta\lesssim \zeta_0+ 0.006$, $\zeta(t)$ continues to decrease. Indeed, using the values of the model parameters and the constraint (\ref{zeta stage 3}) (before the transition to the unitary gauge), we get
\begin{equation}
\frac{d\zeta}{dt}\approx 1.2\cdot 10^{-2}\frac{d}{dt}\left(\frac{\phi^2}{M_P^2}\right)+1.3\cdot 10^{-9}\frac{\lambda}{M_P^2}\frac{d|\check{\Phi}_m|^2}{dt}, 
\label{dz dt 3}
\end{equation}
where  the redefinition  (\ref{Phi and mir measured}) was used, and we  neglected the contribution of the SM Higgs field, which is many orders of magnitude smaller than the contribution of the mirror Higgs field.
Consequently, after the end of inflation, the decrease in $\zeta(t)$ is still mainly  governed by the decrease in the inflaton field. 

Since $\zeta(t)$  decreases from $\zeta\approx \zeta_0+ 0.006$ to   $\zeta_0$,  it is convenient to represent it in the form
\begin{equation}
\zeta(t)=\zeta_0+\delta\zeta(t),  \quad 0<\delta\zeta(t)<0.006.
\label{delta zeta def}
\end{equation} 
 It turns out that, due to the smallness of $\zeta_0-b=1.3\cdot 10^{-12}$, it is convenient to  divide the study of the subsequent   cosmological evolution into two stages: {\em Stage 3 when $\delta\zeta(t)>\zeta_0-b$,  and   Stage 4 when $\delta\zeta(t)<\zeta_0-b$}.

Choose the parameter $c$ so that $1.3\cdot 10^{-12}<c-\zeta_0< 0.006$. At the moment $t=t_1$ when  $\zeta(t_1)=\zeta_1$ and the scalar curvature 
$\tilde{R}(t_1)=\tilde{R}_1$  satisfies the condition 
\begin{equation}
\frac{\zeta_1+b}{1+\zeta_1}m^2+(\zeta_1-c)\xi \tilde{R}_1=0,
\label{cr 1}
\end{equation}
 the phase transition for the mirror Higgs field starts. When the decreasing $\zeta(t)$ reaches the value  $\zeta=c$, the TMT effective potential of the mirror Higgs field acquires a minimum at
\begin{equation}
|\check{\Phi}_m|^2_{min}(\zeta)|_{\zeta=c}=\frac{\zeta_0-b}{c-b}\frac{c+b}{\zeta_0+b}\cdot \frac{\check{v}^2}{2},
 \label{min mir Higgs gen}
\end{equation}
where Eq.(\ref{v under mir 2}) and    the redefinition  (\ref{v both 2}) were used. After $\zeta(t)$ becomes less than $c$, the non-minimal coupling term in Eq.(\ref{Higgs mirror eqom Ein cosm}) changes sign and, therefore,  additionally increases  $|\check{\Phi}_m|^2_{min}(\zeta)$. To get an idea of how fast $|\check{\Phi}_m|^2_{min}(\zeta)$ increases during the second order phase transition at stage 3, it is enough to make rough estimates, ignoring the contribution of the non-minimal coupling. For illustration, let us choose $c=\zeta_0+0.005$ and compare the value of $|\check{\Phi}_m|^2_{min}(c)$ with  value
$|\check{\Phi}_m|^2_{min}(\zeta)$ at the moment close to the end of stage 3, say, at $\zeta_{(end 3)}=\zeta_0+1.3\cdot 10^{-11}$:
\begin{equation}
|\check{\Phi}_m|^2_{min}(c)\approx 10^{-10}\cdot\frac{\check{v}^2}{2}; \qquad |\check{\Phi}_m|^2_{min}(\zeta_{(end 3)})\approx 0.1\frac{\check{v}^2}{2}.
 \label{min mir Higgs stage 3 interval}
\end{equation}
Therefore,  during the second order phase transition at stage 3, the instantaneous equilibrium values of the mirror Higgs field can increase from  $\sim 10^{-5}\check{v}$ to about $\check{v}/3$.

\item
At Stage 4, i.e. when 
\begin{equation}
0<\delta\zeta<\zeta_0-b=1.3\cdot 10^{-12},
\label{delta z in  stage 4}
\end{equation}
 the phase transition for the SM Higgs field begins when 
\begin{equation}
\frac{\zeta-b}{(1+\zeta)(\zeta+b)}m^2+\frac{\zeta+c}{\zeta+b}\xi \tilde{R}
 \label{phase trans Higgs  stage 4}
\end{equation}
changes sign from negative to positive.  
   {\em After the phase transition also occurs with the SM Higgs field}, the smallness of   $\delta\zeta$ at stage 4 allows one to go to the unitary gauge, and we will use the definitions and notations of Sec.IV.B.
Then Eqs.(\ref{Higgs eqom Ein cosm}) and (\ref{Higgs mirror eqom Ein cosm}) are represented as follows
\begin{equation}
\ddot{\varphi}+3H\dot{\varphi}+\frac{1+b}{(\zeta+b)(1+\zeta)}\dot{\zeta}\dot{\varphi}+\frac{1+\zeta_0}{1+\zeta}\lambda\varphi^3 -
\frac{\zeta-b}{(\zeta+b)(1+\zeta)}m^2\varphi=0,
 \label{phi vicinty vac eq}
\end{equation}
\begin{equation}
\ddot{{\check{\varphi}}}+3H\dot{{\check{\varphi}}}+\frac{1-b}{(\zeta-b)(1+\zeta)}\dot{\zeta}\dot{{\check{\varphi}}}+\frac{1+\zeta_0}{1+\zeta}\lambda{\check{\varphi}}^3 -
\frac{\zeta+b}{(\zeta-b)(1+\zeta)}m^2{\check{\varphi}}=0.
 \label{phi mirror vicinty vac eq}
\end{equation}
The smallness of $\delta\zeta$ at stage 4 also means that  the classical fields 
$\varphi(t)$ and $\check{\varphi}(t)$ are very close to their VEV's, and it is convenient to describe them in terms 
of their time-dependent deviations from $v$ and $\check{v}$, respectively:
\begin{equation}
\varphi(t)=v+\delta\varphi(t), \qquad |\delta\varphi(t)|< v; \qquad \check{\varphi}(t)=\check{v}+\delta\check{\varphi}(t), \qquad 
|\delta\check{\varphi}(t)|<\check{v},
 \label{form of Higgs near vac}
\end{equation} 
This representation for $\varphi(t)$ is true only after the phase transition also occurs with the SM Higgs field.
The type of functions $\delta\varphi(t)$ and $\delta\check{\varphi}(t)$ is very important for understanding how the process of approaching a vacuum occurs.
Unlike stage 3, asymptotic solutions of Eqs.(\ref{phi vicinty vac eq}) and (\ref{phi mirror vicinty vac eq}) can be found analytically (see  below Subsec.B).

\item
Stage 5, at which the cosmological evolution of the inflaton and classical Higgs fields is complete, the inflaton is in the vacuum state $\phi\equiv 0$, and the Higgs fields are represented as in Eq.(\ref{Higgs doublets mesured}). In Subsec.D we will return to studying stage 5.

\end{itemize}

\subsection{Stage 4. A new type of transition to the vacuum state. Preliminary consideration}

Using the representation (\ref{form of Higgs near vac}) for the Higgs fields 
we get an expansion of the constraint (\ref{zeta stage 3}) to the first degree in $\phi^2(t)$, $\delta\check{\varphi}(t)$ and $\delta\varphi(t)$  
\begin{equation}
\delta\zeta(t)\approx 2.4\cdot 10^{-2}\, \frac{\phi^2(t)}{M_P^2}+4.8\cdot 10^{-9}\lambda \frac{\delta\check{\varphi}(t)}{\check{v}},
\label{delta zeta 3b}
\end{equation}
The term with $\delta\varphi/v$ (which could appear after the phase transition for $\varphi$) was omitted because the coefficient before it is 35 orders of magnitude less than that before $\delta\check{\varphi}/\check{v}$. It also used the value of the parameter $q$ found in Eq.(\ref{q}). Recall that in  the constraint (\ref{zeta stage 3}) and, therefore, in (\ref{delta zeta 3b}), we used the value of the integration constant $\mathcal{M}$ determined by Eq.(\ref{M4 vac final}), which means that we consider the case of zero vacuum energy density. 

In stage 4, Eq.(\ref{phi  Ein}) for the inflaton can be represented by Eq.(\ref{phi  cosm vac eq}), since the corrections caused by the deviation of $\zeta$ from $\zeta_0$ lead to higher powers of $\phi^2$, which, as usual, we  will not take into account.
Expanding the TMT effective  potential, Eq.(\ref{Wmn})-(\ref{Ueff})), near the zero  energy vacuum  to the first degrees of $\phi^2$ and deviations of the Higgs fields from their VEV's, we get 
\begin{equation}
U_{eff}|_{(\text{near vac})}=\frac{1}{2}m_{\phi}^2M_P^2\left(\frac{\phi^2}{M_P^2}+1.9\cdot 10^{-7}\lambda \frac{\delta\check{\varphi}(t)}{\check{v}}\right)
\label{phi eff pot vicinty vac}
\end{equation}
where the contribution of the SM Higgs can again be neglected. As we will see below, despite the small factors before $\delta\check{\varphi}(t)/\check{v}$, the second terms in $\delta\zeta(t)$ and $U_{eff}|_{(\text{near vac})}$ can become dominant over the inflaton contributions when $m_{\phi}t\gg 10^5$.  This effect requires special research, which will be carried out elsewhere. However, 
near the vacuum, but before the second term becomes dominant,  solving the system of  Einstein's equations and  Eq.(\ref{phi  cosm vac eq})  for the homogeneous inflaton field leads to the  following well-known  asymptotic regime
\begin{equation}
\phi(t)=\tilde{\phi}(t)\sin(m_{\phi}t); \qquad \tilde{\phi}(t)=\sqrt{\frac{8}{3}}\frac{M_P}{m_{\phi}t}.
\label{phi oscillating}
\end{equation}
From Eqs.(\ref{delta zeta 3b}) and  (\ref{phi oscillating})  it follows  that in the course of damped oscillations of the inflaton field around zero, $\zeta(t)=\zeta_0+\delta\zeta(t)$ approaches $\zeta_0$ according to 
\begin{equation}
\delta\zeta(t)\approx\frac{3.2\cdot 10^{-2}}{(m_{\phi}t)^2}\left[1-\cos(2m_{\phi}t)\right],
\label{delta zeta oscillating}
\end{equation}
that is the approach of $\zeta(t)$ to $\zeta_0$ occurs as {\em a superposition of monotonic displacement and damped oscillations around these instantaneous values}. 
The upper constraint on $\zeta$  in stage 4, Eq.(\ref{delta z in  stage 4}), is satisfied when $m_{\phi}t >2\cdot 10^{5}$.

Studying the process of transition to vacuum, in this paper I restrict myself to a preliminary 
consideration sufficient to demonstrate its inherent properties that qualitatively distinguish
 it from the generally accepted idea of how the transition to vacuum should occur.

\subsubsection{Asymptotic solution for the mirror Higgs field}

{\em Leaving aside the very important issue of preheating} let us start with the equation
 for the mirror Higgs field (\ref{phi mirror vicinty vac eq})  in the stage 4. Using the linear 
in $\delta\zeta$ expansion  in Eq.(\ref{phi mirror vicinty vac eq}), we obtain the following equation for
the {\em classical} $\delta\check{\varphi}(t)$
\begin{equation}
\frac{d^2(\delta\check{\varphi})}{dt^2}+\frac{2}{t}\frac{d(\delta\check{\varphi})}{dt}+
2\lambda \check{v}^2\delta\check{\varphi}+\frac{\lambda\check{v}^3}{\zeta_0-b}\delta\zeta=0,
\label{deltavarphi mirr eq delta zeta}
\end{equation}
which describes the cosmological evolution of the mirror Higgs field 
$\check{\varphi}(t)=\check{v}+\delta\check{\varphi}(t)$ {\em before it reaches its VEV $\check{v}$}.
Since we are interested in  the transition to the vacuum in the asymptotic regime, 
we can simplify the solution by neglecting  terms that decay faster than $\sim t^{-2}$. 
This is why the term proportional to $\dot{\zeta}\dot{{\check{\varphi}}}$ has been omitted. 
Consider a regime when the inflaton contribution dominates  in $\delta\zeta(t)$
 and $U_{eff}|_{(\text{near vac})}$. The  interval of $\zeta$ corresponding to the
 condition $\delta\zeta(t)<\zeta_0-b\approx 10^{-12}$  is realized when
 $m_{\phi}t\gtrapprox 10^5$. Substitution of $\delta\zeta(t)$ from 
Eq.(\ref{delta zeta oscillating}) into Eq.(\ref{deltavarphi mirr eq delta zeta})  
 results in the interaction of the inflaton with the Higgs mirror field, which was
 absent in the underlying action. The corresponding term in the TMT effective Lagrangian
 near  vacuum has the form
 \begin{equation}
 L^{(eff)}_{\phi^2\check{\varphi}} \approx -\frac{2.5\cdot 10^{-2}}{\zeta_0-b}\lambda\check{v}^3\frac{\phi^2}{M_P^2}(\check{\varphi}-\check{v})
=-2.5\cdot 10^{10}\frac{1-\cos(2m_{\phi}t)}{(m_{\phi}t)^2}\lambda\check{v}^3(\check{\varphi}-\check{v}),
\label{L int eff mirr}
\end{equation}
which is linear in $\check{\varphi}-\check{v}$.
It is very important to note here the difference between the structure of this interaction 
and the interaction of the inflaton with a scalar field $\varphi$ of the type $\sim \phi^2\varphi^2$, 
which is usually introduced into the Lagrangian to obtain a mechanism for preheating the Universe\cite{KLS}.
 This difference is the reason for the discovered and described below a new way of transition of the mirror
 Higgs field to the vacuum state. Below we will see that a similar effect also takes place for the SM Higgs field. 
And this is the key point that will allow us to formulate a new concept of vacuum.

 Using Eq.(\ref{delta zeta 3b}), we reduce Eq.(\ref{deltavarphi mirr eq delta zeta}) to the following 
\begin{equation}
\frac{d^2(\delta\check{\varphi})}{dt^2}+\frac{2}{t}\frac{d(\delta\check{\varphi})}{dt}
+m_{\check{h}}^2\left[\delta\check{\varphi}
+k_1\cdot\frac{\check{v}}{(m_{\phi}t)^2}\right]=
2.5\cdot10^{10}\lambda  \check{v}^3\frac{\cos(2m_{\phi}t)}{(m_{\phi}t)^2},
\label{deltavarphi mirr eq asymp}
\end{equation} 
where $m_{\check{h}}$ is the mass of the mirror Higgs boson (see below, Eq.(\ref{m2 h mir})) and
\begin{equation}
 k_1=\frac{1.25\cdot 10^{10}}{1+1.9\cdot 10^3\lambda}
\label{notations mirror eq}
\end{equation} 
The equation shows that due to the linearity of the interaction  (\ref{L int eff mirr}) 
in $\check{\varphi}$, the transition of the Higgs mirror field to the vacuum state occurs
 in a very unusual way: the values of $\delta\check{\varphi}(t)$ shift over time due to the second term in parentheses, and forced oscillations around these instantaneous values caused by the inflaton are superimposed.
Using the notations 
\begin{equation}
  k_2=\frac{2.5\cdot10^{10}\check{v}^2}{m_{\phi}^2}\approx 1.1\cdot 10^{12}, \quad 
\check{\omega}_0^2=\frac{m_{\check{h}}^2}{m_{\phi}^2}
\label{notations mirror Higgs}
\end{equation}
and neglecting terms that decrease faster than $t^{-2}$, the general solution for $\check{\varphi}(t)=\check{v}+\delta\check{\varphi}(t)$ looks as follows:
\begin{equation}
\check{\varphi}(t)=\check{v}\left[1+\frac{\check{a}}{m_{\check{h}}t}\cos(m_{\check{h}}t+\check{\alpha})
-\frac{1}{(m_{\phi}t)^2}\left(k_1- \lambda k_2 
\frac{\cos(2m_{\phi}t+\check{\beta})}{\sqrt{(\check{\omega}_0^2  -4)^2+\frac{16}{(m_{\phi}t)^2}}}\right)\right],
\label{solution near vac mir Higgs}
\end{equation}
where
\begin{equation}
\tan\check{\beta}=\frac{4}{(\check{\omega}_0^2 -4)m_{\phi}t}.
\label{tan mir beta}
\end{equation}
Estimates in the tree approximation give $\check{\omega}_0^2\approx 2.8\cdot 10^3$ and
$k_1 \approx  5\cdot 10^7$. Note that $\check{\beta}\approx 0$.
The second term in (\ref{solution near vac mir Higgs}) is the solution to the homogeneous equation (that is, in the absence of the inflaton contribution).
The integration constants $\check{a}$ and $\check{\alpha}$ could be determined by 
matching if we knew $\check{\varphi}(t)$ and $\dot{\check{\varphi}}(t)$ at the very 
end of stage 3. It can be expected that the amplitude of these free oscillations with friction
 caused by the expansion of the Universe should be sufficiently small so that in the interval 
$10^5\lesssim m_{\phi}t\lesssim 10^{10}$, the inflaton-governed behavior of $\delta\check{\varphi}$ is dominant.
 Nevertheless, due to the  dependence $\sim (m_{\check{h}}t)^{-1}$, for sufficiently 
large $t$ this term becomes dominant in  $\delta\check{\varphi}$, and contributions
 of $\frac{\delta\check{\varphi}(t)}{\check{v}}$  in $\delta\zeta$ and 
$U_{eff}|_{(\text{near vac})}$  also become dominant.
 In this paper, I will not study these processes, since the results obtained are sufficient for the first qualitative conclusion: 
the classical mirror Higgs field approaches its VEV  $\check{v}$ 
so that the monotonic increase of $\check{v}(1-\frac{k_1}{(m_{\phi}t)^2})$
 is accompanied by damping oscillations around these instantaneous values.

\subsubsection{Preliminary consideration of the evolution of the SM Higgs field at stage 4}

In the regime  described by Eqs.(\ref{phi eff pot vicinty vac}), (\ref{phi oscillating}) and discussed in the paragraph involving these equations, the energy density can be found as $\rho\approx \frac{1}{2}\dot{\phi}^2+\frac{1}{2}m_{\phi}^2\phi^2$. 
Then the expression (\ref{phase trans Higgs  stage 4}), the change in sign of which causes the beginning of the phase transition, is reduced to the following
\begin{equation}
\frac{1}{2}m_h^2-\frac{4}{3}\xi m_{\phi}^2\frac{1}{(m_{\phi}t)^2},
 \label{phase trans Higgs  stage 4 R}
\end{equation}
where Eqs.(\ref{v2 with choice}) and (\ref{m h2}) were used; it was taken into account that the parameter $c$ is chosen such that $c\approx\zeta_0$, and at stage 4 $\zeta=\zeta_0+\delta\zeta$ and $\delta\zeta<1.3\cdot 10^{-12}$.
 Therefore, the phase transition  for the SM Higgs field occurs when
$m_{\phi}t\geq 2\sqrt{\xi}\cdot 10^{11}$.
 The solution of Eq.(\ref{Higgs  eqom Ein cosm}) in stage 4 before and immediately after the phase transition is of particular interest, since it can include the oscillation mode responsible for production of SM particles. The study of this issue is  beyond the scope of this paper.

Turn now to the solution of Eq.(\ref{phi vicinty vac eq}) for the SM Higgs field in stage 4 after the phase transition in an  asymptotic regime when $m_{\phi}t\gg \sqrt{\xi}\cdot 10^{11}$. In this case we can use the representation (\ref{form of Higgs near vac}) for $\varphi(t)$.
Acting in a similar way to what we did when studying the asymptotic solution for the  mirror Higgs field, we note that the substitution of $\delta\zeta(t)$ into the expansion of  Eq.(\ref{phi vicinty vac eq}) near the vacuum not only leads to the interaction of the 
inflaton with the SM Higgs field (of the form $\phi^2\varphi$), but also an interaction arises between the  SM and the mirror classical Higgs fields of the form 
$\sim\varphi\check{\varphi}$.
Therefore, the behavior of the SM Higgs field can be influenced not only by the inflaton,
 but also by the mirror Higgs field. Using Eqs.(\ref{form of Higgs near vac}), 
(\ref{delta zeta 3b}), (\ref{phi oscillating}) and  (\ref{solution near vac mir Higgs}), 
we arrive at the equation
\begin{eqnarray}
&&\frac{d^2(\delta\varphi)}{dt^2}+\frac{2}{t}\frac{d(\delta\varphi)}{dt}+m_{h}^2\left[\delta\varphi
-v\frac{(1.25-9\lambda)\cdot 10^{10}}{(m_{\phi}t)^2}\right]
\nonumber
\\
&=&
-m_{h}^2 v\left[\left(1.25\cdot 10^{10}-6.2\cdot 10^{12}\lambda \right)\frac{\cos(2m_{\phi}t)}{(m_{\phi}t)^2}+1.8\cdot 10^3 \lambda\check{a}
\frac{\cos(m_{\check{h}}t+\check{\alpha})}{m_{\check{h}}t}\right],
\label{deltavarphi eq asymp}
\end{eqnarray}
Again neglecting terms decreasing faster than $t^{-2}$, the general solution of this equation reads
\begin{equation}
\varphi(t)=v\left(1+\frac{a}{m_h t}\cos(m_h  t+\alpha)+F(t)\right)
\label{Higgs sol vac}
\end{equation}
\begin{equation}
F(t)=\frac{0.9\cdot 10^9}{(m_{\phi}t)^2}
+\frac{5.3\cdot 10^{-10}}{(m_{\phi}t)^2}\lambda\frac{\cos(2m_{\phi}t+\beta)}{\sqrt{(\omega_0^2 -4)^2
+\frac{16}{(m_{\phi}t)^2}}}
+1.2\cdot 10^{-19}\frac{\check{a}}{m_{\check{h}}t}\lambda 
\frac{\cos(m_{\check{h}}t+\check{\alpha})}{\check{\omega}_0\sqrt{\check{\omega}_0^2
 +\frac{4}{(m_{\phi}t)^2}}},
\label{Higgs F(t) sol vac}
\end{equation} 
 where 
 \begin{equation}
\omega_0^2=\frac{m_h^2}{m_{\phi}^2}\approx 6.9\cdot 10^{-23}; 
\quad \tan\beta =\frac{4}{(\omega_0^2 -4)m_{\phi}t}\approx -\frac{1}{m_{\phi}t},
\label{tan omega 0}
\end{equation}
and used the estimates in the tree approximation similar to those we did after Eq.(\ref{tan mir beta}).
The integration constants $a$ and $\alpha$  could be determined by matching if we knew 
$\varphi(t)$ and $\dot{\varphi}(t)$  immediately after the phase transition.
 For now, one can formulate the second qualitative conclusion:
the classical SM Higgs field approaches its VEV $v$ so that the monotonic 
decrease of $v[1+0.9\cdot 10^9(m_{\phi}t)^{-2}]$ is accompanied by damping 
oscillations around these instantaneous values.

To better understand why the results obtained are fundamentally new, 
it is worth recalling that usually the transition of the SM Higgs field to the vacuum state 
 occurs {\em in the form of  oscillations of the Higgs field around its VEV}.
The solutions obtained show  that  in the considered model
the transition of the Higgs fields to their VEV's occurs  {\em in a radically different way}. 
As was formulated in the two above qualitative conclusions, 
the solutions for both the SM Higgs field and the mirror Higgs  field asymptotically
 approach their VEV's {\em in the regime of composition of monotonic displacements 
of their instantaneous values  and oscillations around these instantaneous values}.
Apparently, it is worth emphasizing once again that the described evolution 
at stage 4 occurs before the Higgs fields reach their VEV's.

\subsection{Vacuum: a new concept consistent with general physical principles and avoiding the cosmological constant problem}

As we begin to analyze the implications of the results formulated in the last  paragraph of the previous subsection,
 we must take into account that the values of $\zeta_0$ and $\mathcal{M}$ were obtained 
in Sec.VIII.B.1 using the vacuum energy postulate. 
In the studied model, the  vacuum energy postulate implies that the following conditions are fulfilled in the vacuum state: 1) the inflaton and Higgs fields have values at the minima of their TMT effective potentials; 2) the vacuum value of the scalar $\zeta$ satisfies the constraint; 3) the value of the vacuum energy density is selected based on some additional requirement not contained in the model. As we found out earlier, the implementation of these three items allows us to find the value of the integration constant $\mathcal{M}$,  the value of the scalar $\zeta=\zeta_0$ and the values of the classical Higgs fields in the vacuum.
 For the sake of simplicity, in Sec.VII.B.1,  the zero value of the vacuum energy density is chosen. Any other choice is also possible which will naturally lead to different physical results. For example, if the vacuum energy density  is chosen to be non-zero, but tiny $\Lambda_0$,  as in the current cosmological epoch, then Eq.(\ref{rho vac 0}) is modified. In this case we again use Eq.(\ref{Wmn}) but instead of equating $V_2-B/(1+\zeta)^2$ to zero, we must equate it to $\Lambda_0$.  Of course, this leads to corresponding tiny changes in  the values of $\zeta_0$ and $\mathcal{M}$.

In the light of  finding the previously unknown behavior of the Higgs fields at the  stage 4 of cosmological evolution, it becomes possible to understand what physical meaning has an additional requirement imposed on the value of the vacuum energy density. Asymptotic solutions to the Higgs field equations allow us to find out how the oscillation amplitudes of the classical Higgs fields (around their instanteneous monotonically shifting values) decrease due to the expansion of the universe. But we left aside the fact that the oscillatory regime of transition to a vacuum  is usually accompanied by particle  production, and thus the energy of scalar fields is converted into energy of matter and heats up the Universe. In the case of studied MESM, oscillations of the classical Higgs fields at stage 4 of cosmological evolution obviously cause oscillations of the effective masses of both SM fermions and mirror fermions with a frequency equal to the inflaton mass, or with frequencies that can be equal to the masses of the Higgs bosons. Investigation of the efficiency of these and other possible preheating mechanisms (such as e.g. in Refs.\cite{Kolb 1}-\cite{Riotto}) is beyond the scope of this paper.
But if effective mechanisms for preheating/reheating  exist, then besides the friction caused by the expansion of the Universe, the energy of the classical Higgs fields will also decrease due to a production of matter particles. Monotonic displacements of the classical Higgs fields continue until all the energy of the inflaton and Higgs fields is converted into the energy of the expanded Universe filled with matter.
The corresponding final states of the classical Higgs fields are their VEV's.  Therefore, {\em  vacuum is realized as the limiting state of the inflaton and classical Higgs fields, in which the energy stored in them becomes minimal, providing the maximum possible contribution of these fields to the increase in the entropy of the Universe} (by the time the vacuum is reached). In this context, the vacuum energy postulate appears to be quite justified: in the course of the described continuous process, {\em the system of scalar fields selects and stops in that state (from the continuous set of their instanteneous equilibrium values) which is most probable from the point of view of general physical principles}. 

Based on what we explained in Sec.VII.B.1, it is very important to emphasize once again that in the model under study, there is no need to fine-tune the parameters  to implement the vacuum energy postulate. Thus, it is possible to circumvent the cosmological constant  problem.

Let us return to the previously noted analogy between the temperature of the universe and the scalar $\zeta(x)$ from the point of view of their role in the second order phase transitions. Due to the constraint, the value of the scalar $\zeta(x)$ is the function of the inflaton and the Higgs fields at the same space-time point $x$. Therefore, in contrast to the temperature, $\zeta(x)$ reacts to any fluctuation of the inflaton $\phi(x)$ and Higgs fields $\varphi(x)$ and $\check{\varphi}(x)$ with a corresponding fluctuation $\delta\zeta(x)$. Thus, the decreasing scalar $\zeta(t)$, reacting to the oscillations of the inflaton and Higgs fields, acquires an oscillating addition, which is a function of these fields. Since the field equations contain $\zeta$, the feedback appears in the form of interactions that are not present in the underlying action.
 In particular, one of the results of this effect is the emergence of  interactions of the Higgs fields with the inflaton field (like in Eq.(\ref{L int eff mirr}) for $\phi^2\check{\varphi}$ and similar for $\phi^2\varphi$ coupling), which provide the possibility of the new concept of vacuum formulated above.

 I should note that within the framework of the presented model, there are doubts that oscillations of the classical SM Higgs field with a very small amplitude at Stage 4, as in Eqs.({\ref{Higgs sol vac}), (\ref{Higgs F(t) sol vac}), can be effective for the production of the SM Higgs bosons. It is also not obvious whether the mass oscillations of SM fermions caused by these oscillations of the classical SM Higgs field can lead to efficient  production of the SM fermions at Stage 4. In any case, the proposed concept of a vacuum can be acceptable only if it is possible to build a model that ensures efficient preheating of the Universe.

\subsection{Stage 5. Masses and mixing of the SM and mirror Higgs bosons in Minkowski space}

Let us continue with stage 5, which we have already begun  to study in Sec.IV.B (see Eqs.(\ref{Higgs doublets mesured})-(\ref{h mir eq})).
The constraint in  stage 5 can be obtained from Eq.(\ref{zeta stage 3}) by setting $\phi\equiv 0$ and using the representations    $\varphi =v+h$, 
$\check{\varphi}=\check{v}+\check{h}$ instead of $\varphi(t)$, \, $\check{\varphi}(t)$. Then for $\delta\zeta(h, \check{h})$ introduced in Eq.(\ref{delta zeta vac}) we get the following linear expansion  in quantum Higgs fields $h$ and $\check{h}$
\begin{eqnarray}
\delta\zeta(h, \check{h})&=&\frac{2(1+\zeta_0)^2}{\sqrt{V_2}(\sqrt{V_1}+\sqrt{V_2})}\left[1+b-\frac{(\sqrt{V_1}+\sqrt{V_2})^2}{2\sqrt{V_2}(\sqrt{V_1}+\sqrt{V_2})}\right]\left(\lambda\check{v}^3\check{h}+\frac{\zeta_0+b}{\zeta_0-b}\lambda v^3h\right)
\nonumber
\\
&\approx & \frac{1.6\cdot10^9\lambda}{M_P^4}\left(\check{v}^3\check{h}+\frac{\zeta_0+b}{\zeta_0-b} v^3h\right),
\label{delta z h mirr h}
\end{eqnarray}
where, again, as in stage 4, considering the case of zero vacuum energy density.

We can now  find some important properties of the Higgs fields in Minkowski space.
Inserting the found  dependence of $\delta\zeta$ on $h$ and $\check{h}$ into Eqs.(\ref{h eq}) and (\ref{h mir eq}), we obtain
\begin{equation}
 \frac{\partial V_{eff}(h;\zeta_0)}{\partial h}=2\lambda v^2\left(1-1.5\cdot 10^{-55}\lambda \right) h+\lambda(3vh^2+h^3)
-\Delta_{mix}\cdot  \lambda^2 v\check{v}\check{h},
\label{dVdh}
\end{equation}
 \begin{equation}
 \frac{\partial V_{eff}(\check{h};\zeta_0)}{\partial \check{h}}=2\lambda\check{v}^2\left(1+1.9\cdot 10^{3}\lambda \right) \check{h}+\lambda(3\check{v}\check{h}^2+\check{h}^3)+\frac{\zeta_0+b}{\zeta_0-b}\, \Delta_{mix}\cdot  \lambda^2v\check{v}h.
\label{dVdcheckh}
\end{equation}
\begin{equation}
\Delta_{mix}=\frac{(1+\zeta_0)^2}{\zeta_0-b}\, 10^9\,\frac{v^2\check{v}^2}{M_P^4}\approx 2.2\cdot 10^{-20},
\label{Delta mix}
\end{equation}
As seen, the squared mass of the SM Higgs field has a negligible correction compared to the standard expression $2\lambda v^2$. However, the mass $m_{\check{h}}$ of the mirror Higgs boson acquires an addition that can be significant
 \begin{equation}
 m_{ \check{h}}^2=2\lambda\check{v}^2\left(1+1.9\cdot 10^{3}\lambda \right) 
\label{m2 h mir}
\end{equation}
 Let us consider this issue at the  tree approximation where  $\lambda$ is the same in the SM and in the mirror world. From Eq.(\ref{dVdcheckh}), using the known values of $v$, $m_h$ and the predicted VEV $\check{v}=10^{14}GeV$ we obtain $m_{\check{h}}\approx 7.9\cdot 10^{14}GeV$, which is 53 times greater than the inflaton mass predicted 
by Eq.(\ref{mphi}). Although not directly related to the model under study, it is interesting to note that, as shown in  paper\cite{KLS}, the  production of superheavy particles with mass $M$  can be efficient during preheating 
if $M\gg m_{\phi}$.
 However, if  quantum corrections are taken into account, then these estimates may turn out to be incorrect, since there is no reason for the effective values of $\lambda$ in the SM and in the mirror world to remain equal.

Here I have limited myself to only linear in $h$ and $\check{h}$ contributions arising from $\delta\zeta(h, \check{h})$  in the derivatives of the TMT effective potentials, that is to quadratic contributions in the TMT effective potentials. Keeping the quadratic terms in the expansion of $\delta\zeta(h, \check{h})$, we can get changes in the trilinear  self couplings $\lambda_3h^3$ and $\check{\lambda}_3\check{h}^3$   in the TMT potentials. 
The corresponding correction in $\lambda_3$  as compared with the SM value at the tree level $\lambda_3^{SM}=\lambda v$ turns out to be much less than the correction that we observed in $m_h^2$ (see the first term on the right side of Eq.(\ref{dVdh})). Therefore, they cannot serve as  a beyond SM indication\cite{Degrassi 12}. The correction in 
$\check{\lambda}_3$ may be significant, but not observable.

The terms with $\Delta_{mix}$ in Eqs.(\ref{dVdh}) and (\ref{dVdcheckh}) describe the effect of mixing $h$ and $\check{h}$ resulting from SB of the MESM symmetries. Finding the appropriate  effective mixing potential using these terms turns out to be difficult due to the appearance of additional multipliers when performing the TMT procedure. 
 A more direct and controllable method consists in the expansion in powers of $h$ and $\check{h}$ of the effective TMT potential (\ref{Ueff}) near the vacuum with zero energy density. Then one can see that  the effective TMT potential contains the following term  describing the interaction between   $h$ and $\check{h}$:
\begin{equation}
U_{h\check{h}}^{(eff)}=4 \frac{(1+b)(\zeta_0+b)}{1+\zeta_0}\, \Delta_{mix}\cdot \lambda^2v\check{v}h \check{h}\approx
4\cdot 10^{-20}\lambda^2v\check{v}h \check{h}.
\label{U mix}
\end{equation}

Despite the weakness of the described effect, the very fact of its occurrence is of fundamental importance.  First, it demonstrates the special role that the constraint and the scalar $\zeta$  play in the theory. Due to the constraint, quantum fluctuations $h$ and $\check{h}$ of the Higgs fields  lead to fluctuations of the scalar $\zeta$, expressed in terms of $h$ and $\check{h}$, Eq.(\ref{delta z h mirr h}). And as a feedback, these oscillations of the scalar $\zeta$ lead to the appearance of the effects described above. Secondly, apparently this is the only interaction between the particles of the visible and mirror worlds that exists in the MESM. Third, this is the only interaction that violates  paritation conservation.  And if dark matter consists of mirror particles, then, in addition to gravity, it is also the only interaction between visible and dark matter.

In conclusion of this section, it is necessary to make an important general theoretical remark.
To obtain the expression (\ref{delta z h mirr h})  for $\delta\zeta(h, \check{h})$, we actually made a very non-trivial assumption that requires special discussion. The representation of a scalar field as a sum of a classical field (condensate) and a quantum scalar field describing excitation over a condensate is generally accepted in field theory and cosmology. This concept is understandable within the framework of quantum field theory with spontaneous symmetry breaking, where the role of the condensate is played by a nonzero VEV.
Recall, however,  that the scalar $\zeta(x)$  is defined by Eqs.(\ref{Phi}) and (\ref{zeta}) where both $\sqrt{-g}$ and $\Upsilon$ are geometric in nature. Therefore,  representing $\zeta(x)$ in the form of the sum of the constant (vacuum) value $\zeta_0$ and the combination (\ref{delta z h mirr h}) of the quantum Higgs fields we must assume that either $\sqrt{-g}$ or $\Upsilon$ (or both) have quantum components. Since we are confident that the gravitational field in the model under study remains classical, we must accept the idea that $\Upsilon$ and, therefore, the 4-form on the space-time manifold may have some properties of quantum fields. Although this idea is highly controversial, it seems to me that it deserves serious study.

\section{SUMMARY AND DISCUSSION}

In Sec.II, motivation was given for what I would call the restoration of "mathematical justice" in the basic concepts of field theory. The recognition of the need to use the volume measure $dV_{\Upsilon}$, Eq.(\ref{Phi}), which is an attribute of space-time manifold, along with the generally accepted metric-dependent volume measure $dV_{g}$, has far-reaching consequences, as demonstrated in this paper.
The construction of a model with spontaneously broken left-right symmetry became possible, since we took into account that  $\Upsilon$ and $\sqrt{-g}$ have opposite parity.

The use of both volume elements leads to the appearance of the  function $\zeta(x)\equiv\frac{\Upsilon}{\sqrt{-g}}$, which is a scalar relative to general coordinate transformations and a pseudoscalar relative to spatial reflections. $\zeta(x)$ is present in all field equations, and its value, due to the constraint, locally depend on the matter fields. The emergence of this function is critical for all TMT models, including the model studied in this paper. This can be verified once again by discussing the results of this work.

\subsection{MESM answers  to three fundamental questions of SM}

There are several  questions in SM that have no satisfactory answers.  Among them there are three puzzles, without an answer to which, in my opinion, it is extremely difficult to recognize SM as complete.
These are the following: 1) What is the nature of the giant interval of Yukawa coupling constants required to obtain the mass hierarchy of SM fermions; 
2) Why the mass term of the Higgs field potential has a "wrong" sign; 
3) What is the origin of tiny neutrino masses. 

It is important to note that preserving all symmetries, one can simplify the model by choosing in the underlying action the same sign in front of $\zeta$ in all Lagrangian terms of the SM fields (say plus) and the same sign (minus) in front of $\zeta$ in all Lagrangian terms of mirror fields. After spontaneous symmetry breaking, such a model turns out to be somewhat similar to the model studied in Refs.\cite{Foot Volkas}-\cite{Foot 4}: mirror counter-partners have not only the same quantum numbers, but also the same masses (however, it should be noted that, unlike the models \cite{Foot Volkas}-\cite{Foot 4}, there is no need to postulate the presence of mixing of some mirror counter-partners  in the unerlying action to break the left-right symmetry).
 The cost of this simplification of the model turns out to be the impossibility of finding satisfactory answers to the listed open questions of SM.

The answers that the  MESM can give to these puzzles have been explored in the present paper. 
But there are some aspects that need to be discussed in order to better understand how this became possible.

\subsubsection{Some important aspects of the implementation of the idea of the  universal Yukawa coupling constant}

As we have seen, choosing a universal and quite reasonable value of the Yukawa coupling constant 
for charged leptons and up quarks, it is possible to obtain the masses of all these SM fermions
 and to predict the masses of their mirror partners. This turns out to be possible due to the appropriate
 choice of signs in front of $\zeta$ in the linear combinatins $b_l\pm\zeta$ for charged leptons and 
$b_q\pm\zeta$ for up quarks. Of course, the suitable values of the parameters  
$b_l$ \, ($l=e, \mu, \tau $), $b_q$ \, ($q=u, c, t$) are also selected. If, for example, we choose 
the same universal Yukawa coupling constant for charged leptons and up quarks
 $y^{(ch)}=y^{(up)}=10^{-3}$ then six parameters  $b_l$,  $b_q$ have values in the interval  
$-0.242\lesssim (b_l, b_q)\lesssim 0.247$. Comparing this with the need to adopt in SM that
 the Yukawa coupling constants have a scatter of values that differ by 5 orders of magnitude,
 it can be argued that in MESM it is possible to overcome the problem of the hierarchy of fermion masses.

The origin of the correlations between the signs in front of $\zeta$ is explained in detail in Appendix A.

 In the underlying Lagrangian for charged SM leptons, Eq.(\ref{L 3}),
a linear combination $b_l+\zeta$ was chosen  in the kinetic terms and $b_l-\zeta$ in 
terms describing the Yukawa couplings. In order to guarantee  the (mirror$\times$parity)
 invariance of MESM, in the underlying Lagrangian for mirror charged leptons, Eq.(\ref{L mirror 3}), 
we were forced to choose  opposite signs.

To implement the idea of the universal Yukawa coupling constant for up quarks, we were forced to 
choose the underlying Lagrangians with signs in $b_q\pm\zeta$ opposite to those that were in the case of charged leptons.

There are no obstacles to using the idea of the universal Yukawa coupling constant to obtain the masses
 of the down quarks, in the same way as was done for the up quarks. But this would require accounting
 for mixing, which would lead to the appearance of a large number of additional details that are not directly
 related to the main ideas of this paper. In the future, this, as well as the inclusion of three generations of neutrinos,
 will need to be done.

\subsubsection{Miracle of "wrong" sign mass term in the Higgs potential}

When constructing the  Higgs fields sector of MESM, Eq.(\ref{S-H 12}), I abandoned the standard requirement
 for mass terms $\sqrt{-g}m^2|\Phi|^2$ and $\sqrt{-g}m^2|\check{\Phi}|^2$  to have “wrong” signs. 
But to ensure the (mirror$\times$parity) invariance  of the underlying MESM Lagrangian density,
 some terms containing  $\Upsilon$ are chosen with "wrong" signs.
All this leads to a non-trivial structure of the underlying action (\ref{S-H 12}), the role of which apparently deserves additional clarification.
 To start this discussion, it is convenient to introduce the following notations
\begin{equation}
 S_H=\int \sqrt{-g}d^4x(\mathbb{L}_{H}+\mathbb{L}_{\check{H}}),
\label{S HH}
\end{equation}
\begin{equation}
\mathbb{L}_{H}=(b+\zeta) 
\left(g^{\alpha\beta}\left(\mathcal{D}_{\alpha}\Phi\right)^{\dag}\mathcal{D}_{\beta}\Phi
 -\lambda |\Phi|^4\right)
 -\left(b-\zeta\right)m^2 |\Phi|^2
\label{Lagr H}
\end{equation}
\begin{equation}
\mathbb{L}_{\check{H}}=(b-\zeta) 
\left(g^{\alpha\beta}\left(\mathcal{D}_{\alpha}\check{\Phi}\right)^{\dag}\mathcal{D}_{\beta}\check{\Phi}
 -\lambda |\check{\Phi}|^4\right)
 -\left(b+\zeta\right)m^2 |\check{\Phi}|^2.
\label{Lagr mir H}
\end{equation}
As we already know, in the model studied in this paper, throughout the evolution of the universe, from inflation to the transition to vacuum, the condition 
$\zeta>b$ is satisfied. Therefore, in the underlying action for the SM Higgs field, the mass term $\left(b-\zeta\right)m^2 |\Phi|^2$ always has  a "wrong" sign, and the first two terms have "correct" signs. In contrast, in the  underlying action
for the mirror Higgs field, the mass term has a "correct" sign, but the first two terms have  "wrong" sign. However, after varying the action (\ref{S HH}) with respect to $\Phi$ and $\check{\Phi}$ and passing to the Einstein frame, we must bring the equations to the form where the left-hand side provides the canonical Green's functions. To do this, we must divide the found equations for the SM and mirror Higgs fields by $(b+\zeta)$ and $(b-\zeta)$, respectively. As a result, the equations of the Higgs fields in the Einstein frame acquire the standard form with the "wrong" sign mass terms.

The choice of signs in front of $\zeta$ in $\mathbb{L}_{H}$ and $\mathbb{L}_{\check{H}}$ was motivated by a similar correlation that was found and applied to fermions. The true source of this in the case of Higgs fields remains unclear. But compared to the mystery of the "wrong" sign of the Higgs mass term in SM, the proposed structure of the Higgs Lagrangians in MESM seems to be much more attractive.

\subsubsection{Problem of tiny neutrino masses}

  The upper  bound\cite{ParticleDataGroup} on the SM neutrino mass $m_{\nu}\lesssim 1.1eV$ means that the Yukawa coupling constant must be incredibly small: $y^{(\nu)}_{SM}\lesssim 6\cdot 10^{-12}$. As shown in Sec.VI.A.4, in MESM, when choosing the value of the neutrino Yukawa coupling constant $y^{(\nu)}$ in the underlying action, we find arbitrariness in a fairly wide range.  For the mass of a mirror neutrino $m_{\check{\nu}}$ do not exceed  $10^{17}GeV$, it is necessary that $y^{(\nu)}\lesssim 10^{-4}$. If $m_{\check{\nu}}\sim 10^{13}GeV$ then $y^{(\nu)}\sim 10^{-6}$
However, if there is a mirror neutrino lighter than charged mirror leptons,  $m_{\check{\nu}}\lesssim 10^{10}GeV$, then $y^{(\nu)}$ must  be very small: $y^{(\nu)}\lesssim 10^{-8}$. It is important that for all values of $y^{(\nu)}$ in the described interval, the corresponding values of the parameter $b_{\nu}$ are very  close to $\zeta_0$.

In conclusion, it is probably worth noting once again that although the tiny mass of an ordinary neutrino is associated with the existence of a heavy mirror neutrino, in MESM this mechanism is common to all particles: not only all SM fermions, but also all SM bosons have superheavy mirror partners. Therefore, although the general idea is similar to the see-saw mechanism, its implementation differs in principle: there is no need specifically for neutrino to postulate the existence of a sterile heavy neutrino.

Summarizing the details of how the three problems discussed above are solved in MESM, it is important to emphasize once again that this was done at close  values of the parameters $b$,  $b_l$, ($l=e, \mu,  \tau$),  $b_{\nu}$,  $b_{q}$, ($q=u, c, t$).

\subsection{Cosmology and vacuum state}

In order to provide the simplest,  single-field slow-rolling inflation that guarantees good agreement with recent Planck observations, in this paper we have used the T-model\cite{KL T-model}. But the implementation of this inflationary scenario within the framework of the TMT  turns out to be impossible without a very nontrivial change in our ideas about which vacuum-like terms can be in the underlying  action. In addition to the usual vacuum-like contribution to the action in the form $-\int d^4x\sqrt{-g}V_1$, the vacuum-like term $-\int d^4x\sqrt{-g}\zeta^2V_2$ associated with the volume element $dV_{\Upsilon}$, also required. The presence of {\bf both} of these vacuum-like terms allows one to obtain the  TMT effective potential of the T-model, in which, as is known, the cosmological constant is assumed to be zero. 
To implement the latter in the constructed TMT model, {\bf both} of these vacuum-like terms are also required. This fundamentally distinguishes our model from the usual ones where there is only one vacuum parameter of dimension $(mass)^4$ in the form of a cosmological constant. It should be emphasized 
once again that the introduction of two vacuum-like parameters is possible due to the presence of two measures of volume.

In addition, the  choice of  the parameters $V_1^{1/4}$ and $V_2^{1/4}$ of the order $\sim 10^{16}GeV$, which is dictaited by the need to satisfy the Planck observations in the context of the T-model inflation, turns out to be very important  also for the proposed particle model. 
 Indeed, due to this,  the obtained values of $\zeta$ in the vacuum and the integration constant ${\mathcal M}$ are insensitive to the details of the studied particle physics model (see Sec.VII.B.1). This feature of the "gravity $+$ inflaton $+$ particle fields" model plays a decisive role in the possibility of constructing the  mirror-extended electroweak standard model. However, when generalizing the developed approach, for example, to the grand unified theory (GUT), a certain correlation may arise between the cosmological parameters and the parameters of GUT.

As far as the problem of the cosmological constant is concerned, the TMT models are radically different from the rest of the models. This is primarily due to the inherent structural feature of TMT, which implies that gravity is fully accounted for. Therefore, applying the action principle  and solving the equations, we cannot ignore  the cosmological constant  problem or push it aside in the hope that it will be solved by some alternative method or theory.

   The definition of vacuum of the  inflaton and classical Higgs fields  and its energy density was one of the main topics of the present paper. In Sec.VII.B.1, as a preliminary tool, we started with the formulation of the vacuum energy postulate, which changes the usual definition of vacuum in a significant way. Then, in Secs.VIII.B and VIII.C, we investigated the unusual features of the transition to a vacuum. The results obtained mean that the vacuum is the limiting state, in the transition to which the energy of the inflaton and classical Higgs fields is converted into the energy of the expanded universe, filled with matter in the maximum possible amount.

The unusual scenario of the transition to vacuum is one of the key  results of the model: the approach of the classical Higgs fields to the vacuum state occurs in the form of monotonic changes in their instantaneous equilibrium values, accompanied by oscillations around these instantaneous equilibrium values; this complex behavior is a superposition of free modes and forced modes governed by inflaton. This feature, like many other results of the model, have a common source - a pseudoscalar $\zeta=\Upsilon /\sqrt{-g}$ of geometric nature.

If further study of the model reveals the existence of an effective mechanism for preheating the Universe, the proposed new concept of understanding the vacuum of the classical Higgs fields, including the discovered new mechanism of how it is reached, would mean that the problem of the cosmological constant is absent in its usual sense. It is important to emphasize that this is achieved without the need for fine tuning of the model parameters. 

The investigation of quantum corrections in the context of the studied MESM is important both for testing the renormalizability of the model and its consistency. Since the SM is renormalizable, and the particle model in the mirror world has a structure identical to that of the SM, we could argue that the theory is renormalizable. It can be shown that the quantum corrections to the MESM do not violate the vacuum-related results that were obtained at the tree level. This statement is true both in relation to the applicability of the vacuum energy postulate and in the substantiation of the new concept of vacuum. In addition, in MESM there are possibilities for new interesting results in the problem of electroweak instability of vacuum at
and after inflation. However, proof of these claims would require a significant increase in the length of the paper  and it will be done elsewhere.

\subsection{Some topics for further research}

In this paper, I did not touch on a number of important issues without which the physical picture based on the MESM is incomplete. Here I list in my opinion the main ones.

1)  Extention of MESM by  adding QCD, which also implies the development of QCD for mirror quarks.

2) Study of physical processes and possible bound states of mirror particles. The scale of masses of mirror particles depends mainly on the value of the parameter $b$. In this paper, I have chosen the value of $b$ given by 
Eq.(\ref{bh choice}) as an example, and all other numerical estimates have been derived from this example. Further study of the problems of  particle physics and cosmology should lead to restrictions on the parameters.

3) As shown in Sec.VIII.D, there is practically no interaction of the SM with mirror particles, except for the gravitational interaction. Therefore, {\em DM may be composed of superheavy mirror particles predicted by MESM}. For example, a gas of stable mirror superheavy nonrelativistic neutrinos can be one of the DM components. However, it is also necessary to study which bound states of other mirror particles can constitute DM. It should be noted that the mass spectrum of mirror fermions predicted in this paper  is close in order of magnitude to the masses of hypothetical superheavy long-lived dark matter particles (wimpzillas) \cite{Kolb 2}, \cite{Wimpzillas}. MESM allows one to provide a more accurate correspondence of masses simply by a small change in parameter $b$.
Note that the mirror fermions predicted by the proposed Mirror-Extended Standard Model cannot be responsible for ultrahigh-energy cosmic ray events\cite{ultra high cosmic rays 1}-\cite{ultra high cosmic rays final}, since in their decay or annihilation processes only unobservable mirror photons can arise.

4) A comprehensive study of the cosmological process of transition to a vacuum state, which implies:

a) Solving cosmological equations for the Higgs fields at the end of inflation and at stages 3 and 4 (see Secs.VIII.A,B) without taking into account preheating. 

 b) Possible mechanisms of preheating (as the production of both SM and mirror particles).

 c) Study of the back reaction of the produced matter on cosmological evolution, and in particular on the processes arising during the transition to vacuum.
In particular, it may turn out that in the context of MESM, the results of Ref.\cite{GK6} can explain Dark Energy as a mirror neutrino effect.

5) The possible impact of the scalar $\zeta(x)$ on the perturbation spectrum during inflation may turn out to be significant.

6) In this paper, we have limited ourselves to a  single-field $\alpha$-attractor inflation model. But TMT does not create  significant obstacles to the implementation of a multi-field $\alpha$-attractor model\cite{multi}. For example, the double inflation model of Ref.\cite{Maeda} could be fitted instead of the simplest T-model used in this paper.

7) In this paper we investigated some aspects of  the cosmological scenario that arises if the constant of non-minimal coupling to gravity
$\xi>0$, see Eq.(\ref{S nonmin}) and Sec.VIII.A. It would be interesting to check the possibility of an alternative cosmological scenario that could be obtained if the non-minimal coupling constant $\xi<0$. In this case, there is no need for inflaton at all, and inflation is driven by the SM and  mirror Higgs fields, since the $SU(2)\times U(1)$ gauge symmetry is broken at the very beginning of inflation.  Preliminary estimates show that, as a result of the latter, the realization of the Higgs inflation-type scenario explored in papers\cite{Higgs infl 1}, \cite{Higgs infl 2} may be possible without the need for a strong non-minimal coupling.

\section{Acknowledgements}

I am grateful to Matej Pavsic, whose comments prompted me to improve the description of some aspects of the model.

\appendix
\section{Basic ideas for constructing a fermionic sector with parity conservation. 
 A toy model }

In this appendix I am going to demonstrate the idea of how in  TMT it is possible to build an electroweak interaction with spontaneously broken parity.
At the same time, this model allows one to demonstrate some technical aspects of working with fermions in TMT, without complicating them with irrelevant details arising in a realistic model.
 This will be done by means of a toy model which contains  the left  SU(2) electron doublet  $L$ and the right electron singlet $e_R$
\begin{equation}
L=\frac{1-\gamma_5}{2} \left(\begin{matrix} \nu_e \\ e \end{matrix}\right);
\qquad e_R=\frac{1+\gamma_5}{2} e
\label{GWS fermions}
\end{equation}
and also the model contains  the mirror fermions - the right  SU(2) electron dublet $R$ and the left electron singlet $e_L$
\begin{equation}
R=\frac{1+\gamma_5}{2} \left(\begin{matrix} \nu_e \\ e \end{matrix}\right);
\qquad e_L=\frac{1-\gamma_5}{2} e 
\label{mirror fermions}
\end{equation}
It is assumed that the mirror symmetric pairs of fermions have the same $SU(2)\times U(1)$ quantum numbers.
But in contrast to a more realistic model studied in the main text of the paper, the toy model under consideration contains only one set of gauge fields:
isovector $\bf{A}_{\mu}$ and isoscalar $B_{\mu}$, and only one Higgs $SU(2)$ doublet
\begin{equation}
\Phi= \left(\begin{matrix} \phi^{+} \\ \phi^{0} \end{matrix}\right),
\label{Higgs doublet}
\end{equation} 
that is the toy model does not contain mirror bosons.
The assumption that  the appropriate mirror symmetric pairs of fermions have the same $SU(2)\times U(1)$ quantum numbers  obviously leads to the natural transformation laws: $\hat{\mathcal P}L=R$, $\hat{\mathcal P}R=L$,   $\hat{\mathcal P}e_R=e_L$,  $\hat{\mathcal P}e_L=e_R$.

The postulated exact left-right symmetry of the fermion sector in the underlying action allows to start  without splitting the Dirac spinors by chirality, i.e. to proceed formally with  
\begin{equation}
\Psi =L+R= \left(\begin{matrix} \nu_e \\ e \end{matrix}\right);
\qquad e=e_R+e_L
\label{GWS RL fermions}
\end{equation}

A parity  features of  $\Upsilon $ and $\zeta$, Eq.(\ref{Pzeta}),  open new posssibilities for extention of the SM to a model which possesses a spontaneosly broken  space reflection  symmetry. The novelty  is that multiplying a psuedoscalar and an axial vector  by the pseudoscalar $\zeta$ we get a scalar and vector respectively. Indeed, in addition to the available vector currents
\begin{equation}
\overline{\Psi}\gamma^{\mu}\Psi = \overline{L}\gamma^{\mu}L +  \overline{R}\gamma^{\mu}R; \quad \overline{e}\gamma^{\mu}e = \overline{e}_R\gamma^{\mu}e_R +\overline{e}_L\gamma^{\mu}e_L,
\label{vector currents 1}
\end{equation}
one more type of vectors can be constructed from the axial currents
\begin{equation}
\zeta\overline{\Psi}\gamma^{\mu}\gamma_5\Psi =\zeta\left( -\overline{L}\gamma^{\mu}L +  \overline{R}\gamma^{\mu} \check{R}\right); \quad \zeta\overline{e}\gamma^{\mu}\gamma_5 e = \zeta\left(\overline{e}_R\gamma^{\mu}e_R -\overline{e}_L\gamma^{\mu}\check{e}_L\right) 
\label{vector currents 2}
\end{equation}
Here $\gamma^{\mu}=V_{k}^{\mu}\gamma^{k}$ and $V_{k}^{\mu}$ are vierbeins which must be included here because proceeding with two volume elements it is necessary to ensure that the constraint is satisfied. This suggests that the underlying action involves gravity.
Besides, in addition to the Hermitian, gauge and $\hat{\mathcal P}$ invariant scalar
\begin{equation}
\bar{\Psi}\Phi e+\bar{e}\Phi^{\dag}\Psi
\label{scalar 1 with Higgs}
\end{equation}
 one can construct one more Hermitian, gauge and $\hat{\mathcal P}$ invariant scalar
\begin{equation}
\zeta\left(\bar{\Psi}\gamma_5\Phi e-\bar{e}\gamma_5\Phi^{\dag}\Psi\right),
\label{scalar 2 with Higgs}
\end{equation}
where the minus sign is dictated  by hermiticity.

Now we are able to construct a generally coordinate invariant action which is also $SU(2)\times U(1)$ gauge and $\hat{\mathcal P}$ invariant. To do this, use linear combinations of  vectors, Eqs. (\ref{vector currents 1}) and (\ref{vector currents 2}), and scalars, Eqs.(\ref{scalar 1 with Higgs}) and (\ref{scalar 2 with Higgs}), in a form of the action (\ref{S2}) adapted to the case of fermions.
Namely, in the fermion sector, the requirements of  $SU(2)\times U(1)$ gauge invariance suffice, as usual, to impose on Lagrangian scalars $L_1$ and $L_2$ (recall the general structure (\ref{S2}) of action in TMT); however, {\bf the requirement of $\hat{\mathcal P}$ -invariance is imposed on the corresponding Lagrangian densities} $\mathcal{L}_1=\sqrt{-g}L_1$  and  $\mathcal{L}_2=\Upsilon L_2\equiv\zeta\sqrt{-g}L_2$.   
 The implementation of this idea in our toy model leads to the following action:
\begin{equation}
S_{(toy, total)}=\int \left(\mathcal{L}_{(1,total)}+\mathcal{L}_{(2,total)}\right)d^4x 
\label{toy, total}
\end{equation}
where
\begin{eqnarray}
\mathcal{L}_{(1,total)}=
 b_e\sqrt{-g}\bigg\{\frac{i}{2}\bigg[\overline{\Psi}\gamma^{\mu}\nabla_{\mu}\Psi  + \overline{e}\gamma^{\mu}\nabla_{\mu}e 
                                                    - \Big(\nabla_{\mu}\overline{\Psi})\gamma^{\mu}\Psi + 
(\nabla_{\mu}\overline{e})\gamma^{\mu}\gamma_5e\Big)\bigg]
\nonumber
\\
-y^{(e)}\big(\overline{\Psi}\Phi e + \overline{e}\Phi^{\dag}\Psi\big)\bigg\}
\label{L 1 total}
\end{eqnarray}
\begin{eqnarray}
\mathcal{L}_{(2,total)}=
  -\zeta\sqrt{-g}\bigg\{\frac{i}{2}\bigg[\overline{\Psi}\gamma^{\mu}\gamma_5\nabla_{\mu}\Psi  - 
\overline{e}\gamma^{\mu}\gamma_5\nabla_{\mu}e 
                                - \Big((\nabla_{\mu}\overline{\Psi})\gamma^{\mu}\gamma_5\Psi -
(\nabla_{\mu}\overline{e})\gamma^{\mu}\gamma_5e\Big)\Big]
\nonumber
\\
-y^{(e)}\big(\overline{\Psi}\gamma_5\Phi e - \overline{e}\gamma_5\Phi^{\dag}\Psi\big)\bigg\}
\label{L 2 total}
\end{eqnarray}
To simplify the presentation, I have omitted the corresponding neutrino terms. 
They will be taken into account in the main text of the paper.
The subscript 'total' indicates that the Lagrangian densities are constructed from 
Dirac spinors without splitting (\ref{GWS RL fermions}) into left and right fermions. 
Obviously, these Lagrangian densities are $\hat{\mathcal P}$-invariant.
 The operator 
\begin{equation}
\nabla_{\mu}=\mathcal{D}_{\mu}+\frac{1}{2}\omega_{\mu}^{cd}\sigma_{cd}; 
\qquad  \mathcal{D}_{\mu}=\partial_{\mu}
-ig\mathbf{\hat{T} A}_{\mu}-i\frac{g'}{2}\hat{Y}B_{\mu},
\label{nabla L R}
\end{equation}
involves spin-connection $\omega_{\mu}^{ab}$  and, as usually, $\mathbf{\hat{T}}$ 
stands for the three generators of the $SU(2)$ group and $\hat{Y}$ is the generator 
of the U(1) group; $\bf{A}_{\mu}$ and $B_{\mu}$ are the gauge isovector and isoscalar respectively. 

The explanations for the choice of signs that I made in Eq.(\ref{L 2 total}) are necessary. 

In the second line, opposite signs  before  the terms in parentheses are due to the hermiticity 
requirement (see Eq.(\ref{scalar 2 with Higgs})).
In the first line, the opposite signs in the contributions to the kinetic terms from isodublet and isosinglet 
must be chosen to guarantee  the canonical structure of the  current for a free electron.

When choosing a common sign in $\mathcal{L}_{(2,total)}$, it is necessary to take into account
 that $\zeta$, like $\Upsilon$, with equal right can be both positive or negative, depending on
 the orientation of space-time. Therefore, there is arbitrariness in the choice of the common sign
 in $\mathcal{L}_{(2,total)}$. As will be shown in the main text of the paper, due to the initial 
conditions in the inflation model, an orientation with $\zeta>0$ is spontaneously fixed.
 In this toy model, I have chosen a common minus sign in $\mathcal{L}_{(2,total)}$ to illustrate
 below the implications of this choice when passing to a description in terms of left and right fermions.
In the realistic MESM studied in Secs.V and VI, this allows the idea of a universal Yukawa coupling constant
 for three generations of charged leptons to be realized. To implement this idea with up-quarks,
 one needs to choose the opposite sign.

 Using  decompositions (\ref{GWS RL fermions}) into sums with opposite chirality, the definition of 
 $\zeta$, Eq. (\ref{zeta}), and the algebra of Dirac matrices,
the action (\ref{toy, total})-(\ref{L 2 total})   can be rewritten in the following {\em equivalent form}
\begin{equation}
S_{(toy, total)}=\int \left[\mathbb{L}_{L,R}^{(e)}+\mathbb{L}_{\check{R},\check{L}}^{(\check{e})}\right]\sqrt{-g}d^{4}x,
\label{S= LR + RL}
\end{equation}
where the Lagrangian $\mathbb{L}_{L,R}^{(e)}$  stands for the left isodoublet and right isosinglet, 
while  $\mathbb{L}_{\check{R},\check{L}}^{(\check{e})}$  stands for the right isodoublet 
and left isosinglet and they are as follows
\begin{eqnarray}
\mathbb{L}_{L,R}^{(e)}&=&
  (b_e+\zeta )
\frac{i}{2}\left[\overline{L}\gamma^{\mu}\nabla_{\mu}L -(\nabla_{\mu}\overline{L})\gamma^{\mu}L+
\overline{e}_R\gamma^{\mu}\nabla_{\mu}e_R
-(\nabla_{\mu}\overline{e}_R)\gamma^{\mu}e_R\right]
\nonumber\\
&-& (b_e-\zeta )y^{(e)}\left[\overline{L}\Phi e_R +
\overline{e}_R\Phi^{\dag}L\right], 
 \label{f-L R}
\end{eqnarray}

\begin{eqnarray}
\mathbb{L}_{\check{R},\check{L}}^{(\check{e})}&=&
  (b_e-\zeta )
\frac{i}{2}\left[\overline{\check{R}}\gamma^{\mu}\nabla_{\mu}\check{R}-
(\nabla_{\mu}\check{R})\gamma^{\mu}\check{R} +
\overline{\check{e}}_L\gamma^{\mu}\nabla_{\mu}\check{e}_L 
 -(\nabla_{\mu}\overline{\check{e}}_L)\gamma^{\mu}\check{e}_L\right]
\nonumber\\
&-& (b_e+\zeta )y^{(e)}\left[\overline{\check{R}}\Phi \check{e}_L +
\overline{\check{e}}_L\Phi^{\dag}\check{R}\right].
 \label{f-R L mirror}
\end{eqnarray}
Besides the  $SU(2)\times U(1)$ gauge invariance, the Lagrangian $\mathbb{L}_{L,R}^{(e)}+\mathbb{L}_{\check{R},\check{L}}^{(\check{e})}$
possesses  the $\hat{\mathcal P}$ -invariance, and this is due to the parity property of  $\zeta$:  $\hat{\mathcal P}\zeta=-\zeta$.

As we see, 
 the splitting of the initial action (\ref{toy, total})-(\ref{L 2 total}) into the sum of actions
 with Lagrangians  (\ref{f-L R}) and 
 (\ref{f-R L mirror}) for fermions with opposite chirality automatically entails the occurrence of 
opposite signs, with which  $\zeta$ is present in  corresponding terms in (\ref{f-L R}) and  
(\ref{f-R L mirror}). This means that decompositions (\ref{GWS RL fermions}) of the
 Dirac spinors $\Psi$ and $e$    is accompanied by the appearance of {\em a new quantum number} - the signs 
with which $\zeta$ is present in  appropriate terms of the fermion Lagrangians with opposite chirality. 
The latter circumstance looks as if the mirror-symmetric fermion partners being in the same 
space-time manifold would ''perceive'' it as having the opposite orientation. Since the presence 
of the pseudoscalar $\zeta$ in the action (\ref{toy, total})-(\ref{L 2 total}) ensures its parity invariance, 
I propose to call this new quantum number   "paritation" ($=$ parity $+$ orientation). In the toy model
under study, let us   assign $paritation = 1$ for regular fermions, i.e. for left isodoublet and right isosinglet; 
and correspondingly $paritation =-1$ for mirror fermions, i.e. for right isodoublet and left isosinglets. 
Consequently,  when varying the action with respect to electron field, the electron component of the
 isodoublet $L$ and isosinglet $e_L$, as having opposite paritation, should be considered as independent; 
and similarly for the electron component of the isodoublet $R$ and isosinglet $e_R$. For this reason,  
to  explicitly indicate that we are dealing with different physical degrees of freedom, for fermions that 
are mirror symmetrical with respect to regular fermions  we will use the notations $\check{R}$,
 $\check{e}_L$, $\check{\nu}_L$.

\end{document}